\documentclass[showpacs,aps,prd]{revtex4}
\input epsf

\begin{document}

\title{$\{Q\bar{s}\}\{\bar{Q}^{(')}s\}$ molecular states in QCD sum rules \footnote{Supported by the National Natural Science
Foundation of China (10675167 and 10975184).}}
\author{Jian-Rong Zhang and Ming-Qiu Huang}
\affiliation{Department of Physics, National University of Defense
Technology, Hunan 410073, China}

\begin{abstract}
We systematically investigate the mass spectra of $\{Q\bar{s}\}\{\bar{Q}^{(')}s\}$
molecular states in the framework of QCD sum rules. The
interpolating currents representing the molecular states are
proposed. Technically, contributions of the operators up to
dimension six are included in operator product expansion (OPE). The
masses for molecular states with various
$\{Q\bar{s}\}\{\bar{Q}^{(')}s\}$ configurations are presented.
The result $4.36\pm0.08~\mbox{GeV}$ for the $D_{s}^{*}\bar{D}_{s0}^{*}$
molecular state is consistent with
the mass $4350^{+4.6}_{-5.1}\pm0.7~\mbox{MeV}$ of the newly observed $X(4350)$,
which could support $X(4350)$ interpreted as a $D_{s}^{*}\bar{D}_{s0}^{*}$
molecular state.
\end{abstract}
\maketitle

\section{Introduction}\label{sec1}
The field of heavy hadron spectroscopy is experiencing a rapid
advancement mainly propelled by the continuous observations of
hadronic resonances, for example, $X(3872)$ \cite{X3872}, $Y(3930)$
\cite{Y3930}, $Y(4260)$ \cite{Y4260}, $Z(3930)$ \cite{Z3930},
$X(3940)$ \cite{X3940}, $Z^{+}(4430)$ \cite{Z4430},
$Z^{+}_{1}(4050)$ \cite{Z4050}, $Z^{+}_{2}(4250)$ \cite{Z4050},
$Y(4140)$ \cite{Y4140} etc. (for experimental reviews, e.g., see
\cite{Swanson,PDG}), some of which are not easy to accommodate
within the quark model picture and may not be conventional
charmonium states. Masses for some of these hadrons are very close
to the meson-meson thresholds, for which are interpreted as possible
$\{Q\bar{q}\}\{\bar{Q}^{(')}q\}$ molecular candidates in \cite{theory,theory1,theory2,theory-Y3930,Liu}.
For instance, the charmonium-like state $Y(3930)$ has been interpreted as a $D^{*}\bar{D}^{*}$ molecular
state \cite{theory-Y3930,Liu}. Considering the $SU(3)$ symmetry of the light flavor quarks, there may also exist and
have a rich spectroscopy for $\{Q\bar{s}\}\{\bar{Q}^{(')}s\}$ molecular states
acting as the corresponding partners of $\{Q\bar{q}\}\{\bar{Q}^{(')}q\}$ molecular states.
In fact, some authors have deciphered the newly observed $Y(4140)$ as a $D_{s}^{*}\bar{D}_{s}^{*}$ molecular state \cite{Liu,theory3},
just as the molecular partner of $D^{*}\bar{D}^{*}$.
Furthermore, QCD itself does not exclude the existence of $\{Q\bar{s}\}\{\bar{Q}^{(')}s\}$
molecular states besides conventional mesons and baryons, so studies
of them may deepen one's understanding of the strong
interaction. On all accounts, it is interesting to study mass
spectra for the $\{Q\bar{s}\}\{\bar{Q}^{(')}s\}$ molecular states. However, it is far from clear how
to generate hadron masses from first principles in QCD since it is
highly noperturbative in the low energy region where futile to
attempt perturbative calculations, and then one has to treat a
genuinely strong field in nonperturbative methods. Under such a
circumstance, one could resort to QCD sum rule \cite{svzsum} (for
reviews see \cite{overview,overview1,overview2,overview3} and
references therein), which is a comprehensive and reliable way for
evaluating the nonperturbative effects. Up to now, there have been some works testing the $D_{s}^{*}\bar{D}_{s}^{*}$ state
from QCD sum rules \cite{zgwang,Nielsen,zhang}.
Presently, we extend the work on $(Q\bar{s})^{(*)}(\bar{Q}s)^{(*)}$ molecular states \cite{zhang}
to various
$\{Q\bar{s}\}\{\bar{Q}^{(')}s\}$ molecular states.

The paper is organized as follows. In Sec. \ref{sec2}, QCD sum rules
for the molecular states are introduced, and both the
phenomenological representation and QCD side are derived, followed
by the numerical analysis to extract the hadronic masses in Sec. \ref{sec3},
and a
brief summary  in Sec. \ref{sec4}.
\section{Molecular state QCD sum rules}\label{sec2}
\subsection{interpolating currents}
Following the standard scheme \cite{PDG}, the $Q\bar{s}$ mesons with
$J^{P}=0^{-},~1^{-},~ 0^{+},~ \mbox{and}~1^{+}$ are named $D_{s}$,
$D_{s}^{*}$, $D_{s0}^{*}$, and $D_{s1}$ for charmed mesons, with
$B_{s}$, $B_{s}^{*}$, $B_{s0}^{*}$, and $B_{s1}$ for bottom mesons,
respectively. In this work, the corresponding configurations for
these mesons are represented as $(Q\bar{s})$, $(Q\bar{s})^{*}$,
$(Q\bar{s})_{0}^{*}$, and $(Q\bar{s})_{1}$. In full theory, the
interpolating currents for these mesons can be found in Refs.
\cite{reinders,reinders1}. Presently, one constructs the molecular state current from
meson-meson type of fields, while constructs
the tetraquark state current
from diquark-antidiquark
configuration of fields. The currents constructed from meson-meson type of fields
can be related to those composed
of diquark-antidiquark
type of fields
by Fiertz rearrangements.
However, the relations are suppressed
by a typical color and Dirac factors so that one could
obtain a reliable sum rule only if one has chosen the appropriate current
to have a maximum overlap with the physical
state, which is expected to be particularly true for multiquark
configuration with special molecular
or diquark structures.
Concretely,
it will have a maximum
overlap for the molecular state using the
meson-meson current and
the sum rule can reproduce the physical mass well,
whereas the overlap for the molecular state employing
a diquark-antidiquark type of current will be
small and the sum rule will not be able to reproduce the
mass well. Likewise, the opposite is also true
(there are some more concrete calculations and discussions
in the XII. Appendix in Ref. \cite{overview-MN}). Consequently, the interpolating currents
for the related molecular states are constructed. For one type of hadrons,
with
\begin{eqnarray}
j_{(Q\bar{s})(\bar{Q'}s)}=(\bar{s}_{a}i\gamma_{5}Q_{a})(\bar{Q'}_{b}i\gamma_{5}s_{b}),\nonumber
\end{eqnarray}
coupling to $D_{s}\bar{D}_{s}$, $B_{s}\bar{B}_{s}$, $D_{s}\bar{B}_{s}$, or $B_{s}\bar{D}_{s}$ molecular state,
\begin{eqnarray}
j_{(Q\bar{s})^{*}(\bar{Q'}s)^{*}}=(\bar{s}_{a}\gamma_{\mu}Q_{a})(\bar{Q'}_{b}\gamma^{\mu}s_{b}),\nonumber
\end{eqnarray}
for $D_{s}^{*}\bar{D}_{s}^{*}$, $B_{s}^{*}\bar{B}_{s}^{*}$, $D_{s}^{*}\bar{B}_{s}^{*}$, or $B_{s}^{*}\bar{D}_{s}^{*}$ state,
\begin{eqnarray}
j_{(Q\bar{s})_{0}^{*}(\bar{Q'}s)_{0}^{*}}=(\bar{s}_{a}Q_{a})(\bar{Q'}_{b}s_{b}),\nonumber
\end{eqnarray}
for $D_{s0}^{*}\bar{D}_{s0}^{*}$, $B_{s0}^{*}\bar{B}_{s0}^{*}$, $D_{s0}^{*}\bar{B}_{s0}^{*}$, or $B_{s0}^{*}\bar{D}_{s0}^{*}$ state,
\begin{eqnarray}
j_{(Q\bar{s})_{1}(\bar{Q'}s)_{1}}=(\bar{s}_{a}\gamma_{\mu}\gamma_{5}Q_{a})(\bar{Q'}_{b}\gamma^{\mu}\gamma_{5}s_{b}),\nonumber
\end{eqnarray}
for $D_{s1}\bar{D}_{s1}$, $B_{s1}\bar{B}_{s1}$, $D_{s1}\bar{B}_{s1}$, or $B_{s1}\bar{D}_{s1}$ state,
\begin{eqnarray}
j_{(Q\bar{s})(\bar{Q'}s)_{0}^{*}}=(\bar{s}_{a}i\gamma_{5}Q_{a})(\bar{Q'}_{b}s_{b}),\nonumber
\end{eqnarray}
for $D_{s}\bar{D}_{s0}^{*}$, $B_{s}\bar{B}_{s0}^{*}$, $D_{s}\bar{B}_{s0}^{*}$, or $B_{s}\bar{D}_{s0}^{*}$ state, and
\begin{eqnarray}
j_{(Q\bar{s})^{*}(\bar{Q'}s)_{1}}=(\bar{s}_{a}\gamma_{\mu}Q_{a})(\bar{Q'}_{b}\gamma^{\mu}\gamma_{5}s_{b}),\nonumber
\end{eqnarray}
for $D_{s}^{*}\bar{D}_{s1}$, $B_{s}^{*}\bar{B}_{s1}$, $D_{s}^{*}\bar{B}_{s1}$, or $B_{s}^{*}\bar{D}_{s1}$ state,
where $Q$ and $Q'$ denote heavy quarks ($Q=Q'$ or
$Q\neq Q'$), with $a$ and $b$ are color indices. For another type, with
\begin{eqnarray}
j^{\mu}_{(Q\bar{s})^{*}(\bar{Q'}s)}=(\bar{s}_{a}\gamma^{\mu}Q_{a})(\bar{Q'}_{b}i\gamma_{5}s_{b}),\nonumber
\end{eqnarray}
for $D_{s}^{*}\bar{D}_{s}$, $B_{s}^{*}\bar{B}_{s}$, $D_{s}^{*}\bar{B}_{s}$, or $B_{s}^{*}\bar{D}_{s}$ state,
\begin{eqnarray}
j^{\mu}_{(Q\bar{s})_{1}(\bar{Q'}s)}=(\bar{s}_{a}\gamma^{\mu}\gamma_{5}Q_{a})(\bar{Q'}_{b}i\gamma_{5}s_{b}),\nonumber
\end{eqnarray}
for $D_{s1}\bar{D}_{s}$, $B_{s1}\bar{B}_{s}$, $D_{s1}\bar{B}_{s}$, or $B_{s1}\bar{D}_{s}$ state,
\begin{eqnarray}
j^{\mu}_{(Q\bar{s})^{*}(\bar{Q'}s)_{0}^{*}}=(\bar{s}_{a}\gamma^{\mu}Q_{a})(\bar{Q'}_{b}s_{b}),\nonumber
\end{eqnarray}
for $D_{s}^{*}\bar{D}_{s0}^{*}$, $B_{s}^{*}\bar{B}_{s0}^{*}$, $D_{s}^{*}\bar{B}_{s0}^{*}$, or $B_{s}^{*}\bar{D}_{s0}^{*}$ state,
and
\begin{eqnarray}
j^{\mu}_{(Q\bar{s})_{1}(\bar{Q'}s)_{0}^{*}}=(\bar{s}_{a}\gamma^{\mu}\gamma_{5}Q_{a})(\bar{Q'}_{b}s_{b}),\nonumber
\end{eqnarray}
for $D_{s1}\bar{D}_{s0}^{*}$, $B_{s1}\bar{B}_{s0}^{*}$, $D_{s1}\bar{B}_{s0}^{*}$, or $B_{s1}\bar{D}_{s0}^{*}$ state.

\subsection{the molecular state QCD sum rule}
For the former case, the starting point is the two-point correlator
\begin{eqnarray}\label{correlator}
\Pi(q^{2})=i\int
d^{4}x\mbox{e}^{iq.x}\langle0|T[j(x)j^{+}(0)]|0\rangle.
\end{eqnarray}
The correlator can be phenomenologically expressed as
\begin{eqnarray}
\Pi(q^{2})=\frac{\lambda^{2}_H}{M_{H}^{2}-q^{2}}+\frac{1}{\pi}\int_{s_{0}}
^{\infty}ds\frac{\mbox{Im}\Pi^{\mbox{phen}}(s)}{s-q^{2}}
+\mbox{subtractions},
\end{eqnarray}
where $M_{H}$ denotes the mass of the hadronic resonance, and
$\lambda_{H}$ gives the coupling of the current to the hadron
$\langle0|j|H\rangle=\lambda_{H}$. In the OPE side, the correlator
can be written as
\begin{eqnarray}
\Pi(q^{2})=\int_{(m_{Q}+m_{Q'}+2m_{s})^{2}}^{\infty}ds\frac{\rho^{\mbox{OPE}}(s)}{s-q^{2}}~~(m_{Q}=m_{Q'}
~\mbox{or}~m_{Q}\neq
m_{Q'}),
\end{eqnarray}
where the spectral density is given by
$\rho^{\mbox{OPE}}(s)=\frac{1}{\pi}\mbox{Im}\Pi^{\mbox{OPE}}(s)$.
After equating the two sides, assuming quark-hadron duality, and
making a Borel transform, the sum rule can be written as
\begin{eqnarray}
\lambda_{H}^{2}e^{-M_{H}^{2}/M^{2}}&=&\int_{(m_{Q}+m_{Q'}+2m_{s})^{2}}^{s_{0}}ds\rho^{\mbox{OPE}}(s)e^{-s/M^{2}},
\end{eqnarray}
where $M^2$ indicates Borel parameter. To eliminate the hadronic coupling constant $\lambda_H$, one reckons
the ratio of derivative of the sum rule and itself, and then yields
\begin{eqnarray}\label{sum rule}
M_{H}^{2}&=&\frac{\int_{(m_{Q}+m_{Q'}+2m_{s})^{2}}^{s_{0}}ds\rho^{\mbox{OPE}}s
e^{-s/M^{2}}}{\int_{(m_{Q}+m_{Q'}+2m_{s})^{2}}^{s_{0}}ds\rho^{\mbox{OPE}}e^{-s/M^{2}}}.
\end{eqnarray}

For the latter case, one starts from
\begin{eqnarray}\label{correlator1}
\Pi^{\mu\nu}(q^{2})=i\int
d^{4}x\mbox{e}^{iq.x}\langle0|T[j^{\mu}(x)j^{\nu+}(0)]|0\rangle.
\end{eqnarray}
Lorentz covariance implies that the correlator (\ref{correlator1})
can be generally parameterized as
\begin{eqnarray}
\Pi^{\mu\nu}(q^{2})=(\frac{q^{\mu}q^{\nu}}{q^{2}}-g^{\mu\nu})\Pi^{(1)}(q^{2})+\frac{q^{\mu}q^{\nu}}{q^{2}}\Pi^{(0)}(q^{2}).
\end{eqnarray}
The $\Pi^{(1)}(q^{2})$ of the correlator proportional to $g_{\mu\nu}$ will be
chosen to extract the mass sum rule here. Similarly, one can finally
yield
\begin{eqnarray}\label{sum rule 1}
M_{H}^{2}=\frac{\int_{(m_{Q}+m_{Q'}+2m_{s})^{2}}^{s_{0}}ds\rho^{\mbox{OPE}}s
e^{-s/M^{2}}}{\int_{(m_{Q}+m_{Q'}+2m_{s})^{2}}^{s_{0}}ds\rho^{\mbox{OPE}}e^{-s/M^{2}}},
\end{eqnarray}
with $\rho^{\mbox{OPE}}(s)=\frac{1}{\pi}\mbox{Im}\Pi^{\mbox{(1)}}(s)$.

\subsection{spectral densities}
Calculating the OPE side, one works at leading order in $\alpha_{s}$
and considers condensates up to dimension six, utilizing the similar
techniques in Refs. \cite{technique,technique1}. The $s$ quark is
dealt as a light one. To keep the heavy-quark mass finite, one uses the
momentum-space expression for the heavy-quark propagator. One
calculates the light-quark part of the correlation function in the
coordinate space, which is then Fourier-transformed to the momentum
space in $D$ dimension. The resulting light-quark part is combined
with the heavy-quark part before it is dimensionally regularized at
$D=4$. For the heavy-quark propagator with two and three gluons
attached, the momentum-space expressions given in Ref.
\cite{reinders} are used.
After some tedious calculations, the concrete forms of spectral
densities can be derived, which are collected in the Appendix.
In detail,
some different currents lead to the similar OPE, for example, the terms for $D_{s}\bar{D}_{s}$ and $D_{s0}^{*}\bar{D}_{s0}^{*}$
are similar, the ones for $D_{s}^{*}\bar{D}_{s}^{*}$ and $D_{s1}\bar{D}_{s1}$ are similar and so on.
Although the terms for them are similar respectively, the corresponding signs may be different, such as a term for $D_{s}\bar{D}_{s}$
may be ``plus" sign while the related one for $D_{s0}^{*}\bar{D}_{s0}^{*}$ may be ``minus", which caused by the differences of $\gamma$-matrices in the
interpolating currents and the
differences of the trace results.
Numerically, the two quark condensate $\langle\bar{s}s\rangle$ is the most important condensate correction, the
absolute value of which is bigger than the absolute values of the
four quark condensate $\langle\bar{s}s\rangle^{2}$ as well as the mixed condensate $\langle
g\bar{s}\sigma\cdot G s\rangle$. Meanwhile,
the two gluon condensate $\langle g^{2}G^{2}\rangle$ and the
three gluon condensate $\langle g^{3}G^{3}\rangle$ are very small and almost negligible.

\section{Numerical analysis}\label{sec3}
In this part, the sum rules (\ref{sum rule}) and (\ref{sum rule 1})
will be numerically analyzed. The input values are taken as
$m_{c}=1.23~\mbox{GeV}$, $m_{b}=4.20~\mbox{GeV}$, and
$m_{s}=0.13~\mbox{GeV}$ \cite{PDG}, with
$\langle\bar{q}q\rangle=-(0.23)^{3}~\mbox{GeV}^{3}$,
$\langle\bar{s}s\rangle=0.8~\langle\bar{q}q\rangle$, $\langle
g\bar{s}\sigma\cdot G s\rangle=m_{0}^{2}~\langle\bar{s}s\rangle$,
$m_{0}^{2}=0.8~\mbox{GeV}^{2}$, $\langle
g^{2}G^{2}\rangle=0.88~\mbox{GeV}^{4}$, and $\langle
g^{3}G^{3}\rangle=0.045~\mbox{GeV}^{6}$ \cite{overview2,technique}. Complying with the standard
procedure of sum rule analysis, the threshold $s_{0}$ and Borel
parameter $M^{2}$ are varied to find the optimal stability window.
Namely, we try to consider the Borel curve stability's dependence on the Borel plateaus (the threshold $s_{0}$ and Borel parameter $M^{2}$), and find the Borel windows where the perturbative contribution should be larger than the condensate contributions in the OPE side while the pole contribution should be larger than continuum contribution in the phenomenological side. Thus, the regions of thresholds are taken as
values presented in the related figure captions, with
$M^{2}=3.5\sim4.5~\mbox{GeV}^{2}$ for $D_{s0}^{*}\bar{D}_{s0}^{*}$,
$D_{s1}\bar{D}_{s0}^{*}$, $D_{s1}\bar{D}_{s1}$,
$D_{s}\bar{D}_{s0}^{*}$, $D_{s1}\bar{D}_{s}$,
$D_{s}^{*}\bar{D}_{s0}^{*}$, and $D_{s}^{*}\bar{D}_{s1}$,
$M^{2}=7.5\sim9.0~\mbox{GeV}^{2}$ for $B_{s}\bar{D}_{s}$,
$B_{s}^{*}\bar{D}_{s}$, $B_{s}^{*}\bar{D}_{s}^{*}$,
$B_{s0}^{*}\bar{D}_{s0}^{*}$, $B_{s1}\bar{D}_{s0}^{*}$,
$B_{s1}\bar{D}_{s1}$, $B_{s}\bar{D}_{s0}^{*}$, $B_{s1}\bar{D}_{s}$,
$B_{s}^{*}\bar{D}_{s0}^{*}$, $B_{s}^{*}\bar{D}_{s1}$,
$D_{s}^{*}\bar{B}_{s}$, $D_{s1}\bar{B}_{s0}^{*}$,
$D_{s}\bar{B}_{s0}^{*}$, $D_{s1}\bar{B}_{s}$,
$D_{s}^{*}\bar{B}_{s0}^{*}$, and $D_{s}^{*}\bar{B}_{s1}$, and
$M^{2}=9.5\sim11.0~\mbox{GeV}^{2}$ for $B_{s0}^{*}\bar{B}_{s0}^{*}$,
$B_{s1}\bar{B}_{s0}^{*}$, $B_{s1}\bar{B}_{s1}$,
$B_{s}\bar{B}_{s0}^{*}$, $B_{s1}\bar{B}_{s}$,
$B_{s}^{*}\bar{B}_{s0}^{*}$, and $B_{s}^{*}\bar{B}_{s1}$,
respectively. Tables 1-2 collect all the numerical results.
Note that uncertainties are owing
to the sum rule windows (variation of the threshold $s_{0}$ and
Borel parameter $M^{2}$), not involving the ones from the variation
of quark masses and QCD parameters for which
are appreciably smaller in comparison with
the ones from the sum rule windows here.
The numerical result $4.13\pm0.10~\mbox{GeV}$ for
$D_{s}^{*}\bar{D}_{s}^{*}$ agrees well with the mass
$4143.0\pm2.9\pm1.2~\mbox{MeV}$ for $Y(4140)$ \cite{zhang}, which supports the
$D_{s}^{*}\bar{D}_{s}^{*}$ molecular configuration for $Y(4140)$.
After the completion of the calculations here, evidence for a new resonance (named as  $X(4350)$)
has been observed by the Belle Collaboration \cite{X4350}.
Note that the predicted value $4.36\pm0.08~\mbox{GeV}$ for the $D_{s}^{*}\bar{D}_{s0}^{*}$
molecular state here is consistent with
the mass $4350^{+4.6}_{-5.1}\pm0.7~\mbox{MeV}$ of the newly observed structure.

\section{Summary}\label{sec4}
In summary, QCD sum rules have been employed to compute the masses
of molecular states, including the contributions of operators up to
dimension six in OPE. Ultimately, we have arrived at mass spectra for
molecular states with various $\{Q\bar{s}\}\{\bar{Q}^{(')}s\}$
configurations. The numerical result $4.13\pm0.10~\mbox{GeV}$ for
$D_{s}^{*}\bar{D}_{s}^{*}$ agrees well with the mass
$4143.0\pm2.9\pm1.2~\mbox{MeV}$ for $Y(4140)$ \cite{zhang}, which supports the
interpretation of $Y(4140)$ as a $D_{s}^{*}\bar{D}_{s}^{*}$ molecular state.
The predicted value $4.36\pm0.08~\mbox{GeV}$ for the $D_{s}^{*}\bar{D}_{s0}^{*}$
molecular state is consistent with
the mass $4350^{+4.6}_{-5.1}\pm0.7~\mbox{MeV}$ of the newly observed $X(4350)$,
which could support $X(4350)$ interpreted as a $D_{s}^{*}\bar{D}_{s0}^{*}$
molecular state.
More experimental evidence on $\{Q\bar{s}\}\{\bar{Q}^{(')}s\}$
molecular states besides $Y(4140)$ and $X(4350)$ may appear
if they do exist, and the data on molecular states are expecting further experimental identification,
which may be searched for experimentally at facilities such as Super-B factories
in the $J/\psi\phi$ mass spectrum
in the future.

\appendix
\section*{Appendix}
It is defined that $r(m_{Q},m_{Q'}) =\alpha m_{Q}^2 +\beta m_{Q'}^2 -\alpha
\beta s~~(m_{Q}=m_{Q'}~~\mbox{or}~m_{Q}\neq m_{Q'})$. With

\begin{eqnarray}
\rho^{\mbox{pert}}(s)&=&\frac{3}{2^{11}\pi^{6}}\int_{\alpha_{min}}^{\alpha_{max}}d\alpha\int_{\beta_{min}}^{1-\alpha}d\beta(1-\alpha-\beta)
[\frac{1}{\alpha^{3}\beta^{3}}r(m_{Q},m_{Q'})^{2}\nonumber\\&&{}
-\frac{2^{2}m_{Q'}m_{s}}{\alpha^{3}\beta^{2}}r(m_{Q},m_{Q'})
-\frac{2^{2}m_{Q}m_{s}}{\alpha^{2}\beta^{3}}r(m_{Q},m_{Q'})+\frac{3\cdot2^{2}m_{Q}m_{Q'}m_{s}^{2}}{\alpha^{2}\beta^{2}}]\nonumber\\&&{}\times r(m_{Q},m_{Q'})^{2},\nonumber\\
\rho^{\langle\bar{s}s\rangle}(s)&=&\frac{3\langle\bar{s}s\rangle}{2^{7}\pi^{4}}\int_{\alpha_{min}}^{\alpha_{max}}d\alpha\{\int_{\beta_{min}}^{1-\alpha}d\beta[-\frac{m_{Q'}}{\alpha^{2}\beta}r(m_{Q},m_{Q'})
-\frac{m_{Q}}{\alpha\beta^{2}}r(m_{Q},m_{Q'})\nonumber\\&&{}+\frac{2^{2}m_{Q}m_{Q'}m_{s}}{\alpha\beta}]r(m_{Q},m_{Q'})
+\{\frac{m_{s}}{\alpha(1-\alpha)}[\alpha
m_{Q}^{2}+(1-\alpha) m_{Q'}^{2}-\alpha(1-\alpha)
s]\nonumber\\&&{}-\frac{m_{Q}m_{s}^{2}}{1-\alpha}-\frac{m_{Q'}m_{s}^{2}}{\alpha}\}[\alpha
m_{Q}^{2}+(1-\alpha) m_{Q'}^{2}-\alpha(1-\alpha) s]\},\nonumber\\
\rho^{\langle\bar{s}s\rangle^{2}}(s)&=&\frac{\langle\bar{s}s\rangle^{2}}{2^{5}\pi^{2}}\{[2m_{Q}m_{Q'}-(m_{Q}+m_{Q'})m_{s}]
\sqrt{(s-m_{Q}^{2}+m_{Q'}^{2})^{2}-4m_{Q'}^{2}s}/s\nonumber\\&&{}
+3m_{s}^{2}\int_{\alpha_{min}}^{\alpha_{max}}d\alpha\alpha(1-\alpha)\},\nonumber\\
\rho^{\langle g\bar{s}\sigma\cdot G s\rangle}(s)&=&\frac{3\langle
g\bar{s}\sigma\cdot G
s\rangle}{2^{8}\pi^{4}}\{\int_{\alpha_{min}}^{\alpha_{max}}d\alpha\{(-\frac{m_{Q'}}{\alpha}-\frac{m_{Q}}{1-\alpha})
[\alpha m_{Q}^{2}
+(1-\alpha) m_{Q'}^{2}-\alpha(1-\alpha)s]\nonumber\\&&{}
+\frac{2m_{s}}{3}[2\alpha m_{Q}^{2}
+2(1-\alpha) m_{Q'}^{2}
-3\alpha(1-\alpha) s]\}\nonumber\\&&{}
+[2m_{Q}m_{Q'}m_{s}
-\frac{(m_{Q}+m_{Q'})m_{s}^{2}}{3}]
\sqrt{(s-m_{Q}^{2}+m_{Q'}^{2})^{2}-4m_{Q'}^{2}s}/s\},\nonumber\\
\rho^{\langle g^{2}G^{2}\rangle}(s)&=&\frac{\langle
g^{2}G^{2}\rangle}{2^{11}\pi^{6}}\int_{\alpha_{min}}^{\alpha_{max}}d\alpha\int_{\beta_{min}}^{1-\alpha}d\beta(1-\alpha-\beta)
[\frac{m_{Q'}^{2}}{\alpha^{3}}r(m_{Q},m_{Q'})+\frac{m_{Q}^{2}}{\beta^{3}}r(m_{Q},m_{Q'})\nonumber\\&&{}-\frac{3m_{Q'}m_{s}}{\alpha^{3}}r(m_{Q},m_{Q'})
-\frac{m_{Q'}^{3}m_{s}\beta}{\alpha^{3}}-\frac{3m_{Q}m_{s}}{\beta^{3}}r(m_{Q},m_{Q'})-\frac{m_{Q}^{3}m_{s}\alpha}{\beta^{3}}\nonumber\\&&{}
-\frac{m_{Q}m_{Q'}^{2}m_{s}}{\alpha^{2}}
-\frac{m_{Q}^{2}m_{Q'}m_{s}}{\beta^{2}}
+\frac{3m_{Q}m_{Q'}m_{s}^{2}}{\alpha^{2}}+\frac{3m_{Q}m_{Q'}m_{s}^{2}}{\beta^{2}}],\nonumber\\
\rho^{\langle g^{3}G^{3}\rangle}(s)&=&\frac{\langle
g^{3}G^{3}\rangle}{2^{13}\pi^{6}}\int_{\alpha_{min}}^{\alpha_{max}}d\alpha\int_{\beta_{min}}^{1-\alpha}d\beta(1-\alpha-\beta)[\frac{1}{\alpha^{3}}r(m_{Q},m_{Q'})+\frac{2m_{Q'}^{2}\beta}{\alpha^{3}}\nonumber\\&&{}
+\frac{1}{\beta^{3}}r(m_{Q},m_{Q'})
+\frac{2m_{Q}^{2}\alpha}{\beta^{3}}
-\frac{6m_{Q'}m_{s}\beta}{\alpha^{3}}
-\frac{6m_{Q}m_{s}\alpha}{\beta^{3}}-\frac{m_{Q}m_{s}}{\alpha^{2}}-\frac{m_{Q'}m_{s}}{\beta^{2}}],\nonumber
\end{eqnarray}
for $(Q\bar{s})(\bar{Q'}s)$,

\begin{eqnarray}
\rho^{\mbox{pert}}(s)&=&\frac{3}{2^{12}\pi^{6}}\int_{\alpha_{min}}^{\alpha_{max}}d\alpha\int_{\beta_{min}}^{1-\alpha}d\beta(1-\alpha-\beta)
[\frac{1}{\alpha^{3}\beta^{3}}(1+\alpha+\beta)r(m_{Q},m_{Q'})^{2}\nonumber\\&&{}
-\frac{2^{2}m_{Q'}m_{s}}{\alpha^{3}\beta^{2}}(1+\alpha+\beta)r(m_{Q},m_{Q'})
-\frac{2^{3}m_{Q}m_{s}}{\alpha^{2}\beta^{3}}r(m_{Q},m_{Q'})\nonumber\\&&{}+\frac{3\cdot2^{3}m_{Q}m_{Q'}m_{s}^{2}}{\alpha^{2}\beta^{2}}] r(m_{Q},m_{Q'})^{2},\nonumber\\
\rho^{\langle\bar{s}s\rangle}(s)&=&\frac{3\langle\bar{s}s\rangle}{2^{7}\pi^{4}}\int_{\alpha_{min}}^{\alpha_{max}}d\alpha\{\int_{\beta_{min}}^{1-\alpha}d\beta[-\frac{m_{Q'}}{\alpha^{2}\beta}(\alpha
+\beta)r(m_{Q},m_{Q'})-\frac{m_{Q}}{\alpha\beta^{2}}r(m_{Q},m_{Q'})\nonumber\\&&{}
-\frac{m_{s}}{\alpha\beta}r(m_{Q},m_{Q'})
+\frac{2^{2}m_{Q}m_{Q'}m_{s}}{\alpha\beta}+\frac{m_{Q'}m_{s}^{2}}{\alpha}]r(m_{Q},m_{Q'})\nonumber\\&&{}+\{\frac{m_{s}}{\alpha(1-\alpha)}[\alpha
m_{Q}^{2}+(1-\alpha) m_{Q'}^{2}-\alpha(1-\alpha)
s]-\frac{m_{Q}m_{s}^{2}}{1-\alpha}
-\frac{m_{Q'}m_{s}^{2}}{\alpha}\}[\alpha
m_{Q}^{2}\nonumber\\&&{}+(1-\alpha) m_{Q'}^{2}-\alpha(1-\alpha) s]\},\nonumber\\
\rho^{\langle\bar{s}s\rangle^{2}}(s)&=&\frac{\langle\bar{s}s\rangle^{2}}{2^{6}\pi^{2}}[(2^{2}m_{Q}m_{Q'}-2m_{Q}m_{s})
\sqrt{(s-m_{Q}^{2}+m_{Q'}^{2})^{2}-4m_{Q'}^{2}s}/s\nonumber\\&&{}
+\int_{\alpha_{min}}^{\alpha_{max}}d\alpha(-2m_{Q'}m_{s}+3m_{s}^{2}\alpha)(1-\alpha)],\nonumber\\
\rho^{\langle g\bar{s}\sigma\cdot G s\rangle}(s)&=&\frac{3\langle
g\bar{s}\sigma\cdot G
s\rangle}{2^{8}\pi^{4}}\{\int_{\alpha_{min}}^{\alpha_{max}}d\alpha\{\int_{\beta_{min}}^{1-\alpha}d\beta\frac{m_{Q'}}{\alpha} r(m_{Q},m_{Q'})-(\frac{m_{Q'}}{\alpha}+\frac{m_{Q}}{1-\alpha})[\alpha
m_{Q}^{2}\nonumber\\&&{}+(1-\alpha) m_{Q'}^{2}-\alpha(1-\alpha)
s]+\frac{2m_{s}}{3}[\alpha
m_{Q}^{2}+(1-\alpha) m_{Q'}^{2}
-2\alpha(1-\alpha) s]\}\nonumber\\&&{}+(2m_{Q}m_{Q'}m_{s}-\frac{m_{Q}m_{s}^{2}}{3})
\sqrt{(s-m_{Q}^{2}+m_{Q'}^{2})^{2}-4m_{Q'}^{2}s}/s\nonumber\\&&{}
-\frac{m_{Q'}m_{s}^{2}}{3}\int_{\alpha_{min}}^{\alpha_{max}}d\alpha(1-\alpha)\}
,\nonumber\\
\rho^{\langle g^{2}G^{2}\rangle}(s)&=&\frac{\langle
g^{2}G^{2}\rangle}{2^{12}\pi^{6}}\int_{\alpha_{min}}^{\alpha_{max}}d\alpha\int_{\beta_{min}}^{1-\alpha}d\beta(1-\alpha-\beta)
[\frac{m_{Q'}^{2}}{\alpha^{3}}(1+\alpha+\beta)r(m_{Q},m_{Q'})\nonumber\\&&{}+\frac{m_{Q}^{2}}{\beta^{3}}(1+\alpha+\beta)
r(m_{Q},m_{Q'})
-\frac{3m_{Q'}m_{s}}{\alpha^{3}}(1+\alpha+\beta)r(m_{Q},m_{Q'})\nonumber\\&&{}
-\frac{m_{Q'}^{3}m_{s}\beta}{\alpha^{3}}(1+\alpha+\beta)-\frac{6m_{Q}m_{s}}{\beta^{3}}r(m_{Q},m_{Q'})
-\frac{2m_{Q}^{3}m_{s}\alpha}{\beta^{3}}
-\frac{2m_{Q}m_{Q'}^{2}m_{s}}{\alpha^{2}}\nonumber\\&&{}
-\frac{m_{Q}^{2}m_{Q'}m_{s}}{\beta^{2}}(1+\alpha+\beta)+\frac{6m_{Q}m_{Q'}m_{s}^{2}}{\alpha^{2}}
+\frac{6m_{Q}m_{Q'}m_{s}^{2}}{\beta^{2}}],\nonumber\\
\rho^{\langle g^{3}G^{3}\rangle}(s)&=&\frac{\langle
g^{3}G^{3}\rangle}{2^{14}\pi^{6}}\int_{\alpha_{min}}^{\alpha_{max}}d\alpha\int_{\beta_{min}}^{1-\alpha}d\beta(1-\alpha-\beta)
[\frac{1}{\alpha^{3}}(1+\alpha+\beta)r(m_{Q},m_{Q'})\nonumber\\&&{}+\frac{2m_{Q'}^{2}\beta}{\alpha^{3}}(1+\alpha+\beta)
+\frac{1}{\beta^{3}}(1+\alpha+\beta)r(m_{Q},m_{Q'})+\frac{2m_{Q}^{2}\alpha}{\beta^{3}}(1+\alpha+\beta)\nonumber\\&&{}
-\frac{6m_{Q'}m_{s}\beta}{\alpha^{3}}(1+\alpha+\beta)-\frac{3\cdot2^{2}m_{Q}m_{s}\alpha}{\beta^{3}}
-\frac{2m_{Q}m_{s}}{\alpha^{2}}-\frac{m_{Q'}m_{s}}{\beta^{2}}(1+\alpha+\beta)]
,\nonumber
\end{eqnarray}
for $(Q\bar{s})^{*}(\bar{Q'}s)$,

\begin{eqnarray}
\rho^{\mbox{pert}}(s)&=&\frac{3}{2^{9}\pi^{6}}\int_{\alpha_{min}}^{\alpha_{max}}d\alpha\int_{\beta_{min}}^{1-\alpha}d\beta(1-\alpha-\beta)
[\frac{1}{\alpha^{3}\beta^{3}}r(m_{Q},m_{Q'})^{2}-\frac{2m_{Q'}m_{s}}{\alpha^{3}\beta^{2}}r(m_{Q},m_{Q'})\nonumber\\&&{}
-\frac{2m_{Q}m_{s}}{\alpha^{2}\beta^{3}}r(m_{Q},m_{Q'})
+\frac{3\cdot2^{2}m_{Q}m_{Q'}m_{s}^{2}}{\alpha^{2}\beta^{2}}]r(m_{Q},m_{Q'})^{2},\nonumber\\
\rho^{\langle\bar{s}s\rangle}(s)&=&\frac{3\langle\bar{s}s\rangle}{2^{6}\pi^{4}}\int_{\alpha_{min}}^{\alpha_{max}}d\alpha\{\int_{\beta_{min}}^{1-\alpha}d\beta[-\frac{m_{Q'}}{\alpha^{2}\beta}r(m_{Q},m_{Q'})
-\frac{m_{Q}}{\alpha\beta^{2}}r(m_{Q},m_{Q'})\nonumber\\&&{}+\frac{2^{3}m_{Q}m_{Q'}m_{s}}{\alpha\beta}]r(m_{Q},m_{Q'})
+\{\frac{2m_{s}}{\alpha(1-\alpha)}[\alpha
m_{Q}^{2}+(1-\alpha) m_{Q'}^{2}-\alpha(1-\alpha)
s]\nonumber\\&&{}
-\frac{m_{Q}m_{s}^{2}}{1-\alpha}-\frac{m_{Q'}m_{s}^{2}}{\alpha}\}[\alpha
m_{Q}^{2}
+(1-\alpha) m_{Q'}^{2}
-\alpha(1-\alpha) s]\},\nonumber\\
\rho^{\langle\bar{s}s\rangle^{2}}(s)&=&\frac{\langle\bar{s}s\rangle^{2}}{2^{4}\pi^{2}}\{[2^{2}m_{Q}m_{Q'}-(m_{Q}+m_{Q'})m_{s}]
\sqrt{(s-m_{Q}^{2}+m_{Q'}^{2})^{2}-4m_{Q'}^{2}s}/s\nonumber\\&&{}
+6m_{s}^{2}\int_{\alpha_{min}}^{\alpha_{max}}d\alpha\alpha(1-\alpha)\},\nonumber\\
\rho^{\langle g\bar{s}\sigma\cdot G s\rangle}(s)&=&\frac{3\langle
g\bar{s}\sigma\cdot G
s\rangle}{2^{7}\pi^{4}}\{\int_{\alpha_{min}}^{\alpha_{max}}d\alpha\{(-\frac{m_{Q'}}{\alpha}
-\frac{m_{Q}}{1-\alpha})[\alpha
m_{Q}^{2}+(1-\alpha) m_{Q'}^{2}-\alpha(1-\alpha)
s]\nonumber\\&&{}+\frac{2^{2}m_{s}}{3}[2\alpha
m_{Q}^{2}+2(1-\alpha) m_{Q'}^{2}-3\alpha(1-\alpha) s]\}\nonumber\\&&{}
+[2^{2}m_{Q}m_{Q'}m_{s}
-\frac{(m_{Q}+m_{Q'})m_{s}^{2}}{3}]
\sqrt{(s-m_{Q}^{2}+m_{Q'}^{2})^{2}-4m_{Q'}^{2}s}/s\}
,\nonumber\\
\rho^{\langle g^{2}G^{2}\rangle}(s)&=&\frac{\langle
g^{2}G^{2}\rangle}{2^{9}\pi^{6}}\int_{\alpha_{min}}^{\alpha_{max}}d\alpha\int_{\beta_{min}}^{1-\alpha}d\beta(1-\alpha-\beta)[\frac{m_{Q'}^{2}}{\alpha^{3}}r(m_{Q},m_{Q'})+\frac{m_{Q}^{2}}{\beta^{3}}r(m_{Q},m_{Q'})\nonumber\\&&{}
-\frac{3m_{Q'}m_{s}}{2\alpha^{3}}r(m_{Q},m_{Q'})
-\frac{m_{Q'}^{3}m_{s}\beta}{2\alpha^{3}}
-\frac{3m_{Q}m_{s}}{2\beta^{3}}r(m_{Q},m_{Q'})-\frac{m_{Q}^{3}m_{s}\alpha}{2\beta^{3}}\nonumber\\&&{}
-\frac{m_{Q}m_{Q'}^{2}m_{s}}{2\alpha^{2}}
-\frac{m_{Q}^{2}m_{Q'}m_{s}}{2\beta^{2}}
+\frac{3m_{Q}m_{Q'}m_{s}^{2}}{\alpha^{2}}+\frac{3m_{Q}m_{Q'}m_{s}^{2}}{\beta^{2}}],\nonumber\\
\rho^{\langle g^{3}G^{3}\rangle}(s)&=&\frac{\langle
g^{3}G^{3}\rangle}{2^{11}\pi^{6}}\int_{\alpha_{min}}^{\alpha_{max}}d\alpha\int_{\beta_{min}}^{1-\alpha}d\beta(1-\alpha-\beta)
[\frac{1}{\alpha^{3}}r(m_{Q},m_{Q'})+\frac{2m_{Q'}^{2}\beta}{\alpha^{3}}\nonumber\\&&{}+\frac{1}{\beta^{3}}r(m_{Q},m_{Q'})
+\frac{2m_{Q}^{2}\alpha}{\beta^{3}}
-\frac{3m_{Q'}m_{s}\beta}{\alpha^{3}}
-\frac{3m_{Q}m_{s}\alpha}{\beta^{3}}-\frac{m_{Q}m_{s}}{2\alpha^{2}}
-\frac{m_{Q'}m_{s}}{2\beta^{2}}],\nonumber
\end{eqnarray}
for $(Q\bar{s})^{*}(\bar{Q'}s)^{*}$,

\begin{eqnarray}
\rho^{\mbox{pert}}(s)&=&\frac{3}{2^{11}\pi^{6}}\int_{\alpha_{min}}^{\alpha_{max}}d\alpha\int_{\beta_{min}}^{1-\alpha}d\beta(1-\alpha-\beta)
[\frac{1}{\alpha^{3}\beta^{3}}r(m_{Q},m_{Q'})^{2}\nonumber\\&&{}+\frac{2^{2}m_{Q'}m_{s}}{\alpha^{3}\beta^{2}}r(m_{Q},m_{Q'})
+\frac{2^{2}m_{Q}m_{s}}{\alpha^{2}\beta^{3}}r(m_{Q},m_{Q'})
+\frac{3\cdot2^{2}m_{Q}m_{Q'}m_{s}^{2}}{\alpha^{2}\beta^{2}}]\nonumber\\&&{}\times
r(m_{Q},m_{Q'})^{2},\nonumber\\
\rho^{\langle\bar{s}s\rangle}(s)&=&\frac{3\langle\bar{s}s\rangle}{2^{7}\pi^{4}}\int_{\alpha_{min}}^{\alpha_{max}}d\alpha\{\int_{\beta_{min}}^{1-\alpha}d\beta[\frac{m_{Q'}}{\alpha^{2}\beta}r(m_{Q},m_{Q'})
+\frac{m_{Q}}{\alpha\beta^{2}}r(m_{Q},m_{Q'})\nonumber\\&&{}+\frac{2^{2}m_{Q}m_{Q'}m_{s}}{\alpha\beta}]r(m_{Q},m_{Q'})
+\{\frac{m_{s}}{\alpha(1-\alpha)}[\alpha
m_{Q}^{2}+(1-\alpha) m_{Q'}^{2}-\alpha(1-\alpha)
s]\nonumber\\&&{}
+\frac{m_{Q}m_{s}^{2}}{1-\alpha}+\frac{m_{Q'}m_{s}^{2}}{\alpha}\}[\alpha
m_{Q}^{2}+(1-\alpha) m_{Q'}^{2}-\alpha(1-\alpha)
s]\},\nonumber\\
\rho^{\langle\bar{s}s\rangle^{2}}(s)&=&\frac{\langle\bar{s}s\rangle^{2}}{2^{5}\pi^{2}}\{[2m_{Q}m_{Q'}+(m_{Q}+m_{Q'})m_{s}]\sqrt{(s-m_{Q}^{2}+m_{Q'}^{2})^{2}-4m_{Q'}^{2}s}/s
\nonumber\\&&{}+3m_{s}^{2}\int_{\alpha_{min}}^{\alpha_{max}}d\alpha\alpha(1-\alpha)\},\nonumber\\
\rho^{\langle g\bar{s}\sigma\cdot G s\rangle}(s)&=&\frac{3\langle
g\bar{s}\sigma\cdot G
s\rangle}{2^{8}\pi^{4}}\{\int_{\alpha_{min}}^{\alpha_{max}}d\alpha\{(\frac{m_{Q'}}{\alpha}+\frac{m_{Q}}{1-\alpha})[\alpha
m_{Q}^{2}+(1-\alpha) m_{Q'}^{2}-\alpha(1-\alpha)
s]\nonumber\\&&{}+\frac{2}{3}m_{s}[2\alpha
m_{Q}^{2}+2(1-\alpha) m_{Q'}^{2}-3\alpha(1-\alpha) s]\}+[2m_{Q}m_{Q'}m_{s}\nonumber\\&&{}
+\frac{(m_{Q}+m_{Q'})m_{s}^{2}}{3}]\sqrt{(s-m_{Q}^{2}+m_{Q'}^{2})^{2}-4m_{Q'}^{2}s}/s\},\nonumber\\
\rho^{\langle g^{2}G^{2}\rangle}(s)&=&\frac{\langle
g^{2}G^{2}\rangle}{2^{11}\pi^{6}}\int_{\alpha_{min}}^{\alpha_{max}}d\alpha\int_{\beta_{min}}^{1-\alpha}d\beta(1-\alpha-\beta)
[\frac{m_{Q'}^{2}}{\alpha^{3}}r(m_{Q},m_{Q'})+\frac{m_{Q}^{2}}{\beta^{3}}r(m_{Q},m_{Q'})\nonumber\\&&{}
+\frac{3m_{Q'}m_{s}}{\alpha^{3}}r(m_{Q},m_{Q'})
+\frac{m_{Q'}^{3}m_{s}\beta}{\alpha^{3}}
+\frac{3m_{Q}m_{s}}{\beta^{3}}r(m_{Q},m_{Q'})+\frac{m_{Q}^{3}m_{s}\alpha}{\beta^{3}}\nonumber\\&&{}
+\frac{m_{Q}m_{Q'}^{2}m_{s}}{\alpha^{2}}+\frac{m_{Q}^{2}m_{Q'}m_{s}}{\beta^{2}}
+\frac{3m_{Q}m_{Q'}m_{s}^{2}}{\alpha^{2}}+\frac{3m_{Q}m_{Q'}m_{s}^{2}}{\beta^{2}}],\nonumber\\
\rho^{\langle g^{3}G^{3}\rangle}(s)&=&\frac{\langle
g^{3}G^{3}\rangle}{2^{13}\pi^{6}}\int_{\alpha_{min}}^{\alpha_{max}}d\alpha\int_{\beta_{min}}^{1-\alpha}d\beta(1-\alpha-\beta)
[\frac{1}{\alpha^{3}}r(m_{Q},m_{Q'})+\frac{2m_{Q'}^{2}\beta}{\alpha^{3}}\nonumber\\&&{}+\frac{1}{\beta^{3}}r(m_{Q},m_{Q'})
+\frac{2m_{Q}^{2}\alpha}{\beta^{3}}
+\frac{6m_{Q'}m_{s}\beta}{\alpha^{3}}
+\frac{6m_{Q}m_{s}\alpha}{\beta^{3}}+\frac{m_{Q}m_{s}}{\alpha^{2}}
+\frac{m_{Q'}m_{s}}{\beta^{2}}],\nonumber
\end{eqnarray}
for $(Q\bar{s})_{0}^{*}(\bar{Q'}s)_{0}^{*}$,

\begin{eqnarray}
\rho^{\mbox{pert}}(s)&=&\frac{3}{2^{9}\pi^{6}}\int_{\alpha_{min}}^{\alpha_{max}}d\alpha\int_{\beta_{min}}^{1-\alpha}d\beta(1-\alpha-\beta)
[\frac{1}{\alpha^{3}\beta^{3}}r(m_{Q},m_{Q'})^{2}+\frac{2m_{Q'}m_{s}}{\alpha^{3}\beta^{2}}r(m_{Q},m_{Q'})\nonumber\\&&{}
+\frac{2m_{Q}m_{s}}{\alpha^{2}\beta^{3}}r(m_{Q},m_{Q'})
+\frac{3\cdot2^{2}m_{Q}m_{Q'}m_{s}^{2}}{\alpha^{2}\beta^{2}}]
r(m_{Q},m_{Q'})^{2},\nonumber\\
\rho^{\langle\bar{s}s\rangle}(s)&=&\frac{3\langle\bar{s}s\rangle}{2^{6}\pi^{4}}\int_{\alpha_{min}}^{\alpha_{max}}d\alpha\{\int_{\beta_{min}}^{1-\alpha}d\beta[\frac{m_{Q'}}{\alpha^{2}\beta}r(m_{Q},m_{Q'})
+\frac{m_{Q}}{\alpha\beta^{2}}r(m_{Q},m_{Q'})\nonumber\\&&{}+\frac{2^{3}m_{Q}m_{Q'}m_{s}}{\alpha\beta}]r(m_{Q},m_{Q'})
+\{\frac{2m_{s}}{\alpha(1-\alpha)}[\alpha
m_{Q}^{2}+(1-\alpha) m_{Q'}^{2}-\alpha(1-\alpha)
s]\nonumber\\&&{}
+\frac{m_{Q}m_{s}^{2}}{1-\alpha}+\frac{m_{Q'}m_{s}^{2}}{\alpha}\}[\alpha
m_{Q}^{2}+(1-\alpha) m_{Q'}^{2}-\alpha(1-\alpha)
s]\},\nonumber\\
\rho^{\langle\bar{s}s\rangle^{2}}(s)&=&\frac{\langle\bar{s}s\rangle^{2}}{2^{4}\pi^{2}}\{[2^{2}m_{Q}m_{Q'}+(m_{Q}+m_{Q'})m_{s}]
\sqrt{(s-m_{Q}^{2}+m_{Q'}^{2})^{2}-4m_{Q'}^{2}s}/s\nonumber\\&&{}
+6m_{s}^{2}\int_{\alpha_{min}}^{\alpha_{max}}d\alpha\alpha(1-\alpha)\},\nonumber\\
\rho^{\langle g\bar{s}\sigma\cdot G s\rangle}(s)&=&\frac{\langle
g\bar{s}\sigma\cdot G
s\rangle}{2^{7}\pi^{4}}\{\int_{\alpha_{min}}^{\alpha_{max}}d\alpha\{(\frac{3m_{Q'}}{\alpha}+\frac{3m_{Q}}{1-\alpha})
[\alpha
m_{Q}^{2}+(1-\alpha) m_{Q'}^{2}-\alpha(1-\alpha)
s]\nonumber\\&&{}
+2^{2}m_{s}[2\alpha
m_{Q}^{2}+2(1-\alpha) m_{Q'}^{2}-3\alpha(1-\alpha) s]\}\nonumber\\&&{}
+[3\cdot2^{2}m_{Q}m_{Q'}m_{s}+(m_{Q}+m_{Q'})m_{s}^{2}]
\sqrt{(s-m_{Q}^{2}+m_{Q'}^{2})^{2}-4m_{Q'}^{2}s}/s\},\nonumber\\
\rho^{\langle g^{2}G^{2}\rangle}(s)&=&\frac{\langle
g^{2}G^{2}\rangle}{2^{9}\pi^{6}}\int_{\alpha_{min}}^{\alpha_{max}}d\alpha\int_{\beta_{min}}^{1-\alpha}d\beta(1-\alpha-\beta)
[\frac{m_{Q'}^{2}}{\alpha^{3}}r(m_{Q},m_{Q'})+\frac{m_{Q}^{2}}{\beta^{3}}r(m_{Q},m_{Q'})\nonumber\\&&{}
+\frac{3m_{Q'}m_{s}}{2\alpha^{3}}r(m_{Q},m_{Q'})
+\frac{m_{Q'}^{3}m_{s}\beta}{2\alpha^{3}}
+\frac{3m_{Q}m_{s}}{2\beta^{3}}r(m_{Q},m_{Q'})+\frac{m_{Q}^{3}m_{s}\alpha}{2\beta^{3}}\nonumber\\&&{}
+\frac{m_{Q}m_{Q'}^{2}m_{s}}{2\alpha^{2}}
+\frac{m_{Q}^{2}m_{Q'}m_{s}}{2\beta^{2}}
+\frac{3m_{Q}m_{Q'}m_{s}^{2}}{\alpha^{2}}+\frac{3m_{Q}m_{Q'}m_{s}^{2}}{\beta^{2}}],\nonumber\\
\rho^{\langle g^{3}G^{3}\rangle}(s)&=&\frac{\langle
g^{3}G^{3}\rangle}{2^{11}\pi^{6}}\int_{\alpha_{min}}^{\alpha_{max}}d\alpha\int_{\beta_{min}}^{1-\alpha}d\beta(1-\alpha-\beta)
[\frac{1}{\alpha^{3}}r(m_{Q},m_{Q'})+\frac{2m_{Q'}^{2}\beta}{\alpha^{3}}\nonumber\\&&{}+\frac{1}{\beta^{3}}r(m_{Q},m_{Q'})
+\frac{2m_{Q}^{2}\alpha}{\beta^{3}}
+\frac{3m_{Q'}m_{s}\beta}{\alpha^{3}}
+\frac{3m_{Q}m_{s}\alpha}{\beta^{3}}
+\frac{m_{Q}m_{s}}{2\alpha^{2}}
+\frac{m_{Q'}m_{s}}{2\beta^{2}}],\nonumber
\end{eqnarray}
for $(Q\bar{s})_{1}(\bar{Q'}s)_{1}$,

\begin{eqnarray}
\rho^{\mbox{pert}}(s)&=&\frac{3}{2^{12}\pi^{6}}\int_{\alpha_{min}}^{\alpha_{max}}d\alpha\int_{\beta_{min}}^{1-\alpha}d\beta(1-\alpha-\beta)
[\frac{1}{\alpha^{3}\beta^{3}}(1+\alpha+\beta)r(m_{Q},m_{Q'})^{2}\nonumber\\&&{}
+\frac{2^{2}m_{Q'}m_{s}}{\alpha^{3}\beta^{2}}(1+\alpha+\beta)r(m_{Q},m_{Q'})
+\frac{2^{3}m_{Q}m_{s}}{\alpha^{2}\beta^{3}}r(m_{Q},m_{Q'})\nonumber\\&&{}
+\frac{3\cdot2^{3}m_{Q}m_{Q'}m_{s}^{2}}{\alpha^{2}\beta^{2}}]r(m_{Q},m_{Q'})^{2},\nonumber\\
\rho^{\langle\bar{s}s\rangle}(s)&=&\frac{3\langle\bar{s}s\rangle}{2^{7}\pi^{4}}\int_{\alpha_{min}}^{\alpha_{max}}d\alpha\{\int_{\beta_{min}}^{1-\alpha}d\beta[\frac{m_{Q'}}{\alpha^{2}\beta}(\alpha+\beta)r(m_{Q},m_{Q'})
+\frac{m_{Q}}{\alpha\beta^{2}}r(m_{Q},m_{Q'})\nonumber\\&&{}-\frac{m_{s}}{\alpha\beta}r(m_{Q},m_{Q'})+\frac{2^{2}m_{Q}m_{Q'}m_{s}}{\alpha\beta}
-\frac{m_{Q'}m_{s}^{2}}{\alpha}]r(m_{Q},m_{Q'})\nonumber\\&&{}+\{\frac{m_{s}}{\alpha(1-\alpha)}[\alpha
m_{Q}^{2}+(1-\alpha) m_{Q'}^{2}
-\alpha(1-\alpha)
s]+\frac{m_{Q}m_{s}^{2}}{1-\alpha}+\frac{m_{Q'}m_{s}^{2}}{\alpha}\}[\alpha
m_{Q}^{2}\nonumber\\&&{}
+(1-\alpha) m_{Q'}^{2}-\alpha(1-\alpha) s]\},\nonumber\\
\rho^{\langle\bar{s}s\rangle^{2}}(s)&=&\frac{\langle\bar{s}s\rangle^{2}}{2^{5}\pi^{2}}[(2m_{Q}m_{Q'}+m_{Q}m_{s})\sqrt{(s-m_{Q}^{2}+m_{Q'}^{2})^{2}-4m_{Q'}^{2}s}/s\nonumber\\&&{}
+\int_{\alpha_{min}}^{\alpha_{max}}d\alpha(m_{Q'}m_{s}+\frac{3m_{s}^{2}}{2}\alpha)(1-\alpha)],\nonumber\\
\rho^{\langle g\bar{s}\sigma\cdot G s\rangle}(s)&=&\frac{3\langle
g\bar{s}\sigma\cdot G
s\rangle}{2^{8}\pi^{4}}\{\int_{\alpha_{min}}^{\alpha_{max}}d\alpha\{-\frac{m_{Q'}}{\alpha}\int_{\beta_{min}}^{1-\alpha}d\beta r(m_{Q},m_{Q'})
+(\frac{m_{Q'}}{\alpha}+\frac{m_{Q}}{1-\alpha})[\alpha
m_{Q}^{2}\nonumber\\&&{}
+(1-\alpha) m_{Q'}^{2}
-\alpha(1-\alpha)
s]+\frac{2m_{s}}{3}[\alpha
m_{Q}^{2}+(1-\alpha) m_{Q'}^{2}-2\alpha(1-\alpha) s]\nonumber\\&&{}
+\frac{m_{Q'}m_{s}^{2}}{3}(1-\alpha)\}
+(2m_{Q}m_{Q'}m_{s}
+\frac{m_{Q}m_{s}^{2}}{3})\sqrt{(s-m_{Q}^{2}+m_{Q'}^{2})^{2}-4m_{Q'}^{2}s}/s\},\nonumber\\
\rho^{\langle g^{2}G^{2}\rangle}(s)&=&\frac{\langle
g^{2}G^{2}\rangle}{2^{12}\pi^{6}}\int_{\alpha_{min}}^{\alpha_{max}}d\alpha\int_{\beta_{min}}^{1-\alpha}d\beta(1-\alpha-\beta)
[\frac{m_{Q'}^{2}}{\alpha^{3}}(1+\alpha+\beta)r(m_{Q},m_{Q'})\nonumber\\&&{}
+\frac{m_{Q}^{2}}{\beta^{3}}(1+\alpha+\beta)r(m_{Q},m_{Q'})
+\frac{3m_{Q'}m_{s}}{\alpha^{3}}(1+\alpha+\beta)r(m_{Q},m_{Q'})\nonumber\\&&{}
+\frac{m_{Q'}^{3}m_{s}\beta}{\alpha^{3}}(1+\alpha+\beta)+\frac{6m_{Q}m_{s}}{\beta^{3}}r(m_{Q},m_{Q'})
+\frac{2m_{Q}^{3}m_{s}\alpha}{\beta^{3}}
+\frac{2m_{Q}m_{Q'}^{2}m_{s}}{\alpha^{2}}\nonumber\\&&{}
+\frac{m_{Q}^{2}m_{Q'}m_{s}}{\beta^{2}}(1+\alpha+\beta)
+\frac{6m_{Q}m_{Q'}m_{s}^{2}}{\alpha^{2}}+\frac{6m_{Q}m_{Q'}m_{s}^{2}}{\beta^{2}}],\nonumber\\
\rho^{\langle g^{3}G^{3}\rangle}(s)&=&\frac{\langle
g^{3}G^{3}\rangle}{2^{14}\pi^{6}}\int_{\alpha_{min}}^{\alpha_{max}}d\alpha\int_{\beta_{min}}^{1-\alpha}d\beta(1-\alpha-\beta)
[\frac{1}{\alpha^{3}}(1+\alpha+\beta)r(m_{Q},m_{Q'})\nonumber\\&&{}+\frac{2m_{Q'}^{2}\beta}{\alpha^{3}}(1+\alpha+\beta)
+\frac{1}{\beta^{3}}(1+\alpha+\beta)r(m_{Q},m_{Q'})+\frac{2m_{Q}^{2}\alpha}{\beta^{3}}(1+\alpha+\beta)\nonumber\\&&{}
+\frac{6m_{Q'}m_{s}\beta}{\alpha^{3}}(1+\alpha+\beta)+\frac{3\cdot2^{2}m_{Q}m_{s}\alpha}{\beta^{3}}+\frac{2m_{Q}m_{s}}{\alpha^{2}}
+\frac{m_{Q'}m_{s}}{\beta^{2}}(1+\alpha+\beta)],\nonumber
\end{eqnarray}
for $(Q\bar{s})_{1}(\bar{Q'}s)_{0}^{*}$,

\begin{eqnarray}
\rho^{\mbox{pert}}(s)&=&\frac{3}{2^{11}\pi^{6}}\int_{\alpha_{min}}^{\alpha_{max}}d\alpha\int_{\beta_{min}}^{1-\alpha}d\beta(1-\alpha-\beta)
[\frac{1}{\alpha^{3}\beta^{3}}r(m_{Q},m_{Q'})^{2}\nonumber\\&&{}+\frac{2^{2}m_{Q'}m_{s}}{\alpha^{3}\beta^{2}}r(m_{Q},m_{Q'})
-\frac{2^{2}m_{Q}m_{s}}{\alpha^{2}\beta^{3}}r(m_{Q},m_{Q'})
-\frac{3\cdot2^{2}m_{Q}m_{Q'}m_{s}^{2}}{\alpha^{2}\beta^{2}}]\nonumber\\&&{}\times
r(m_{Q},m_{Q'})^{2},\nonumber\\
\rho^{\langle\bar{s}s\rangle}(s)&=&\frac{3\langle\bar{s}s\rangle}{2^{7}\pi^{4}}\int_{\alpha_{min}}^{\alpha_{max}}d\alpha\{\int_{\beta_{min}}^{1-\alpha}d\beta[\frac{m_{Q'}}{\alpha^{2}\beta}r(m_{Q},m_{Q'})
-\frac{m_{Q}}{\alpha\beta^{2}}r(m_{Q},m_{Q'})\nonumber\\&&{}-\frac{2^{2}m_{Q}m_{Q'}m_{s}}{\alpha\beta}]r(m_{Q},m_{Q'})
+\{\frac{m_{s}}{\alpha(1-\alpha)}[\alpha
m_{Q}^{2}+(1-\alpha) m_{Q'}^{2}-\alpha(1-\alpha)
s]\nonumber\\&&{}
-\frac{m_{Q}m_{s}^{2}}{1-\alpha}+\frac{m_{Q'}m_{s}^{2}}{\alpha}\}[\alpha
m_{Q}^{2}+(1-\alpha) m_{Q'}^{2}-\alpha(1-\alpha)
s]\},\nonumber\\
\rho^{\langle\bar{s}s\rangle^{2}}(s)&=&\frac{\langle\bar{s}s\rangle^{2}}{2^{5}\pi^{2}}\{[-2m_{Q}m_{Q'}-(m_{Q}-m_{Q'})m_{s}]
\sqrt{(s-m_{Q}^{2}+m_{Q'}^{2})^{2}-4m_{Q'}^{2}s}/s\nonumber\\&&{}
+3m_{s}^{2}\int_{\alpha_{min}}^{\alpha_{max}}d\alpha\alpha(1-\alpha)\},\nonumber\\
\rho^{\langle g\bar{s}\sigma\cdot G s\rangle}(s)&=&\frac{3\langle
g\bar{s}\sigma\cdot G
s\rangle}{2^{8}\pi^{4}}\{\int_{\alpha_{min}}^{\alpha_{max}}d\alpha\{(\frac{m_{Q'}}{\alpha}-\frac{m_{Q}}{1-\alpha})[\alpha
m_{Q}^{2}
+(1-\alpha) m_{Q'}^{2}-\alpha(1-\alpha)
s]\nonumber\\&&{}+\frac{2m_{s}}{3}[2\alpha
m_{Q}^{2}+2(1-\alpha) m_{Q'}^{2}
-3\alpha(1-\alpha) s]\}+[-2m_{Q}m_{Q'}m_{s}\nonumber\\&&{}
-\frac{(m_{Q}-m_{Q'})m_{s}^{2}}{3}]\sqrt{(s-m_{Q}^{2}+m_{Q'}^{2})^{2}-4m_{Q'}^{2}s}/s\},\nonumber\\
\rho^{\langle g^{2}G^{2}\rangle}(s)&=&\frac{\langle
g^{2}G^{2}\rangle}{2^{11}\pi^{6}}\int_{\alpha_{min}}^{\alpha_{max}}d\alpha\int_{\beta_{min}}^{1-\alpha}d\beta(1-\alpha-\beta)
[\frac{m_{Q'}^{2}}{\alpha^{3}}r(m_{Q},m_{Q'})+\frac{m_{Q}^{2}}{\beta^{3}}r(m_{Q},m_{Q'})\nonumber\\&&{}
+\frac{3m_{Q'}m_{s}}{\alpha^{3}}r(m_{Q},m_{Q'})
+\frac{m_{Q'}^{3}m_{s}\beta}{\alpha^{3}}
-\frac{3m_{Q}m_{s}}{\beta^{3}}r(m_{Q},m_{Q'})-\frac{m_{Q}^{3}m_{s}\alpha}{\beta^{3}}\nonumber\\&&{}
-\frac{m_{Q}m_{Q'}^{2}m_{s}}{\alpha^{2}}+\frac{m_{Q}^{2}m_{Q'}m_{s}}{\beta^{2}}
-\frac{3m_{Q}m_{Q'}m_{s}^{2}}{\alpha^{2}}-\frac{3m_{Q}m_{Q'}m_{s}^{2}}{\beta^{2}}],\nonumber\\
\rho^{\langle g^{3}G^{3}\rangle}(s)&=&\frac{\langle
g^{3}G^{3}\rangle}{2^{13}\pi^{6}}\int_{\alpha_{min}}^{\alpha_{max}}d\alpha\int_{\beta_{min}}^{1-\alpha}d\beta(1-\alpha-\beta)
[\frac{1}{\alpha^{3}}r(m_{Q},m_{Q'})+\frac{2m_{Q'}^{2}\beta}{\alpha^{3}}\nonumber\\&&{}+\frac{1}{\beta^{3}}r(m_{Q},m_{Q'})
+\frac{2m_{Q}^{2}\alpha}{\beta^{3}}
+\frac{6m_{Q'}m_{s}\beta}{\alpha^{3}}
-\frac{6m_{Q}m_{s}\alpha}{\beta^{3}}
-\frac{m_{Q}m_{s}}{\alpha^{2}}
+\frac{m_{Q'}m_{s}}{\beta^{2}}],\nonumber
\end{eqnarray}
for $(Q\bar{s})(\bar{Q'}s)_{0}^{*}$,

\begin{eqnarray}
\rho^{\mbox{pert}}(s)&=&\frac{3}{2^{12}\pi^{6}}\int_{\alpha_{min}}^{\alpha_{max}}d\alpha\int_{\beta_{min}}^{1-\alpha}d\beta(1-\alpha-\beta)
[\frac{1}{\alpha^{3}\beta^{3}}(1+\alpha+\beta)r(m_{Q},m_{Q'})^{2}\nonumber\\&&{}
-\frac{2^{2}m_{Q'}m_{s}}{\alpha^{3}\beta^{2}}(1+\alpha+\beta)r(m_{Q},m_{Q'})
+\frac{2^{3}m_{Q}m_{s}}{\alpha^{2}\beta^{3}}r(m_{Q},m_{Q'})\nonumber\\&&{}
-\frac{3\cdot2^{3}m_{Q}m_{Q'}m_{s}^{2}}{\alpha^{2}\beta^{2}}]r(m_{Q},m_{Q'})^{2},\nonumber\\
\rho^{\langle\bar{s}s\rangle}(s)&=&\frac{3\langle\bar{s}s\rangle}{2^{7}\pi^{4}}\int_{\alpha_{min}}^{\alpha_{max}}d\alpha\{\int_{\beta_{min}}^{1-\alpha}d\beta[-\frac{m_{Q'}}{\alpha^{2}\beta}(\alpha+\beta)r(m_{Q},m_{Q'})
+\frac{m_{Q}}{\alpha\beta^{2}}r(m_{Q},m_{Q'})\nonumber\\&&{}-\frac{m_{s}}{\alpha\beta}r(m_{Q},m_{Q'})
-\frac{2^{2}m_{Q}m_{Q'}m_{s}}{\alpha\beta}
+\frac{m_{Q'}m_{s}^{2}}{\alpha}]r(m_{Q},m_{Q'})\nonumber\\&&{}
+\{\frac{m_{s}}{\alpha(1-\alpha)}[\alpha
m_{Q}^{2}+(1-\alpha) m_{Q'}^{2}-\alpha(1-\alpha)
s]+\frac{m_{Q}m_{s}^{2}}{1-\alpha}
-\frac{m_{Q'}m_{s}^{2}}{\alpha}\}[\alpha
m_{Q}^{2}\nonumber\\&&{}
+(1-\alpha) m_{Q'}^{2}-\alpha(1-\alpha)
s]\},\nonumber\\
\rho^{\langle\bar{s}s\rangle^{2}}(s)&=&\frac{\langle\bar{s}s\rangle^{2}}{2^{5}\pi^{2}}[(-2m_{Q}m_{Q'}+m_{Q}m_{s})
\sqrt{(s-m_{Q}^{2}+m_{Q'}^{2})^{2}-4m_{Q'}^{2}s}/s\nonumber\\&&{}
+\int_{\alpha_{min}}^{\alpha_{max}}d\alpha(-m_{Q'}m_{s}+\frac{3m_{s}^{2}}{2}\alpha)(1-\alpha)],\nonumber\\
\rho^{\langle g\bar{s}\sigma\cdot G s\rangle}(s)&=&\frac{3\langle
g\bar{s}\sigma\cdot G
s\rangle}{2^{8}\pi^{4}}\{\int_{\alpha_{min}}^{\alpha_{max}}d\alpha\{\frac{m_{Q'}}{\alpha}\int_{\beta_{min}}^{1-\alpha}d\beta r(m_{Q},m_{Q'})
-(\frac{m_{Q'}}{\alpha}-\frac{m_{Q}}{1-\alpha})[\alpha
m_{Q}^{2}\nonumber\\&&{}+(1-\alpha) m_{Q'}^{2}-\alpha(1-\alpha)
s]
+\frac{2m_{s}}{3}[\alpha
m_{Q}^{2}+(1-\alpha) m_{Q'}^{2}-2\alpha(1-\alpha) s]\nonumber\\&&{}
-\frac{m_{Q'}m_{s}^{2}}{3}(1-\alpha)\}+(-2m_{Q}m_{Q'}m_{s}
+\frac{m_{Q}m_{s}^{2}}{3})\nonumber\\&&{}\times\sqrt{(s-m_{Q}^{2}+m_{Q'}^{2})^{2}-4m_{Q'}^{2}s}/s\}
,\nonumber\\
\rho^{\langle g^{2}G^{2}\rangle}(s)&=&\frac{\langle
g^{2}G^{2}\rangle}{2^{12}\pi^{6}}\int_{\alpha_{min}}^{\alpha_{max}}d\alpha\int_{\beta_{min}}^{1-\alpha}d\beta(1-\alpha-\beta)
[\frac{m_{Q'}^{2}}{\alpha^{3}}(1+\alpha+\beta)r(m_{Q},m_{Q'})\nonumber\\&&{}
+\frac{m_{Q}^{2}}{\beta^{3}}(1+\alpha+\beta)r(m_{Q},m_{Q'})
-\frac{3m_{Q'}m_{s}}{\alpha^{3}}(1+\alpha+\beta)r(m_{Q},m_{Q'})\nonumber\\&&{}
-\frac{m_{Q'}^{3}m_{s}\beta}{\alpha^{3}}(1+\alpha+\beta)+\frac{6m_{Q}m_{s}}{\beta^{3}}r(m_{Q},m_{Q'})
+\frac{2m_{Q}^{3}m_{s}\alpha}{\beta^{3}}
+\frac{2m_{Q}m_{Q'}^{2}m_{s}}{\alpha^{2}}\nonumber\\&&{}
-\frac{m_{Q}^{2}m_{Q'}m_{s}}{\beta^{2}}(1+\alpha+\beta)
-\frac{6m_{Q}m_{Q'}m_{s}^{2}}{\alpha^{2}}-\frac{6m_{Q}m_{Q'}m_{s}^{2}}{\beta^{2}}],\nonumber\\
\rho^{\langle g^{3}G^{3}\rangle}(s)&=&\frac{\langle
g^{3}G^{3}\rangle}{2^{14}\pi^{6}}\int_{\alpha_{min}}^{\alpha_{max}}d\alpha\int_{\beta_{min}}^{1-\alpha}d\beta(1-\alpha-\beta)
[\frac{1}{\alpha^{3}}(1+\alpha+\beta)r(m_{Q},m_{Q'})\nonumber\\&&{}+\frac{2m_{Q'}^{2}\beta}{\alpha^{3}}(1+\alpha+\beta)
+\frac{1}{\beta^{3}}(1+\alpha+\beta)r(m_{Q},m_{Q'})+\frac{2m_{Q}^{2}\alpha}{\beta^{3}}(1+\alpha+\beta)\nonumber\\&&{}
-\frac{6m_{Q'}m_{s}\beta}{\alpha^{3}}(1+\alpha+\beta)+\frac{3\cdot2^{2}m_{Q}m_{s}\alpha}{\beta^{3}}+\frac{2m_{Q}m_{s}}{\alpha^{2}}
-\frac{m_{Q'}m_{s}}{\beta^{2}}(1+\alpha+\beta)]
,\nonumber
\end{eqnarray}
for $(Q\bar{s})_{1}(\bar{Q'}s)$,

\begin{eqnarray}
\rho^{\mbox{pert}}(s)&=&\frac{3}{2^{12}\pi^{6}}\int_{\alpha_{min}}^{\alpha_{max}}d\alpha\int_{\beta_{min}}^{1-\alpha}d\beta(1-\alpha-\beta)
[\frac{1}{\alpha^{3}\beta^{3}}(1+\alpha+\beta)r(m_{Q},m_{Q'})^{2}\nonumber\\&&{}
+\frac{2^{2}m_{Q'}m_{s}}{\alpha^{3}\beta^{2}}(1+\alpha+\beta)r(m_{Q},m_{Q'})
-\frac{2^{3}m_{Q}m_{s}}{\alpha^{2}\beta^{3}}r(m_{Q},m_{Q'})\nonumber\\&&{}
-\frac{3\cdot2^{3}m_{Q}m_{Q'}m_{s}^{2}}{\alpha^{2}\beta^{2}}]
r(m_{Q},m_{Q'})^{2},\nonumber\\
\rho^{\langle\bar{s}s\rangle}(s)&=&\frac{3\langle\bar{s}s\rangle}{2^{7}\pi^{4}}\int_{\alpha_{min}}^{\alpha_{max}}d\alpha\{\int_{\beta_{min}}^{1-\alpha}d\beta
[\frac{m_{Q'}}{\alpha^{2}\beta}(\alpha+\beta)r(m_{Q},m_{Q'})
-\frac{m_{Q}}{\alpha\beta^{2}}r(m_{Q},m_{Q'})\nonumber\\&&{}
-\frac{m_{s}}{\alpha\beta}r(m_{Q},m_{Q'})-\frac{2^{2}m_{Q}m_{Q'}m_{s}}{\alpha\beta}
-\frac{m_{Q'}m_{s}^{2}}{\alpha}]r(m_{Q},m_{Q'})\nonumber\\&&{}
+\{\frac{m_{s}}{\alpha(1-\alpha)}[\alpha
m_{Q}^{2}+(1-\alpha) m_{Q'}^{2}-\alpha(1-\alpha)
s]
-\frac{m_{Q}m_{s}^{2}}{1-\alpha}+\frac{m_{Q'}m_{s}^{2}}{\alpha}\}[\alpha
m_{Q}^{2}\nonumber\\&&{}
+(1-\alpha) m_{Q'}^{2}-\alpha(1-\alpha) s]\},\nonumber\\
\rho^{\langle\bar{s}s\rangle^{2}}(s)&=&\frac{\langle\bar{s}s\rangle^{2}}{2^{5}\pi^{2}}[(-2m_{Q}m_{Q'}-m_{Q}m_{s})
\sqrt{(s-m_{Q}^{2}+m_{Q'}^{2})^{2}-4m_{Q'}^{2}s}/s\nonumber\\&&{}
+\int_{\alpha_{min}}^{\alpha_{max}}d\alpha(m_{Q'}m_{s}+\frac{3m_{s}^{2}}{2}\alpha)(1-\alpha)],\nonumber\\
\rho^{\langle g\bar{s}\sigma\cdot G s\rangle}(s)&=&\frac{3\langle
g\bar{s}\sigma\cdot G
s\rangle}{2^{8}\pi^{4}}\{\int_{\alpha_{min}}^{\alpha_{max}}d\alpha\{-\frac{m_{Q'}}{\alpha}\int_{\beta_{min}}^{1-\alpha}d\beta r(m_{Q},m_{Q'})+(\frac{m_{Q'}}{\alpha}-\frac{m_{Q}}{1-\alpha})[\alpha
m_{Q}^{2}\nonumber\\&&{}+(1-\alpha) m_{Q'}^{2}
-\alpha(1-\alpha)
s]+\frac{2m_{s}}{3}[\alpha
m_{Q}^{2}+(1-\alpha) m_{Q'}^{2}
-2\alpha(1-\alpha) s]\nonumber\\&&{}+\frac{m_{Q'}m_{s}^{2}}{3}(1-\alpha)\}+(-2m_{Q}m_{Q'}m_{s}
-\frac{m_{Q}m_{s}^{2}}{3})\nonumber\\&&{}\times\sqrt{(s-m_{Q}^{2}+m_{Q'}^{2})^{2}-4m_{Q'}^{2}s}/s\}
,\nonumber\\
\rho^{\langle g^{2}G^{2}\rangle}(s)&=&\frac{\langle
g^{2}G^{2}\rangle}{2^{12}\pi^{6}}\int_{\alpha_{min}}^{\alpha_{max}}d\alpha\int_{\beta_{min}}^{1-\alpha}d\beta(1-\alpha-\beta)
[\frac{m_{Q'}^{2}}{\alpha^{3}}(1+\alpha+\beta)r(m_{Q},m_{Q'})\nonumber\\&&{}
+\frac{m_{Q}^{2}}{\beta^{3}}(1+\alpha+\beta)r(m_{Q},m_{Q'})
+\frac{3m_{Q'}m_{s}}{\alpha^{3}}(1+\alpha+\beta)r(m_{Q},m_{Q'})\nonumber\\&&{}
+\frac{m_{Q'}^{3}m_{s}\beta}{\alpha^{3}}(1+\alpha+\beta)-\frac{6m_{Q}m_{s}}{\beta^{3}}r(m_{Q},m_{Q'})
-\frac{2m_{Q}^{3}m_{s}\alpha}{\beta^{3}}
-\frac{2m_{Q}m_{Q'}^{2}m_{s}}{\alpha^{2}}\nonumber\\&&{}
+\frac{m_{Q}^{2}m_{Q'}m_{s}}{\beta^{2}}(1+\alpha+\beta)
-\frac{6m_{Q}m_{Q'}m_{s}^{2}}{\alpha^{2}}-\frac{6m_{Q}m_{Q'}m_{s}^{2}}{\beta^{2}}],\nonumber\\
\rho^{\langle g^{3}G^{3}\rangle}(s)&=&\frac{\langle
g^{3}G^{3}\rangle}{2^{14}\pi^{6}}\int_{\alpha_{min}}^{\alpha_{max}}d\alpha\int_{\beta_{min}}^{1-\alpha}d\beta(1-\alpha-\beta)
[\frac{1}{\alpha^{3}}(1+\alpha+\beta)r(m_{Q},m_{Q'})\nonumber\\&&{}+\frac{2m_{Q'}^{2}\beta}{\alpha^{3}}(1+\alpha+\beta)
+\frac{1}{\beta^{3}}(1+\alpha+\beta)r(m_{Q},m_{Q'})
+\frac{2m_{Q}^{2}\alpha}{\beta^{3}}(1+\alpha+\beta)\nonumber\\&&{}+\frac{6m_{Q'}m_{s}\beta}{\alpha^{3}}(1+\alpha+\beta)
-\frac{3\cdot2^{2}m_{Q}m_{s}\alpha}{\beta^{3}}-\frac{2m_{Q}m_{s}}{\alpha^{2}}
+\frac{m_{Q'}m_{s}}{\beta^{2}}(1+\alpha+\beta)]
,\nonumber
\end{eqnarray}
for $(Q\bar{s})^{*}(\bar{Q'}s)_{0}^{*}$, and

\begin{eqnarray}
\rho^{\mbox{pert}}(s)&=&\frac{3}{2^{9}\pi^{6}}\int_{\alpha_{min}}^{\alpha_{max}}d\alpha\int_{\beta_{min}}^{1-\alpha}d\beta(1-\alpha-\beta)
[\frac{1}{\alpha^{3}\beta^{3}}r(m_{Q},m_{Q'})^{2}+\frac{2m_{Q'}m_{s}}{\alpha^{3}\beta^{2}}r(m_{Q},m_{Q'})\nonumber\\&&{}
-\frac{2m_{Q}m_{s}}{\alpha^{2}\beta^{3}}r(m_{Q},m_{Q'})
-\frac{3\cdot2^{2}m_{Q}m_{Q'}m_{s}^{2}}{\alpha^{2}\beta^{2}}]r(m_{Q},m_{Q'})^{2},\nonumber\\
\rho^{\langle\bar{s}s\rangle}(s)&=&\frac{3\langle\bar{s}s\rangle}{2^{6}\pi^{4}}\int_{\alpha_{min}}^{\alpha_{max}}d\alpha\{\int_{\beta_{min}}^{1-\alpha}d\beta[\frac{m_{Q'}}{\alpha^{2}\beta}r(m_{Q},m_{Q'})
-\frac{m_{Q}}{\alpha\beta^{2}}r(m_{Q},m_{Q'})\nonumber\\&&{}-\frac{2^{3}m_{Q}m_{Q'}m_{s}}{\alpha\beta}]r(m_{Q},m_{Q'})
+\{\frac{2m_{s}}{\alpha(1-\alpha)}[\alpha
m_{Q}^{2}+(1-\alpha) m_{Q'}^{2}-\alpha(1-\alpha)
s]\nonumber\\&&{}-\frac{m_{Q}m_{s}^{2}}{1-\alpha}+\frac{m_{Q'}m_{s}^{2}}{\alpha}\}[\alpha
m_{Q}^{2}+(1-\alpha) m_{Q'}^{2}
-\alpha(1-\alpha)
s]\},\nonumber\\
\rho^{\langle\bar{s}s\rangle^{2}}(s)&=&\frac{\langle\bar{s}s\rangle^{2}}{2^{3}\pi^{2}}\{[-2m_{Q}m_{Q'}-\frac{(m_{Q}-m_{Q'})m_{s}}{2}]
\sqrt{(s-m_{Q}^{2}+m_{Q'}^{2})^{2}-4m_{Q'}^{2}s}/s\nonumber\\&&{}
+3m_{s}^{2}\int_{\alpha_{min}}^{\alpha_{max}}d\alpha\alpha(1-\alpha)\},\nonumber\\
\rho^{\langle g\bar{s}\sigma\cdot G s\rangle}(s)&=&\frac{3\langle
g\bar{s}\sigma\cdot G
s\rangle}{2^{7}\pi^{4}}\{\int_{\alpha_{min}}^{\alpha_{max}}d\alpha\{(\frac{m_{Q'}}{\alpha}-\frac{m_{Q}}{1-\alpha})[\alpha
m_{Q}^{2}+(1-\alpha) m_{Q'}^{2}-\alpha(1-\alpha)
s]\nonumber\\&&{}+\frac{2^{2}m_{s}}{3}[2\alpha
m_{Q}^{2}+2(1-\alpha) m_{Q'}^{2}
-3\alpha(1-\alpha) s]\}+[-2^{2}m_{Q}m_{Q'}m_{s}\nonumber\\&&{}
-\frac{(m_{Q}-m_{Q'})m_{s}^{2}}{3}]\sqrt{(s-m_{Q}^{2}+m_{Q'}^{2})^{2}-4m_{Q'}^{2}s}/s\}
,\nonumber\\
\rho^{\langle g^{2}G^{2}\rangle}(s)&=&\frac{\langle
g^{2}G^{2}\rangle}{2^{9}\pi^{6}}\int_{\alpha_{min}}^{\alpha_{max}}d\alpha\int_{\beta_{min}}^{1-\alpha}d\beta(1-\alpha-\beta)
[\frac{m_{Q'}^{2}}{\alpha^{3}}r(m_{Q},m_{Q'})
+\frac{m_{Q}^{2}}{\beta^{3}}r(m_{Q},m_{Q'})\nonumber\\&&{}
+\frac{3m_{Q'}m_{s}}{2\alpha^{3}}r(m_{Q},m_{Q'})
+\frac{m_{Q'}^{3}m_{s}\beta}{2\alpha^{3}}
-\frac{3m_{Q}m_{s}}{2\beta^{3}}r(m_{Q},m_{Q'})-\frac{m_{Q}^{3}m_{s}\alpha}{2\beta^{3}}\nonumber\\&&{}
-\frac{m_{Q}m_{Q'}^{2}m_{s}}{2\alpha^{2}}+\frac{m_{Q}^{2}m_{Q'}m_{s}}{2\beta^{2}}
-\frac{3m_{Q}m_{Q'}m_{s}^{2}}{\alpha^{2}}-\frac{3m_{Q}m_{Q'}m_{s}^{2}}{\beta^{2}}],\nonumber\\
\rho^{\langle g^{3}G^{3}\rangle}(s)&=&\frac{\langle
g^{3}G^{3}\rangle}{2^{11}\pi^{6}}\int_{\alpha_{min}}^{\alpha_{max}}d\alpha\int_{\beta_{min}}^{1-\alpha}d\beta(1-\alpha-\beta)[\frac{1}{\alpha^{3}}r(m_{Q},m_{Q'})
+\frac{2m_{Q'}^{2}\beta}{\alpha^{3}}\nonumber\\&&{}
+\frac{1}{\beta^{3}}r(m_{Q},m_{Q'})
+\frac{2m_{Q}^{2}\alpha}{\beta^{3}}
+\frac{3m_{Q'}m_{s}\beta}{\alpha^{3}}
-\frac{3m_{Q}m_{s}\alpha}{\beta^{3}}-\frac{m_{Q}m_{s}}{2\alpha^{2}}+\frac{m_{Q'}m_{s}}{2\beta^{2}}],\nonumber
\end{eqnarray}
for $(Q\bar{s})^{*}(\bar{Q'}s)_{1}$. The integration limits are
given by $\alpha_{min}=\frac{s-m_{Q}^{2}+m_{Q'}^{2}-\sqrt{(s-m_{Q}^{2}+m_{Q'}^{2})^{2}-4m_{Q'}^{2}s}}{2s}$,
$\alpha_{max}=\frac{s-m_{Q}^{2}+m_{Q'}^{2}+\sqrt{(s-m_{Q}^{2}+m_{Q'}^{2})^{2}-4m_{Q'}^{2}s}}{2s}$,
and $\beta_{min}=\frac{\alpha m_{Q}^{2}}{s\alpha-m_{Q'}^{2}}$.

\section*{Acknowledgments}
J. R. Zhang is very indebted to Ming Zhong for helpful discussions.


\begin{figure}[b]
\centerline{\epsfysize=5.2truecm
\epsfbox{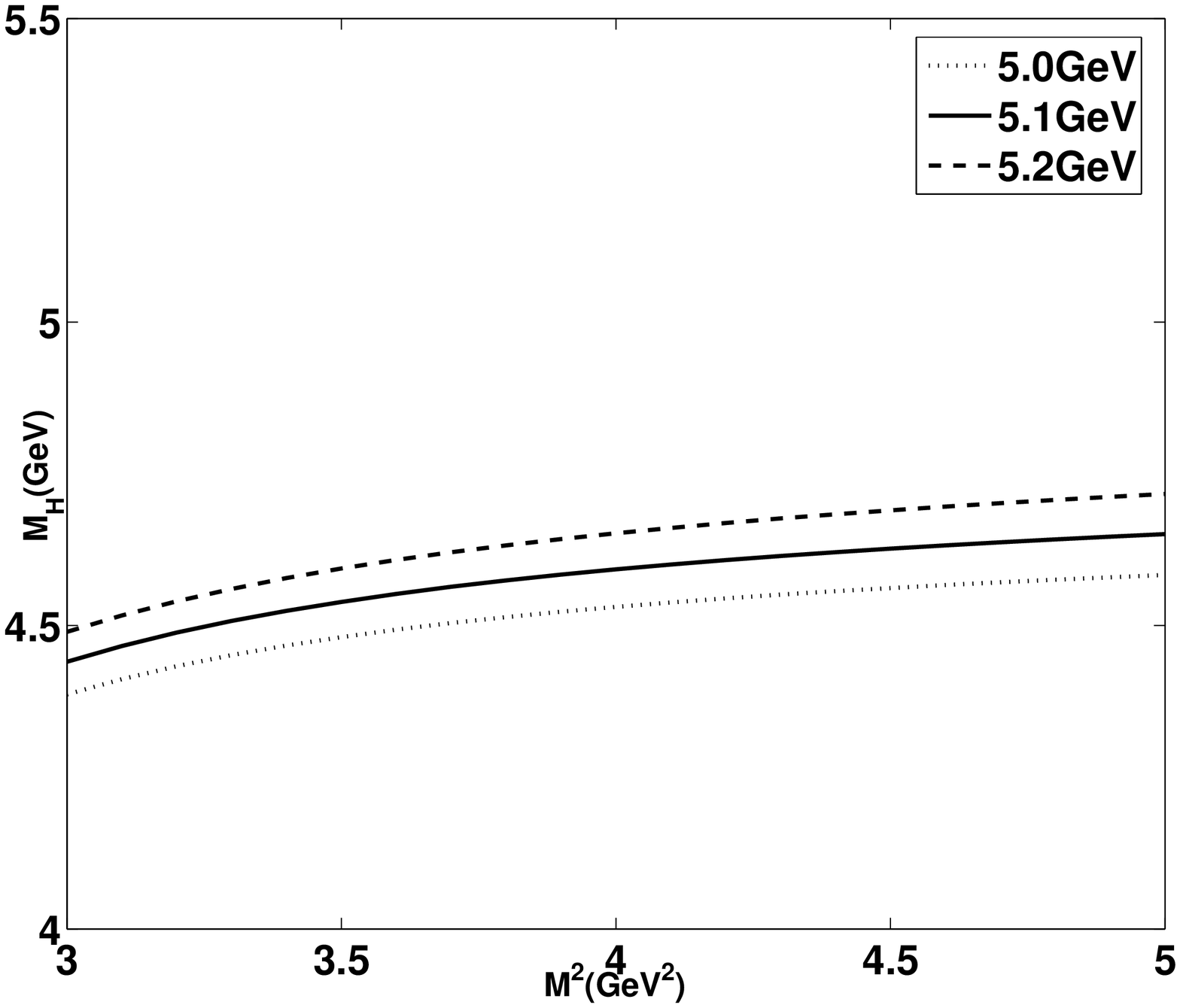}\epsfysize=5.2truecm\epsfbox{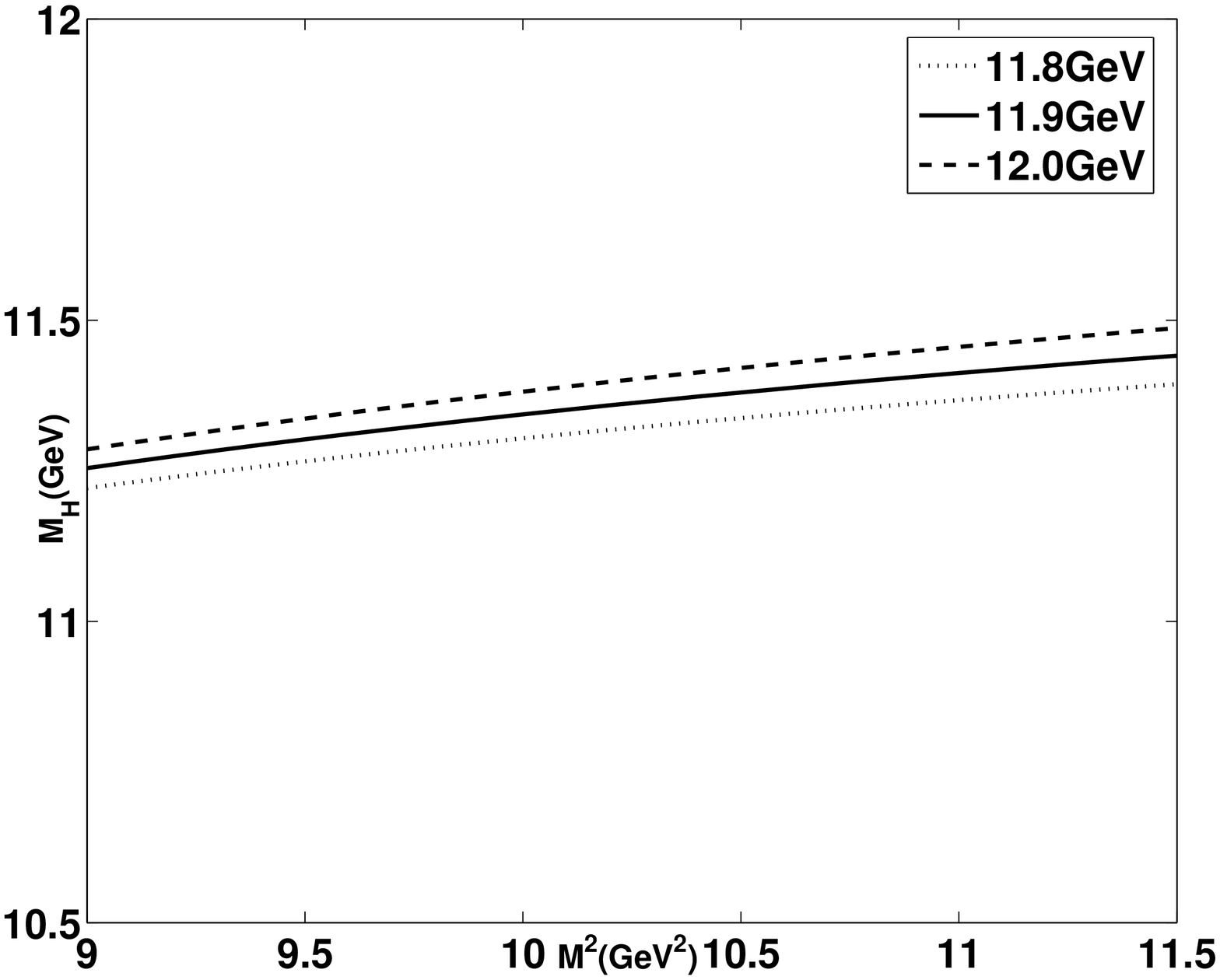}}\caption{The
dependence on $M^2$ for the masses of $D_{s0}^{*}\bar{D}_{s0}^{*}$
and $B_{s0}^{*}\bar{B}_{s0}^{*}$ from sum rule (\ref{sum rule}). The
continuum thresholds are taken as $\sqrt{s_0}=5.0\sim5.2~\mbox{GeV}$
and $\sqrt{s_0}=11.8\sim12.0~\mbox{GeV}$, respectively.}
\label{fig:1}
\end{figure}

\begin{figure}
\centerline{\epsfysize=5.2truecm
\epsfbox{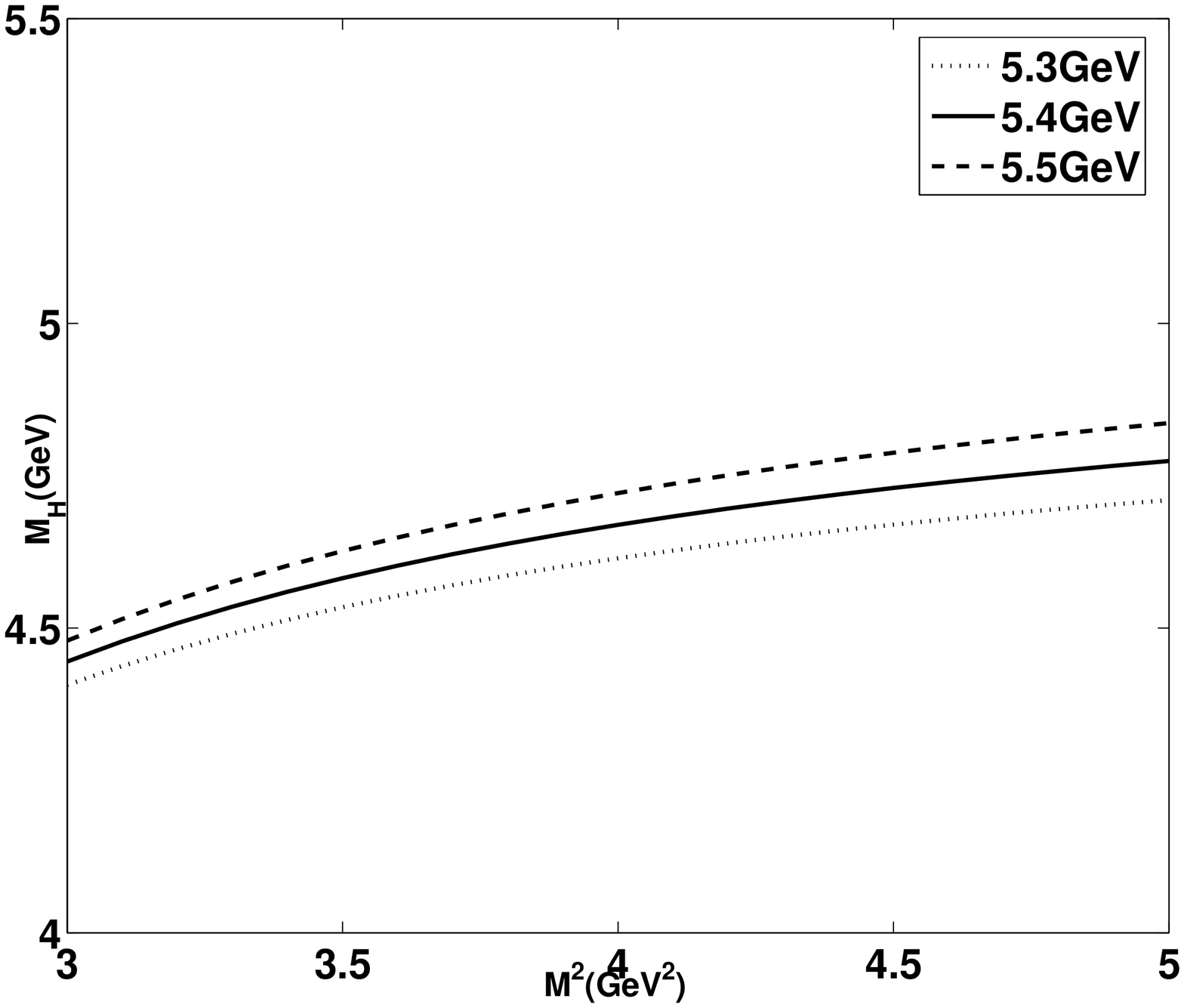}\epsfysize=5.2truecm\epsfbox{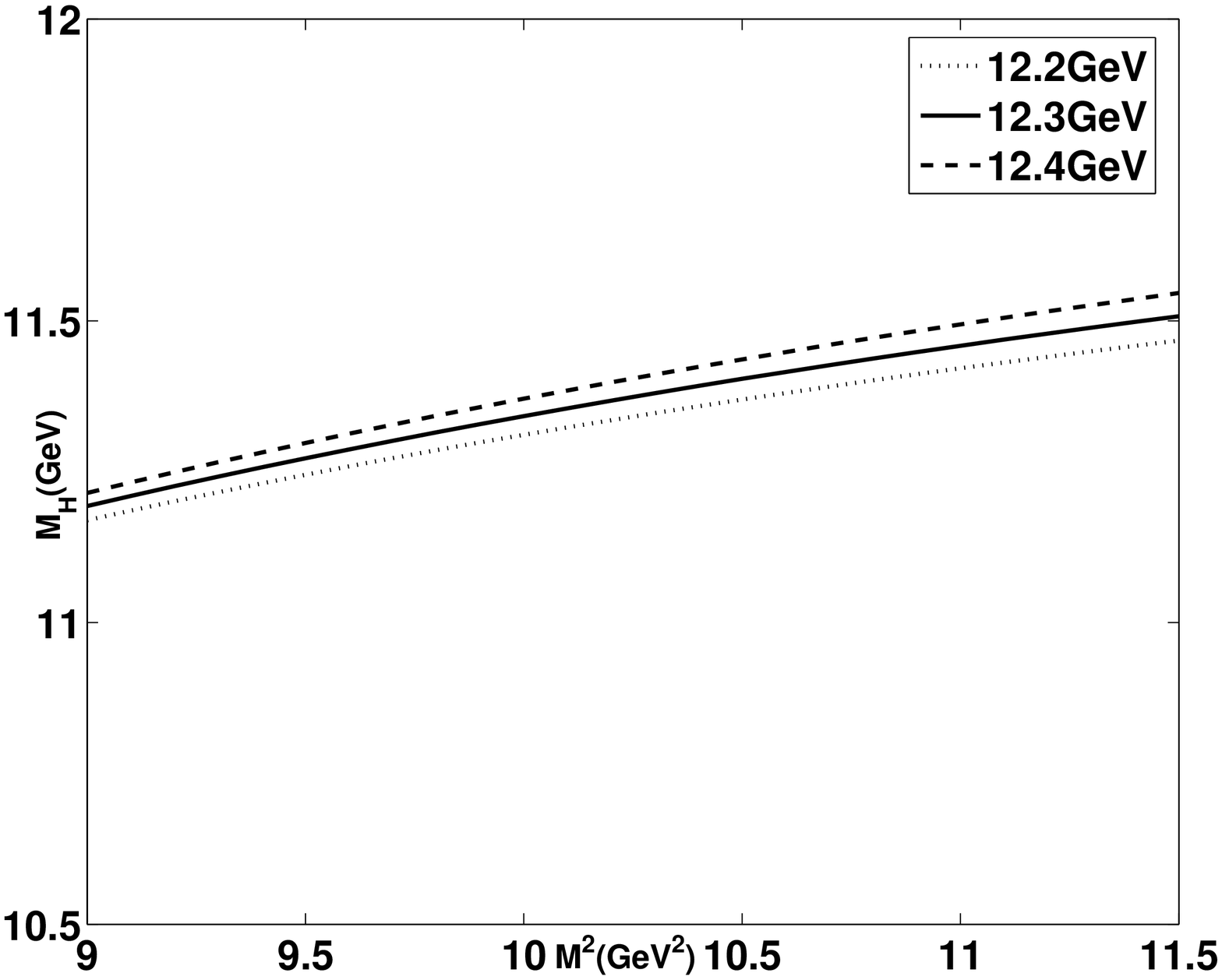}}\caption{The
dependence on $M^2$ for the masses of $D_{s1}\bar{D}_{s1}$ and
$B_{s1}\bar{B}_{s1}$ from sum rule (\ref{sum rule}). The continuum
thresholds are taken as $\sqrt{s_0}=5.3\sim5.5~\mbox{GeV}$ and
$\sqrt{s_0}=12.2\sim12.4~\mbox{GeV}$, respectively.} \label{fig:2}
\end{figure}

\begin{figure}
\centerline{\epsfysize=5.2truecm
\epsfbox{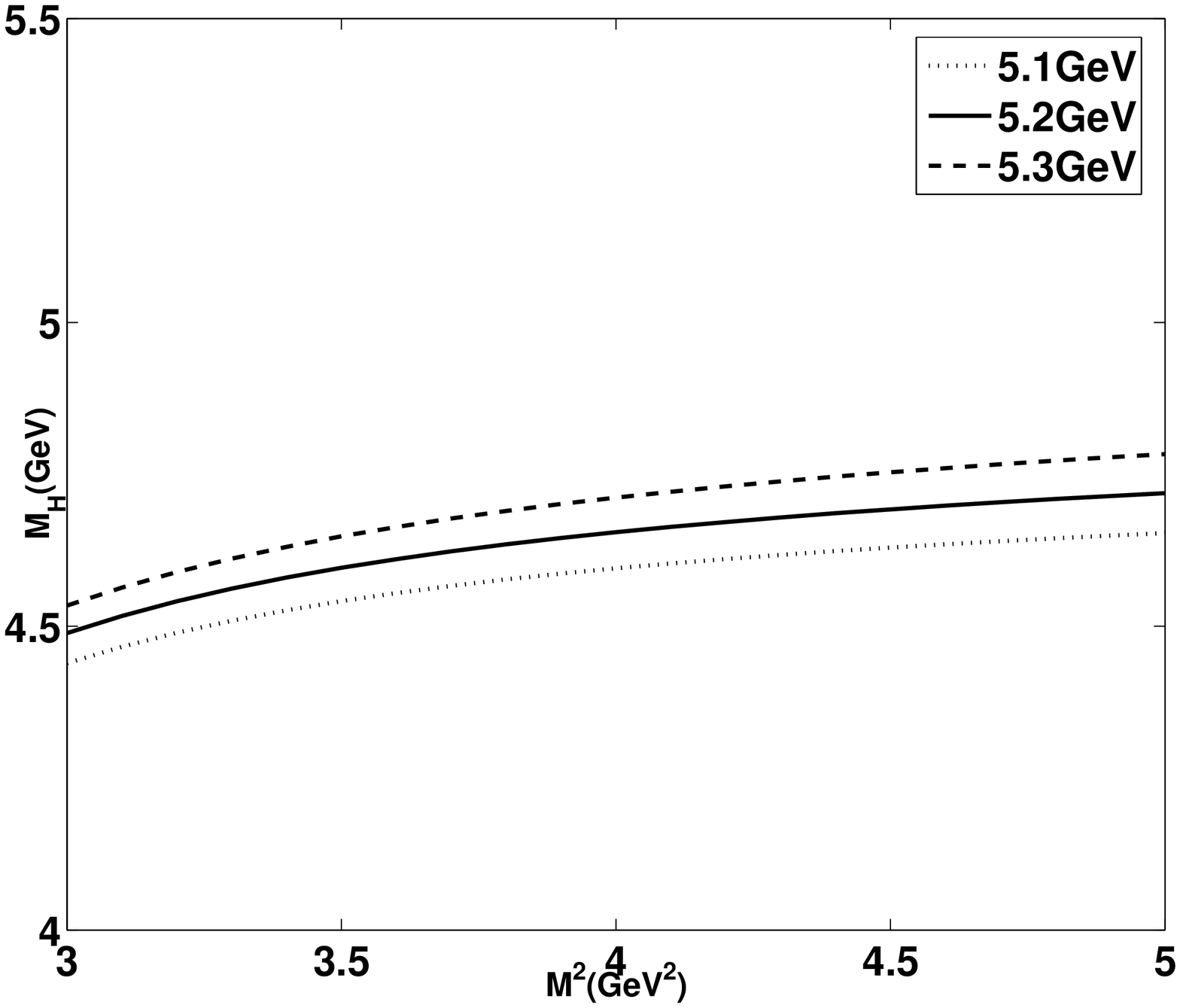}\epsfysize=5.2truecm\epsfbox{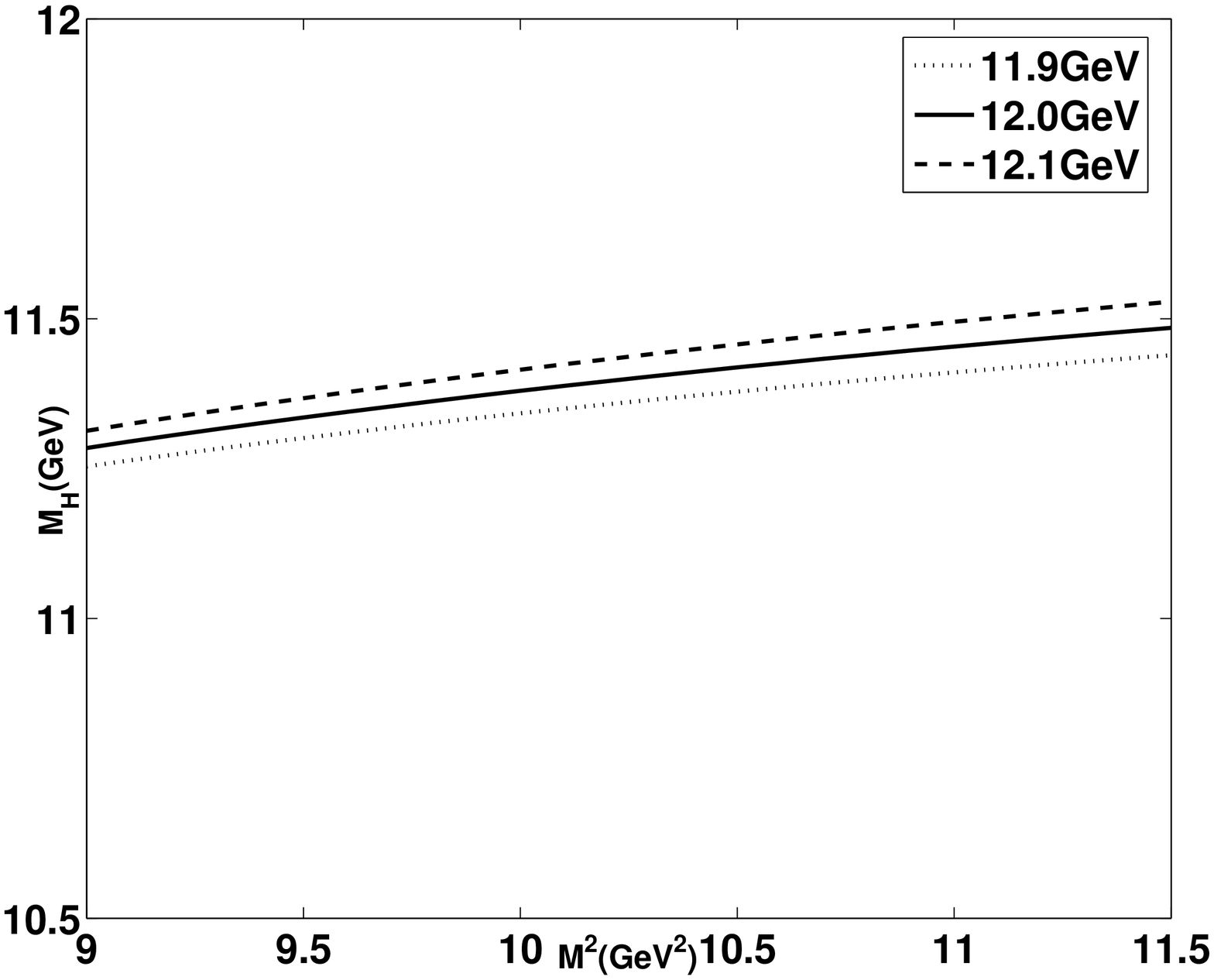}}\caption{The
dependence on $M^2$ for the masses of $D_{s1}\bar{D}_{s0}^{*}$ and
$B_{s1}\bar{B}_{s0}^{*}$ from sum rule (\ref{sum rule 1}). The
continuum thresholds are taken as $\sqrt{s_0}=5.1\sim5.3~\mbox{GeV}$
and $\sqrt{s_0}=11.9\sim12.1~\mbox{GeV}$, respectively.}
\label{fig:3}
\end{figure}

\begin{figure}
\centerline{\epsfysize=5.2truecm
\epsfbox{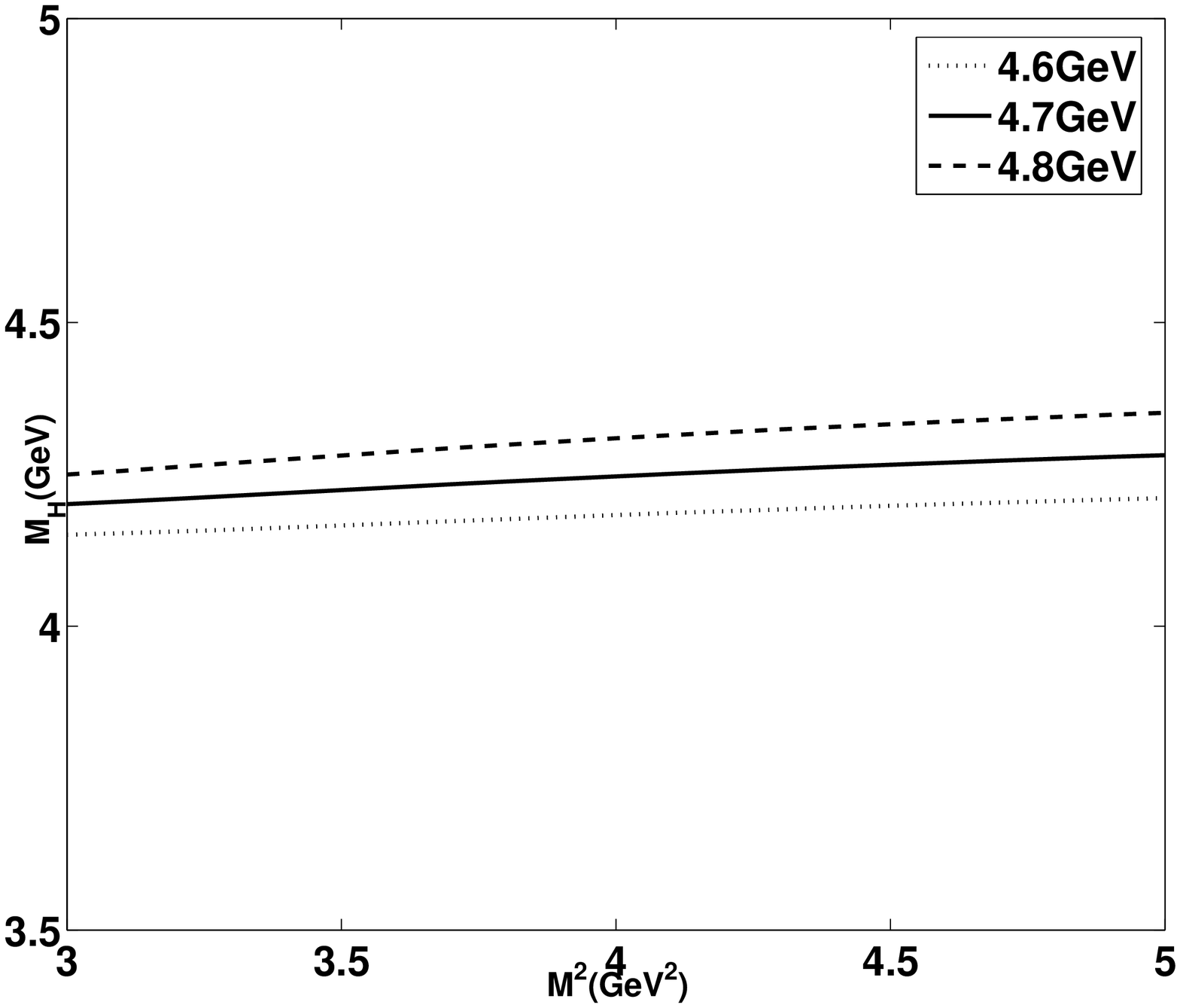}\epsfysize=5.2truecm\epsfbox{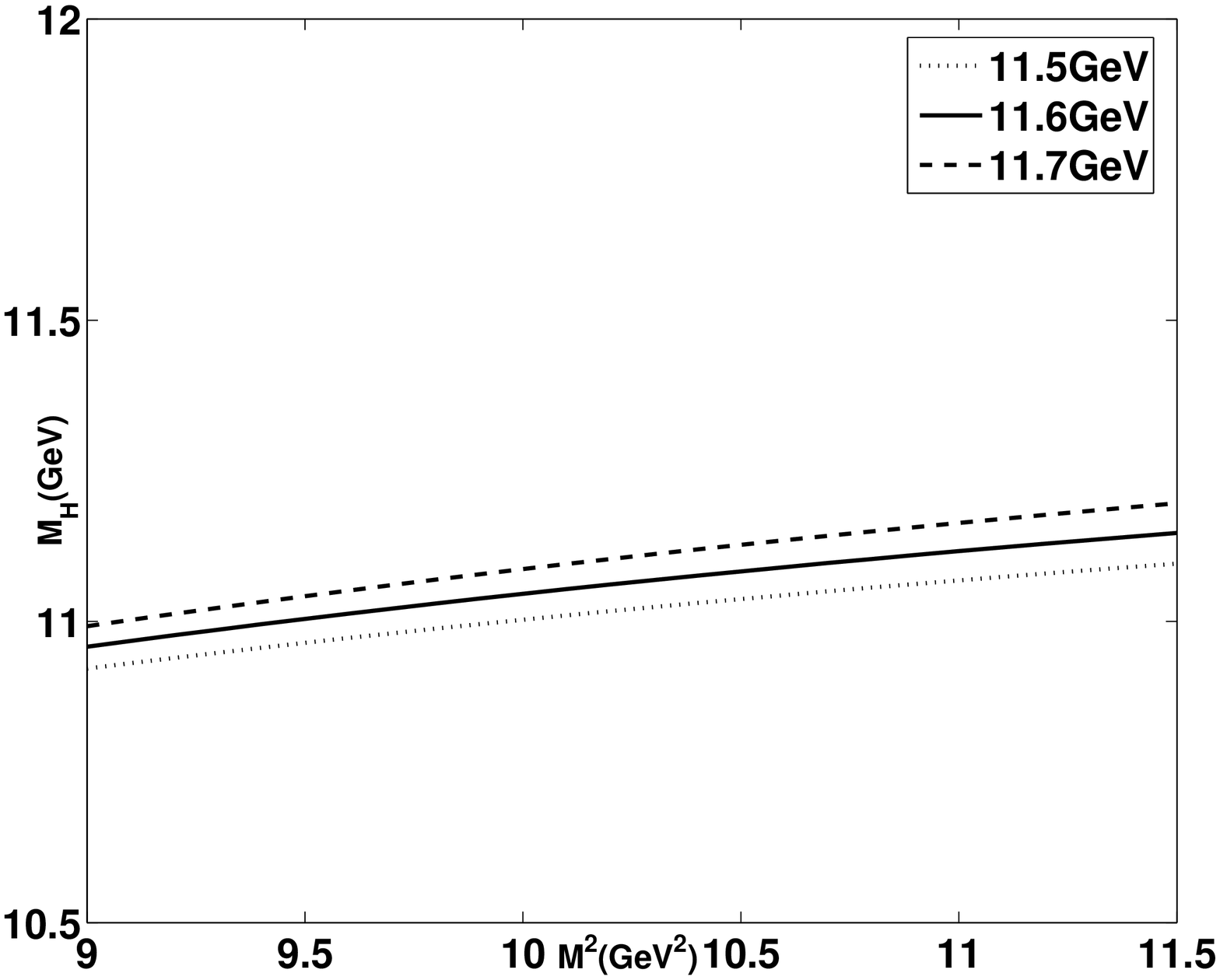}}\caption{The
dependence on $M^2$ for the masses of $D_{s}\bar{D}_{s0}^{*}$ and
$B_{s}\bar{B}_{s0}^{*}$ from sum rule (\ref{sum rule}). The
continuum thresholds are taken as $\sqrt{s_0}=4.6\sim4.8~\mbox{GeV}$
and $\sqrt{s_0}=11.5\sim11.7~\mbox{GeV}$, respectively.}
\label{fig:4}
\end{figure}

\begin{figure}
\centerline{\epsfysize=5.2truecm
\epsfbox{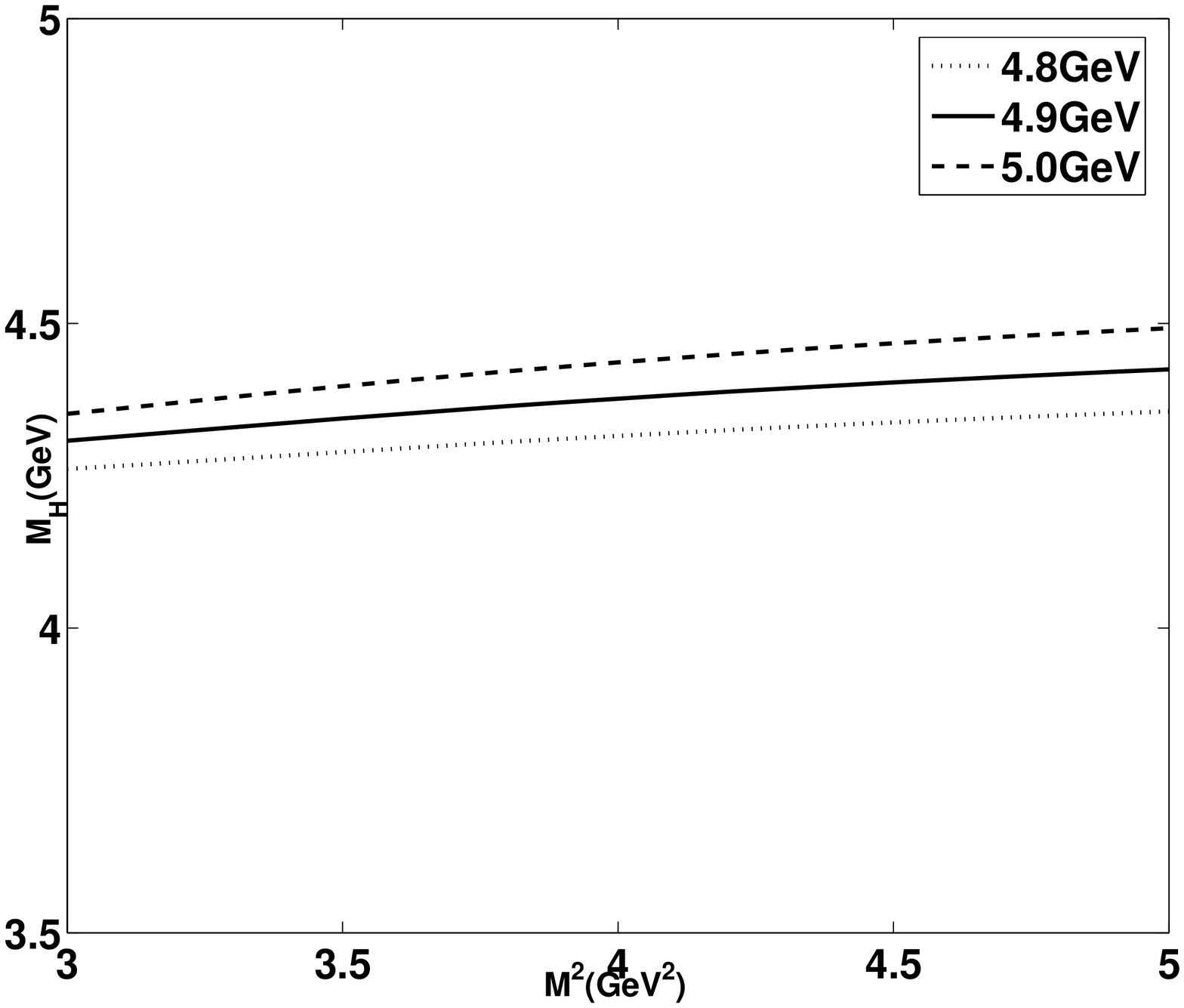}\epsfysize=5.2truecm\epsfbox{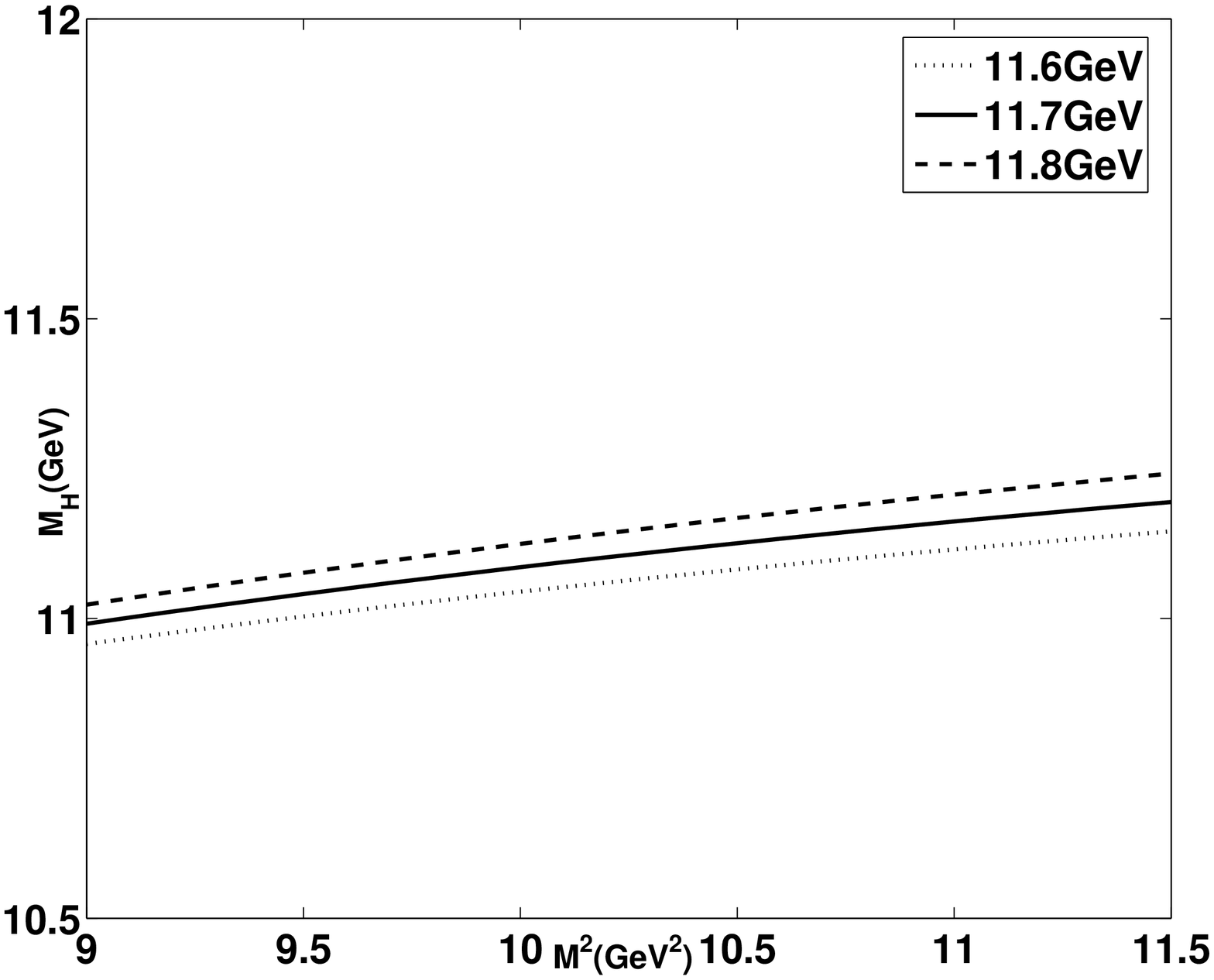}}\caption{The
dependence on $M^2$ for the masses of $D_{s1}\bar{D}_{s}$ and
$B_{s1}\bar{B}_{s}$ from sum rule (\ref{sum rule 1}). The continuum
thresholds are taken as $\sqrt{s_0}=4.8\sim5.0~\mbox{GeV}$ and
$\sqrt{s_0}=11.6\sim11.8~\mbox{GeV}$, respectively.} \label{fig:5}
\end{figure}

\begin{figure}
\centerline{\epsfysize=5.2truecm
\epsfbox{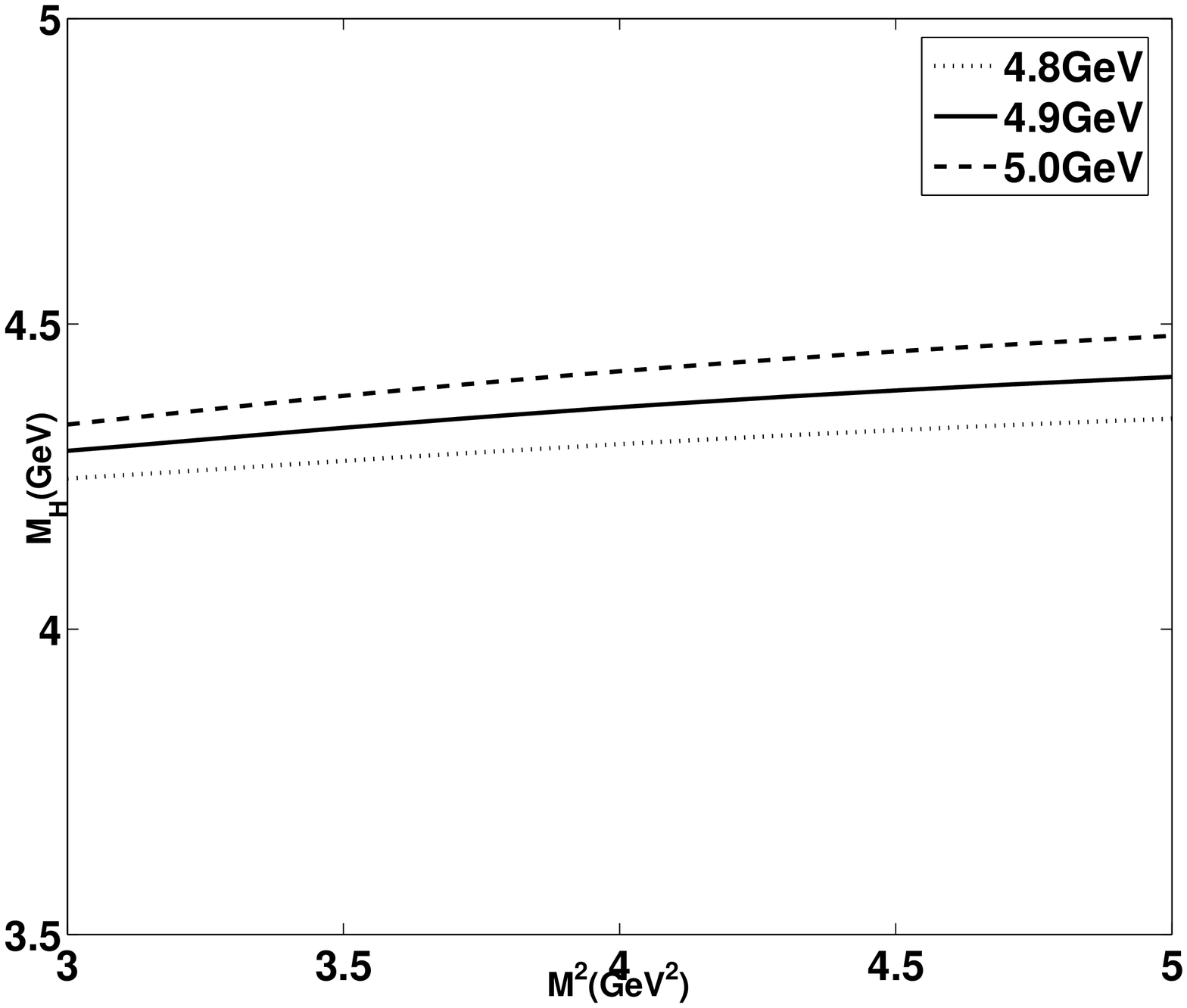}\epsfysize=5.2truecm\epsfbox{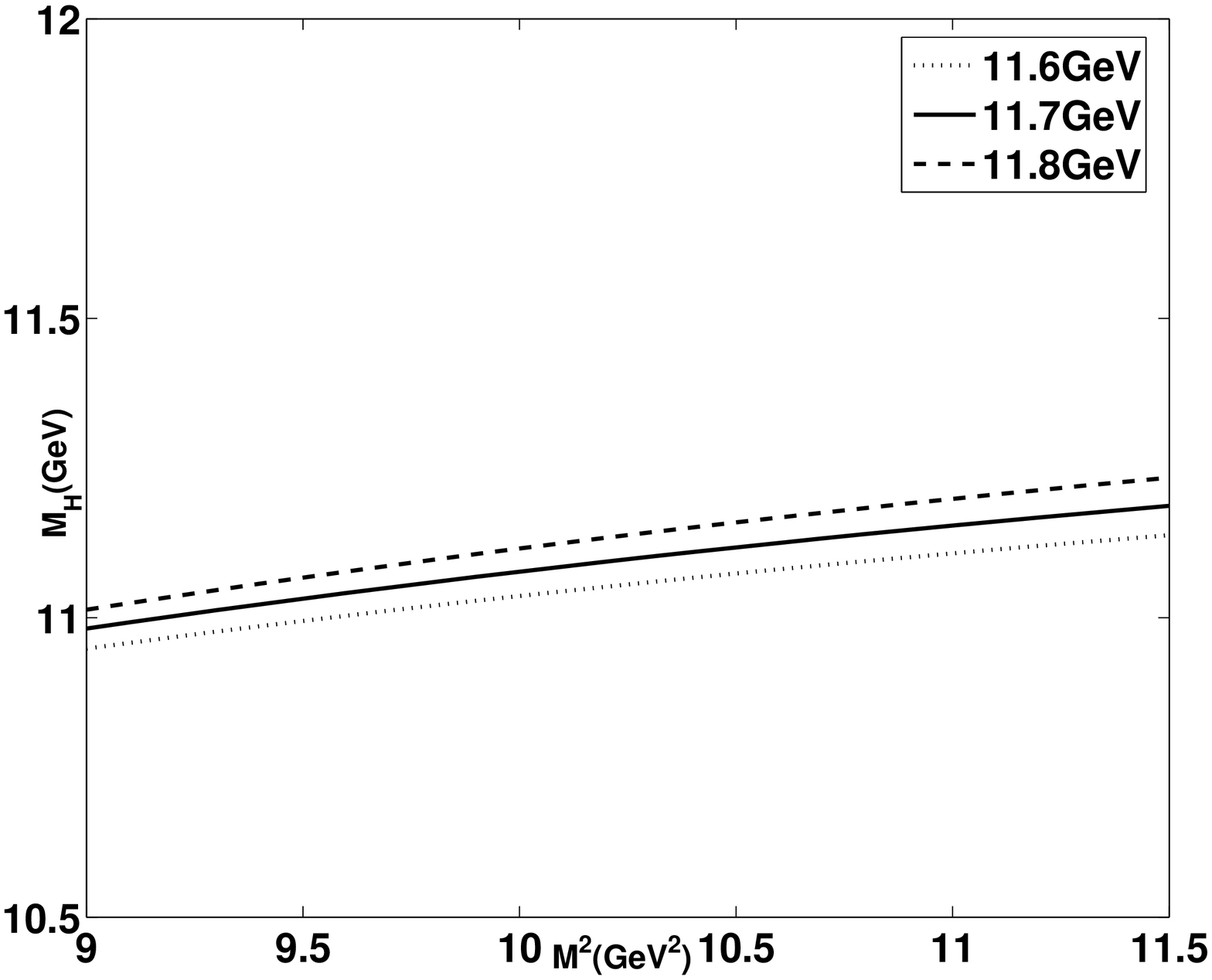}}\caption{The
dependence on $M^2$ for the masses of $D_{s}^{*}\bar{D}_{s0}^{*}$
and $B_{s}^{*}\bar{B}_{s0}^{*}$ from sum rule (\ref{sum rule 1}).
The continuum thresholds are taken as
$\sqrt{s_0}=4.8\sim5.0~\mbox{GeV}$ and
$\sqrt{s_0}=11.6\sim11.8~\mbox{GeV}$, respectively.} \label{fig:6}
\end{figure}

\begin{figure}
\centerline{\epsfysize=5.2truecm
\epsfbox{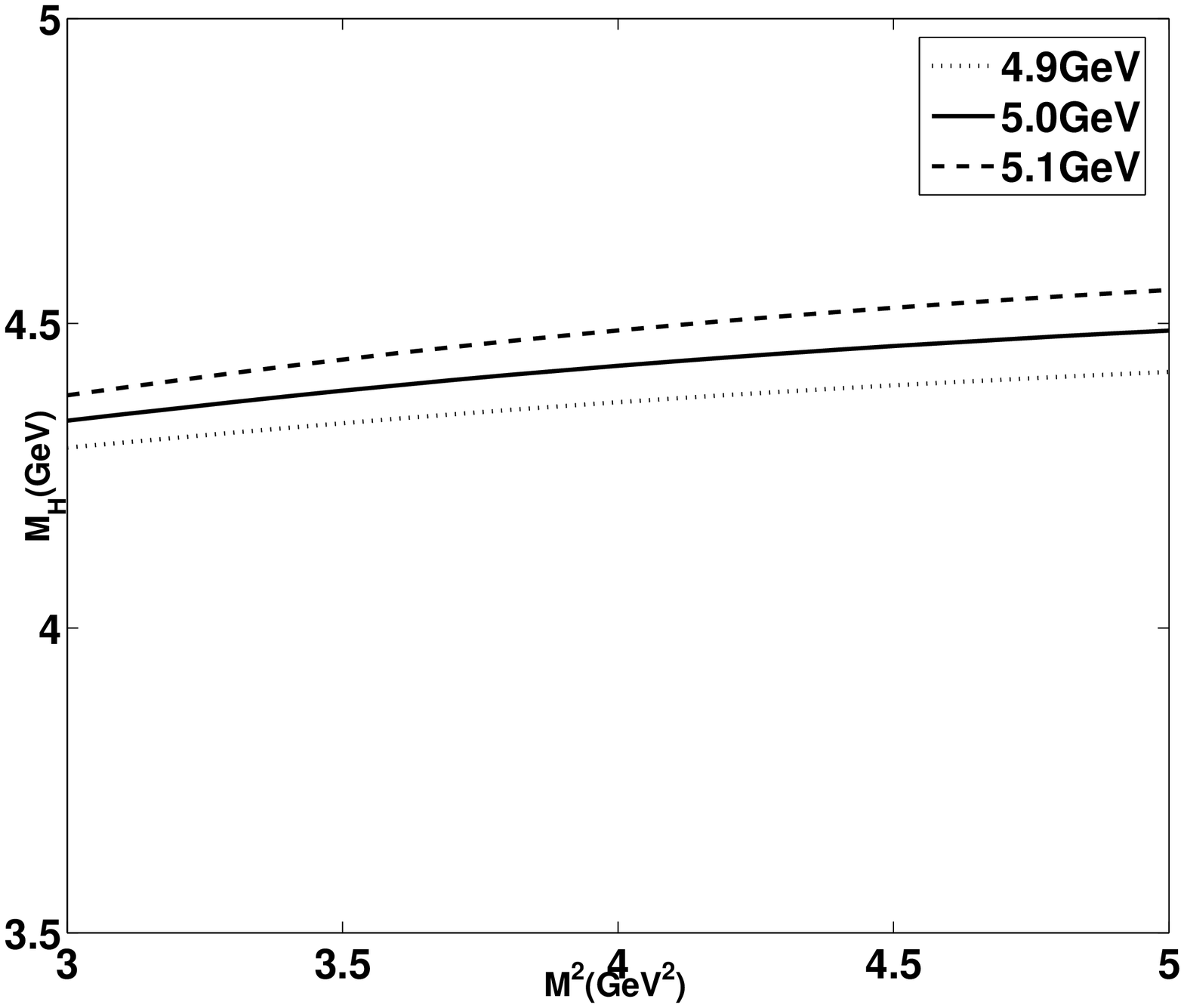}\epsfysize=5.2truecm\epsfbox{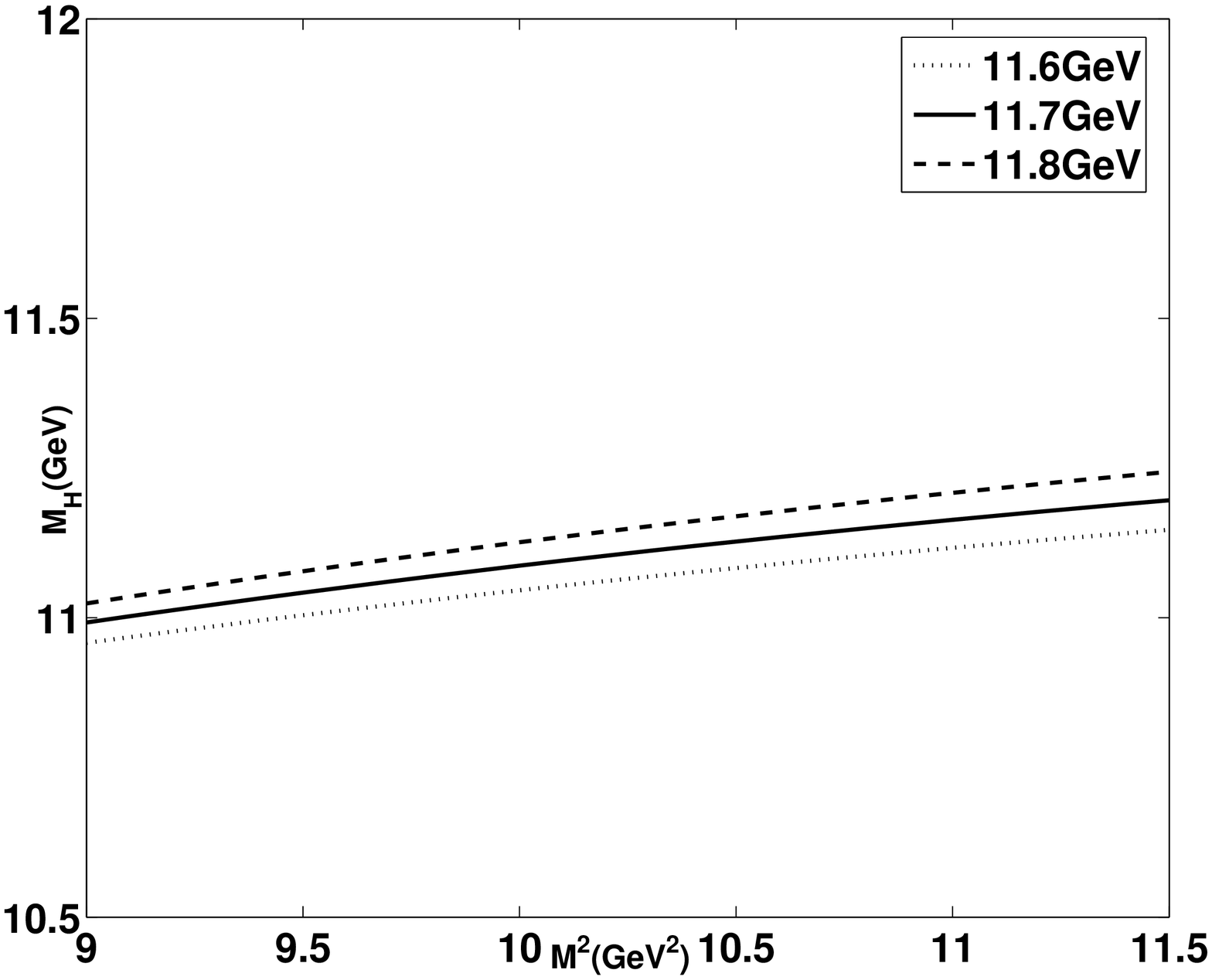}}\caption{The
dependence on $M^2$ for the masses of $D_{s}^{*}\bar{D}_{s1}$ and
$B_{s}^{*}\bar{B}_{s1}$ from sum rule (\ref{sum rule}). The
continuum thresholds are taken as $\sqrt{s_0}=4.9\sim5.1~\mbox{GeV}$
and $\sqrt{s_0}=11.6\sim11.8~\mbox{GeV}$, respectively.}
\label{fig:7}
\end{figure}

\begin{figure}
\centerline{\epsfysize=5.2truecm
\epsfbox{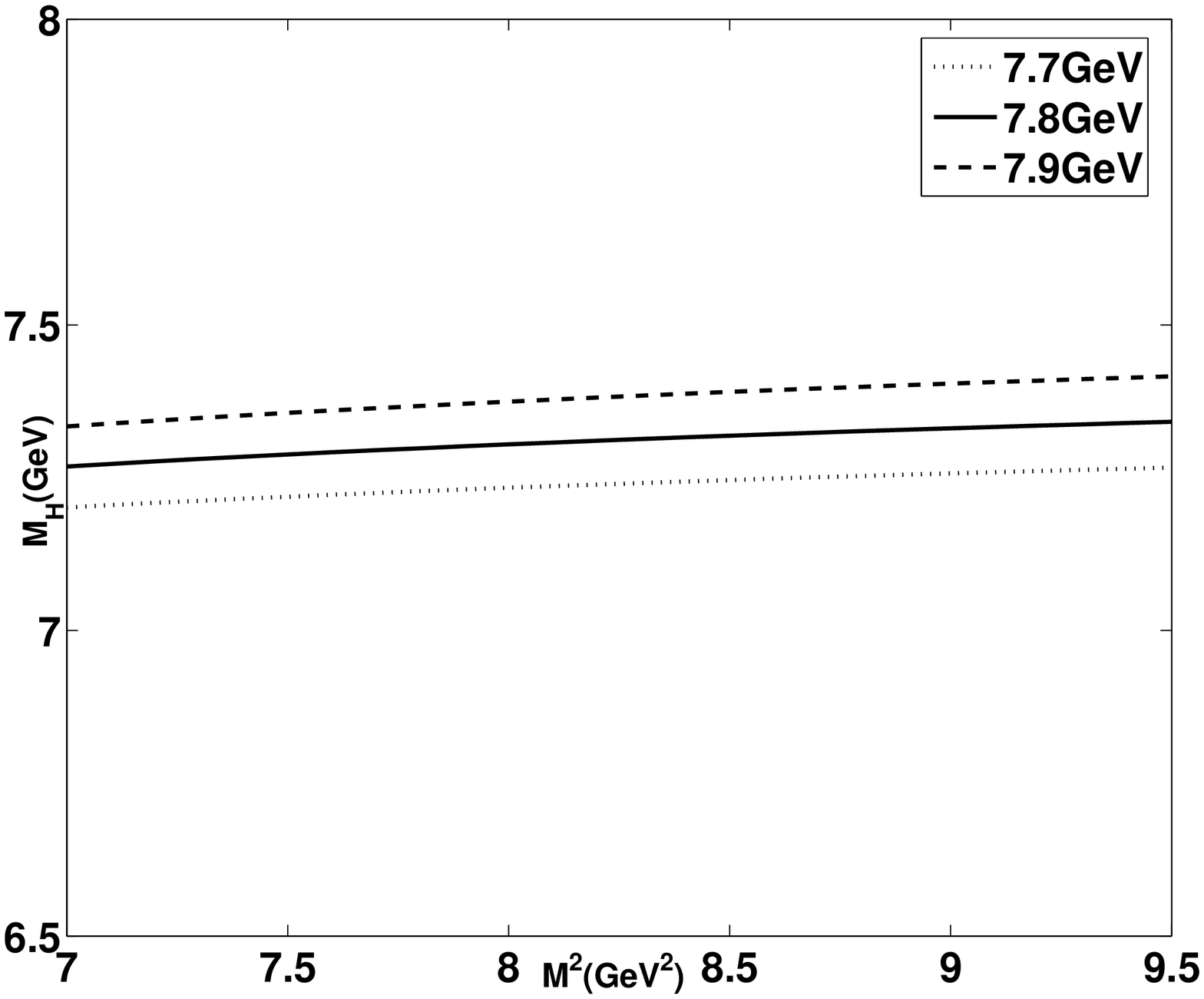}\epsfysize=5.2truecm\epsfbox{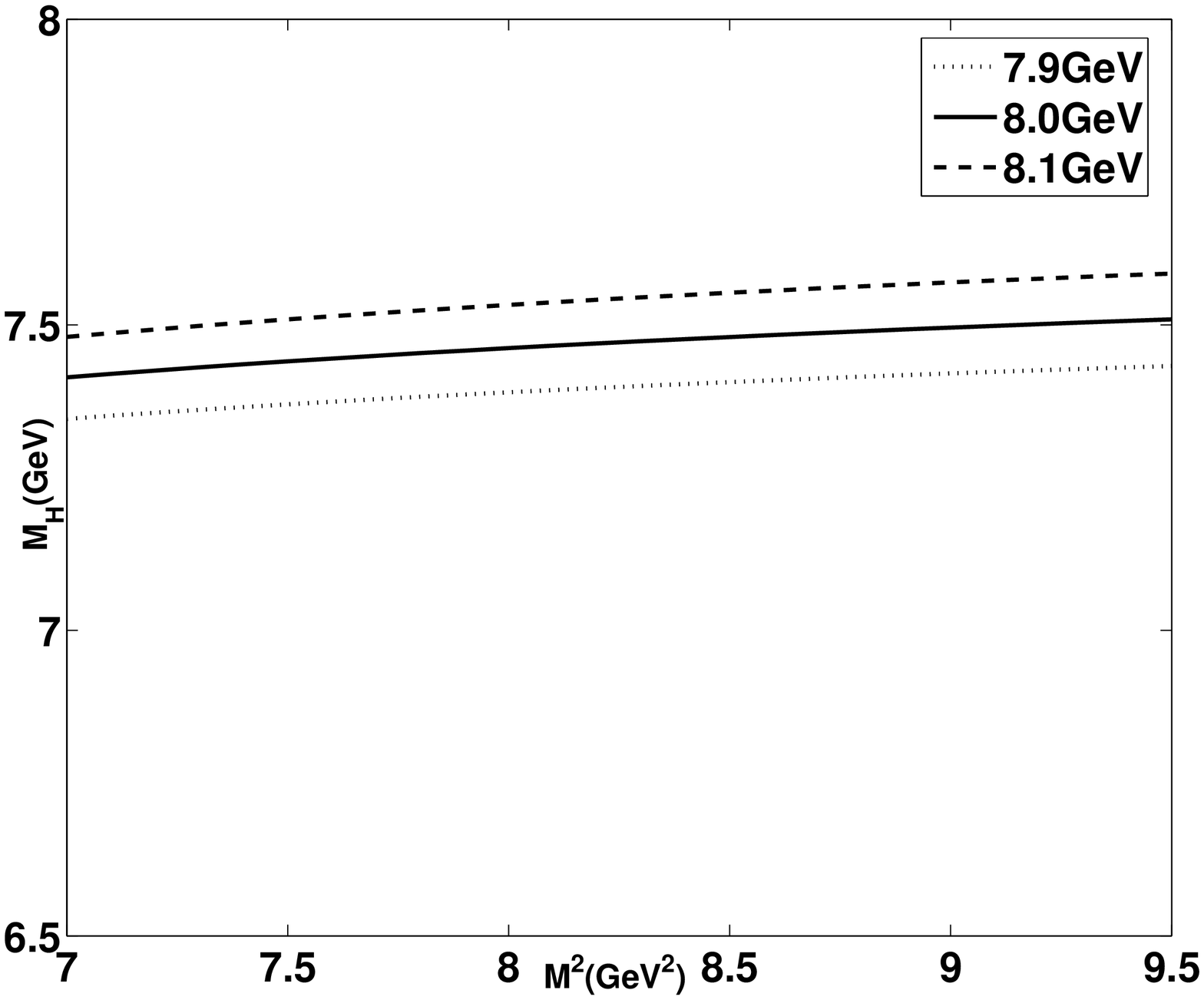}}\caption{The
dependence on $M^2$ for the masses of $B_{s}\bar{D}_{s}$ and
$B_{s}^{*}\bar{D}_{s}^{*}$ from sum rule (\ref{sum rule}). The
continuum thresholds are taken as $\sqrt{s_0}=7.7\sim7.9~\mbox{GeV}$
and $\sqrt{s_0}=7.9\sim8.1~\mbox{GeV}$, respectively.} \label{fig:8}
\end{figure}

\begin{figure}
\centerline{\epsfysize=5.2truecm
\epsfbox{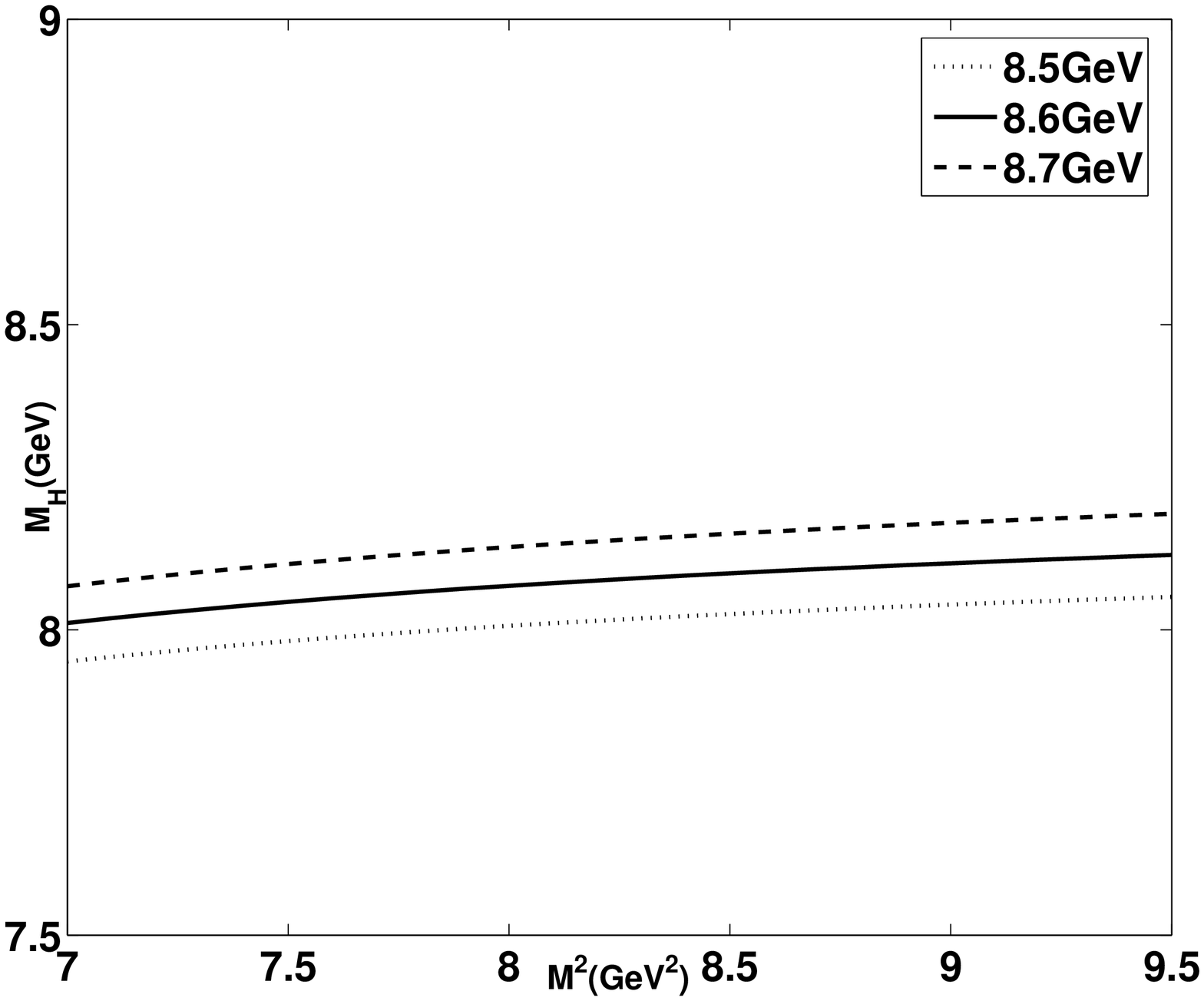}\epsfysize=5.2truecm\epsfbox{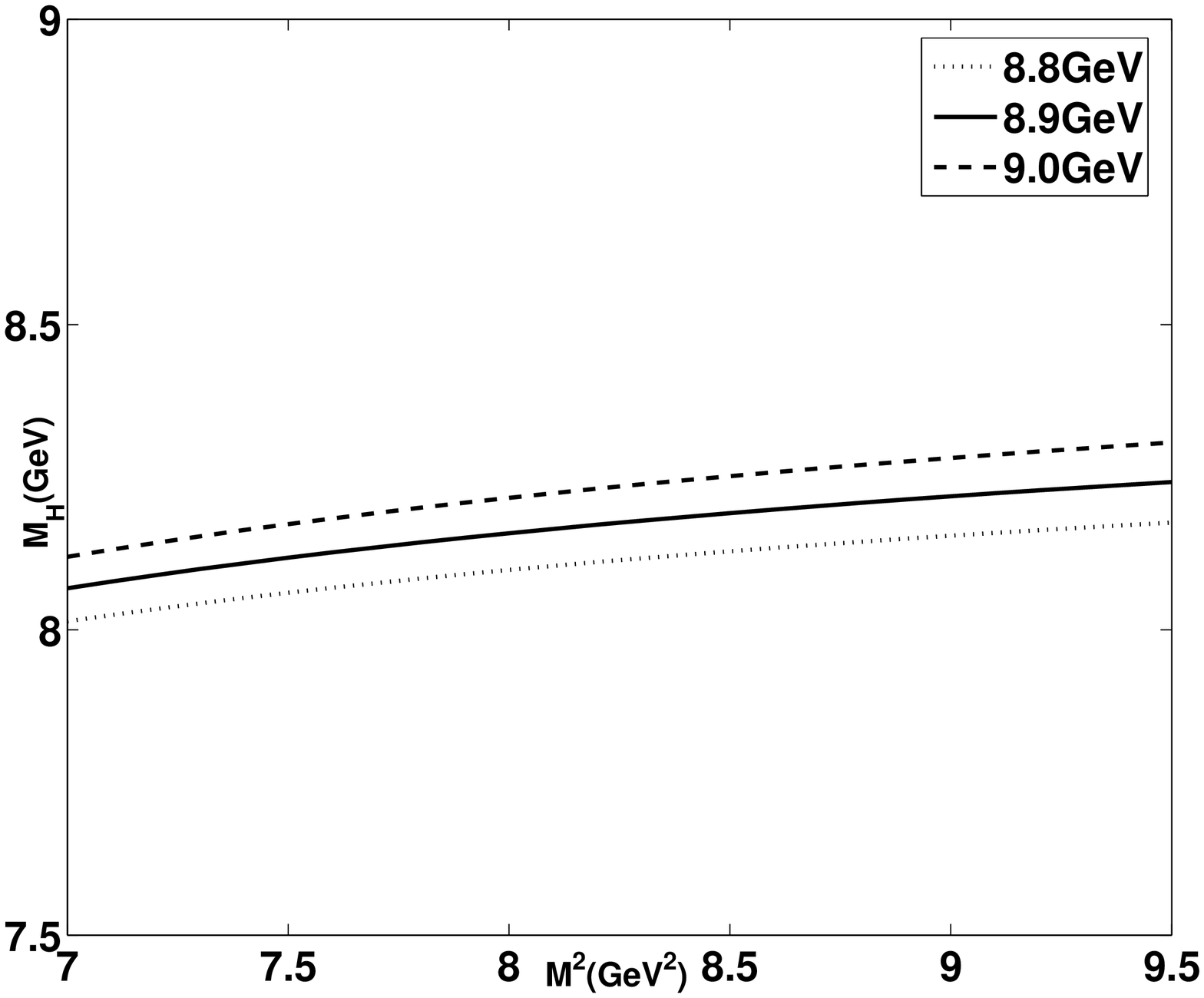}}\caption{The
dependence on $M^2$ for the masses of $B_{s0}^{*}\bar{D}_{s0}^{*}$
and $B_{s1}\bar{D}_{s1}$ from sum rule (\ref{sum rule}). The
continuum thresholds are taken as $\sqrt{s_0}=8.5\sim8.7~\mbox{GeV}$
and $\sqrt{s_0}=8.8\sim9.0~\mbox{GeV}$, respectively.} \label{fig:9}
\end{figure}

\begin{figure}
\centerline{\epsfysize=5.2truecm
\epsfbox{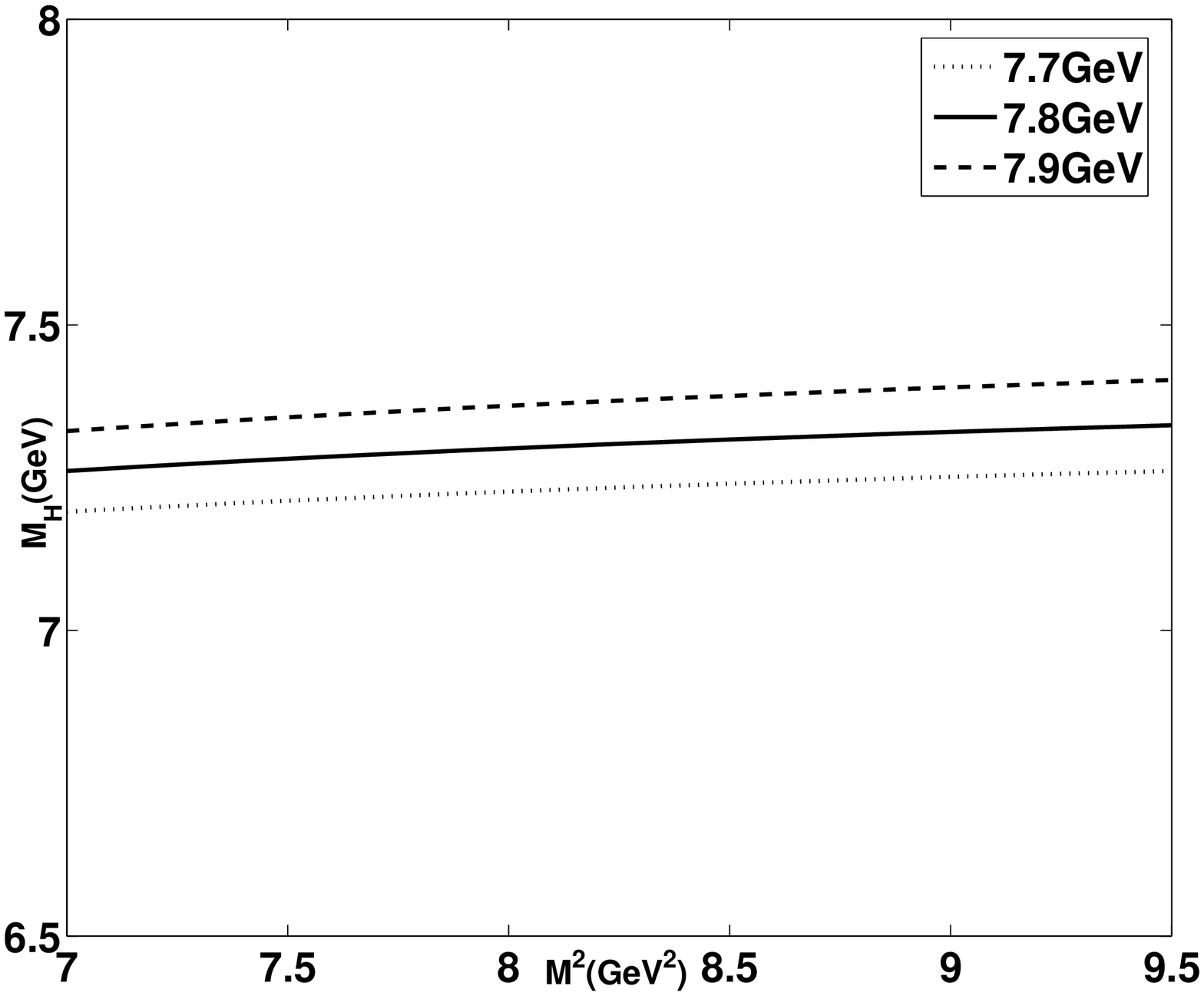}\epsfysize=5.2truecm\epsfbox{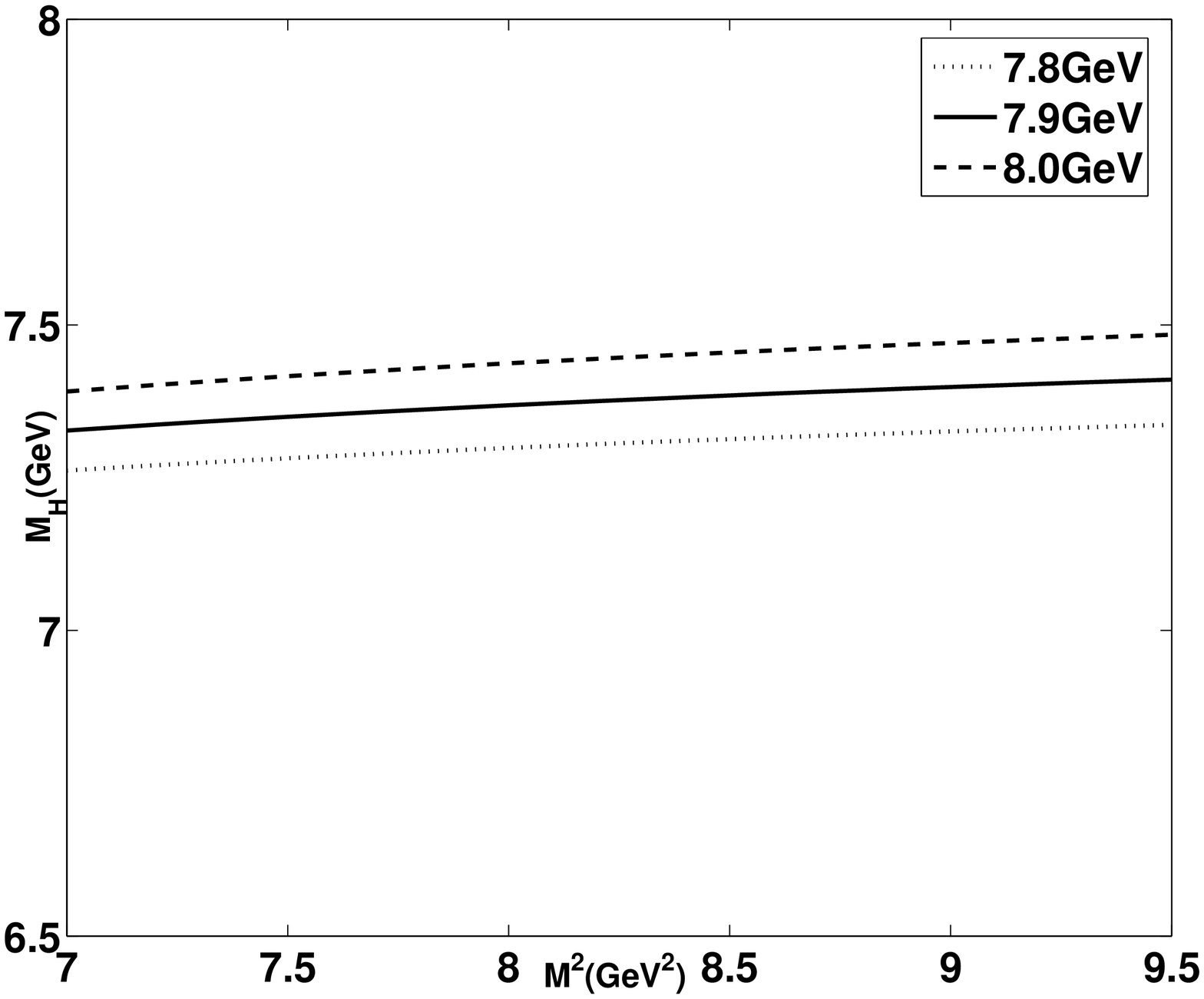}}\caption{The
dependence on $M^2$ for the masses of $D_{s}^{*}\bar{B}_{s}$ and
$B_{s}^{*}\bar{D}_{s}$
 from sum rule (\ref{sum rule 1}). The continuum
thresholds are taken as $\sqrt{s_0}=7.7\sim7.9~\mbox{GeV}$ and
$\sqrt{s_0}=7.8\sim8.0~\mbox{GeV}$, respectively.} \label{fig:10}
\end{figure}

\begin{figure}
\centerline{\epsfysize=5.2truecm
\epsfbox{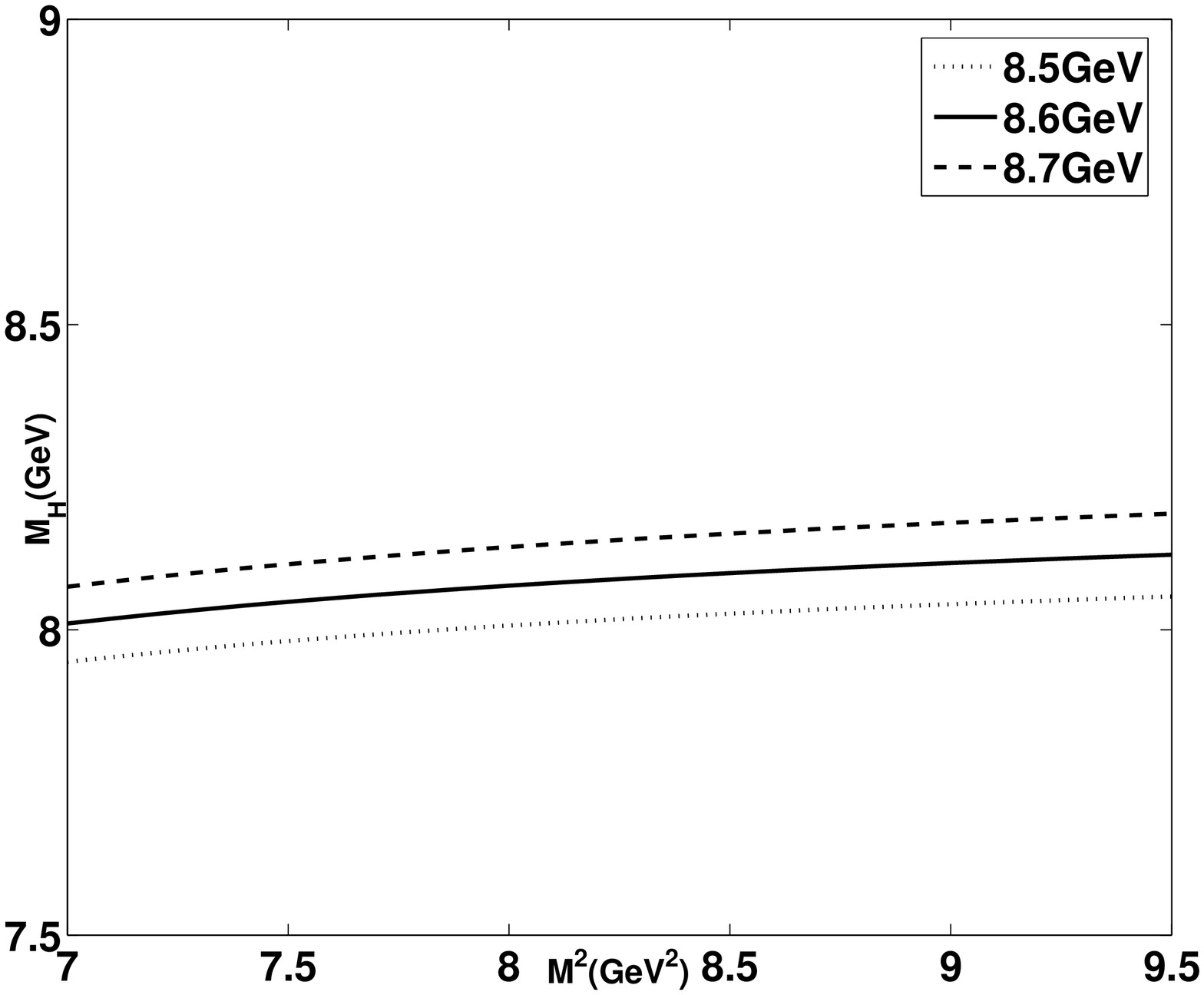}\epsfysize=5.2truecm\epsfbox{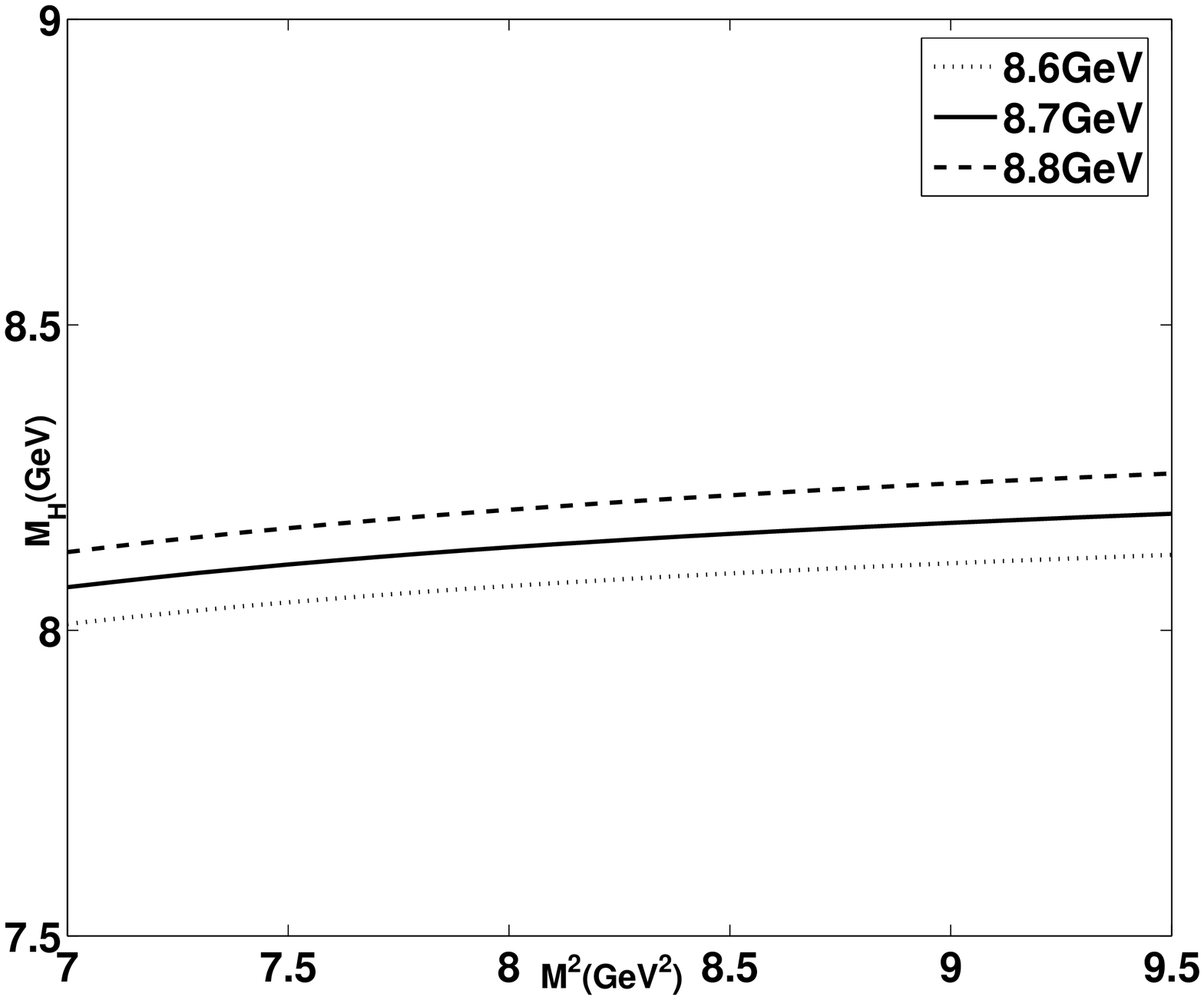}}\caption{The
dependence on $M^2$ for the masses of $D_{s1}\bar{B}_{s0}^{*}$ and
$B_{s1}\bar{D}_{s0}^{*}$ from sum rule (\ref{sum rule 1}). The
continuum thresholds are taken as $\sqrt{s_0}=8.5\sim8.7~\mbox{GeV}$
and $\sqrt{s_0}=8.6\sim8.8~\mbox{GeV}$, respectively.}
\label{fig:11}
\end{figure}

\begin{figure}
\centerline{\epsfysize=5.2truecm
\epsfbox{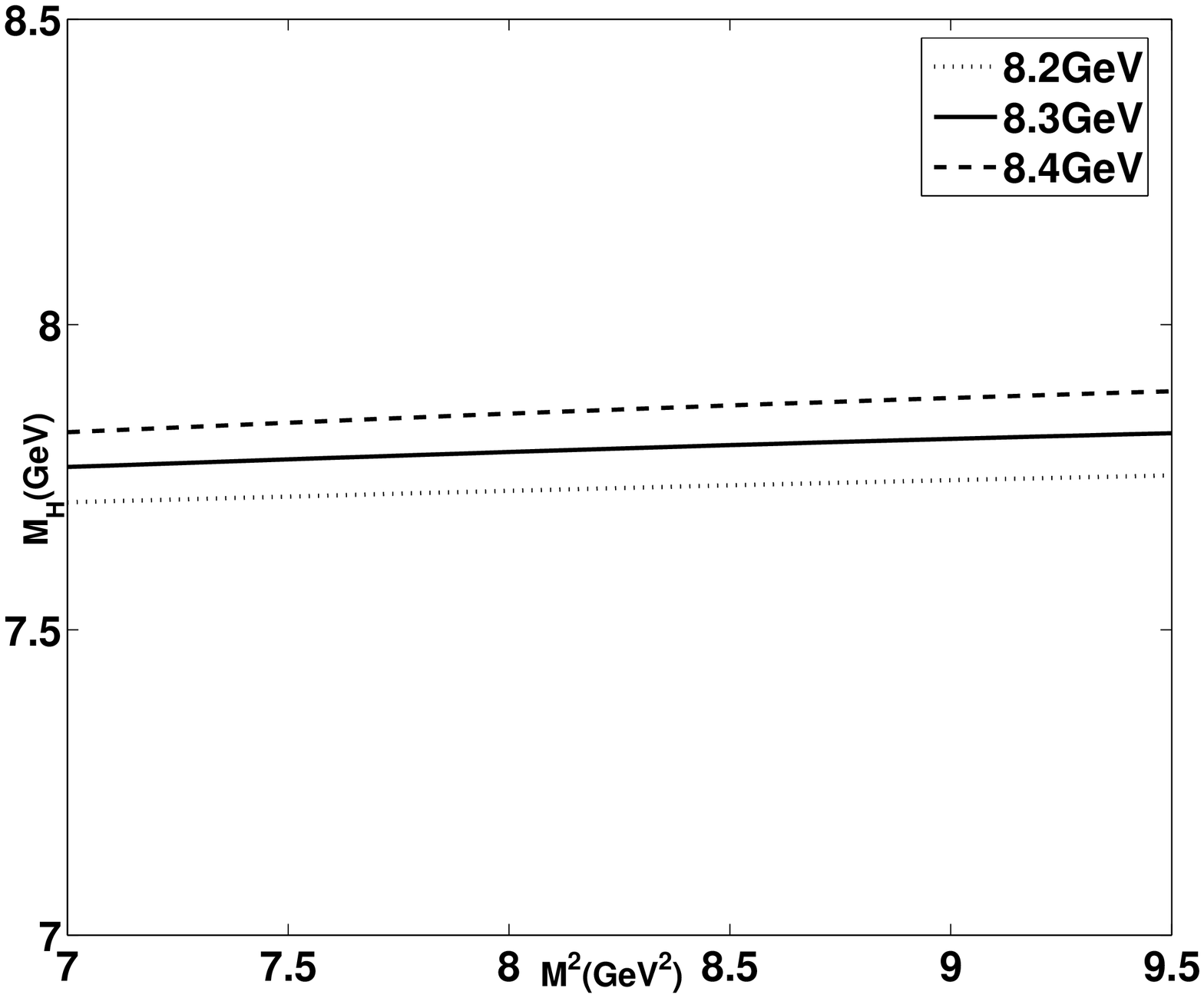}\epsfysize=5.2truecm\epsfbox{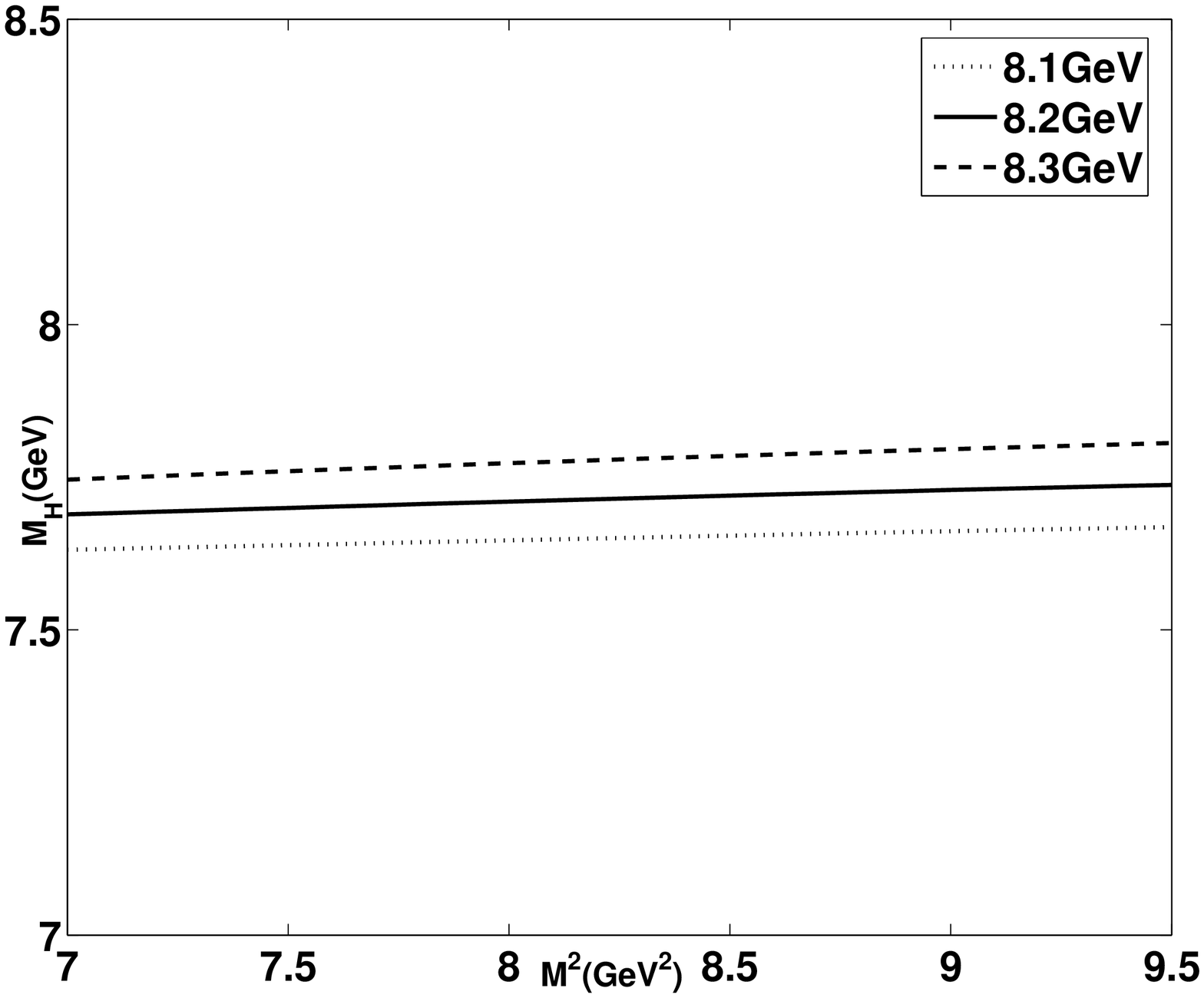}}\caption{The
dependence on $M^2$ for the masses of $D_{s}^{*}\bar{B}_{s0}^{*}$
and $B_{s}^{*}\bar{D}_{s0}^{*}$ from sum rule (\ref{sum rule 1}).
The continuum thresholds are taken as
$\sqrt{s_0}=8.2\sim8.4~\mbox{GeV}$ and
$\sqrt{s_0}=8.1\sim8.3~\mbox{GeV}$, respectively.} \label{fig:12}
\end{figure}

\begin{figure}
\centerline{\epsfysize=5.2truecm
\epsfbox{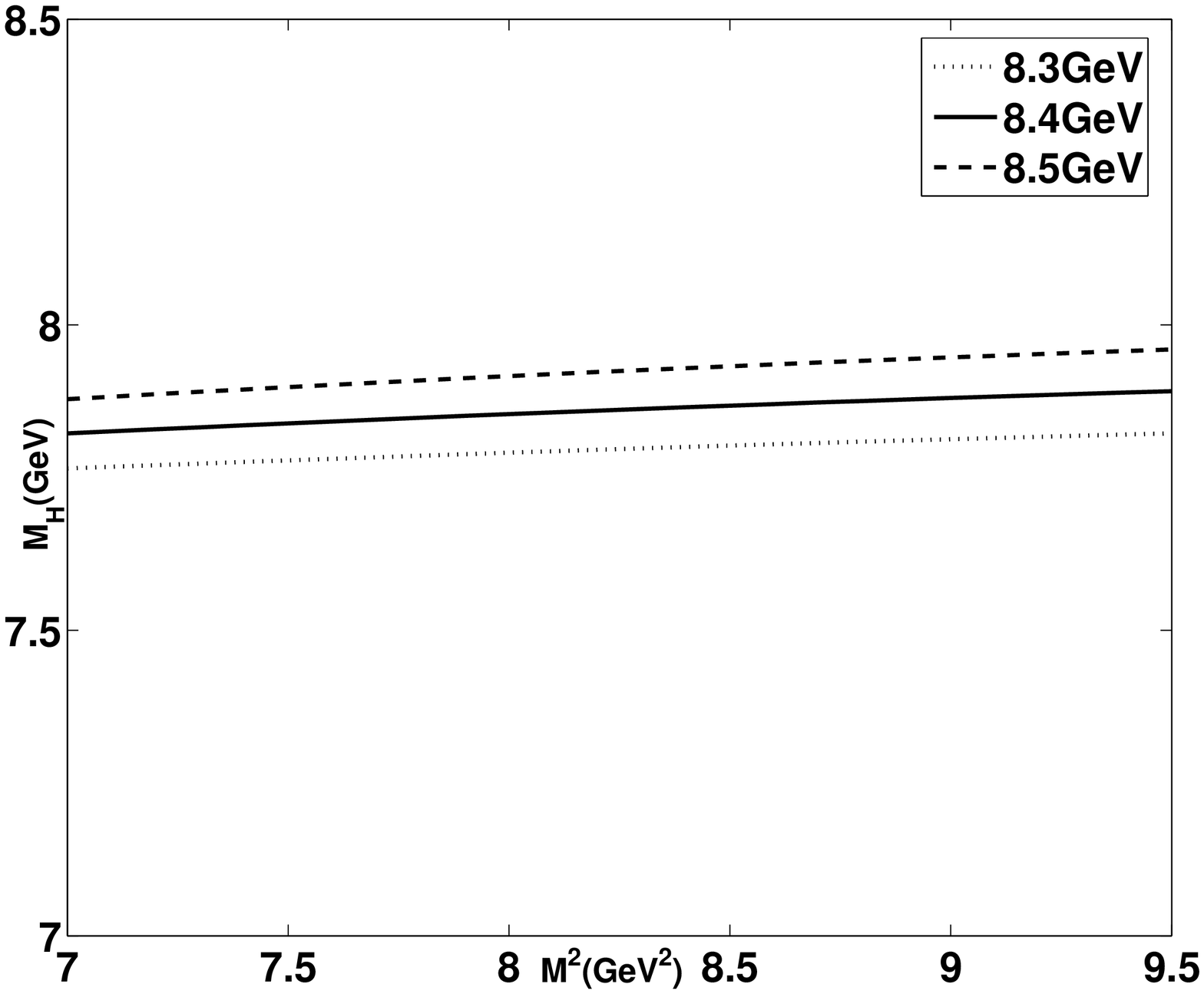}\epsfysize=5.2truecm\epsfbox{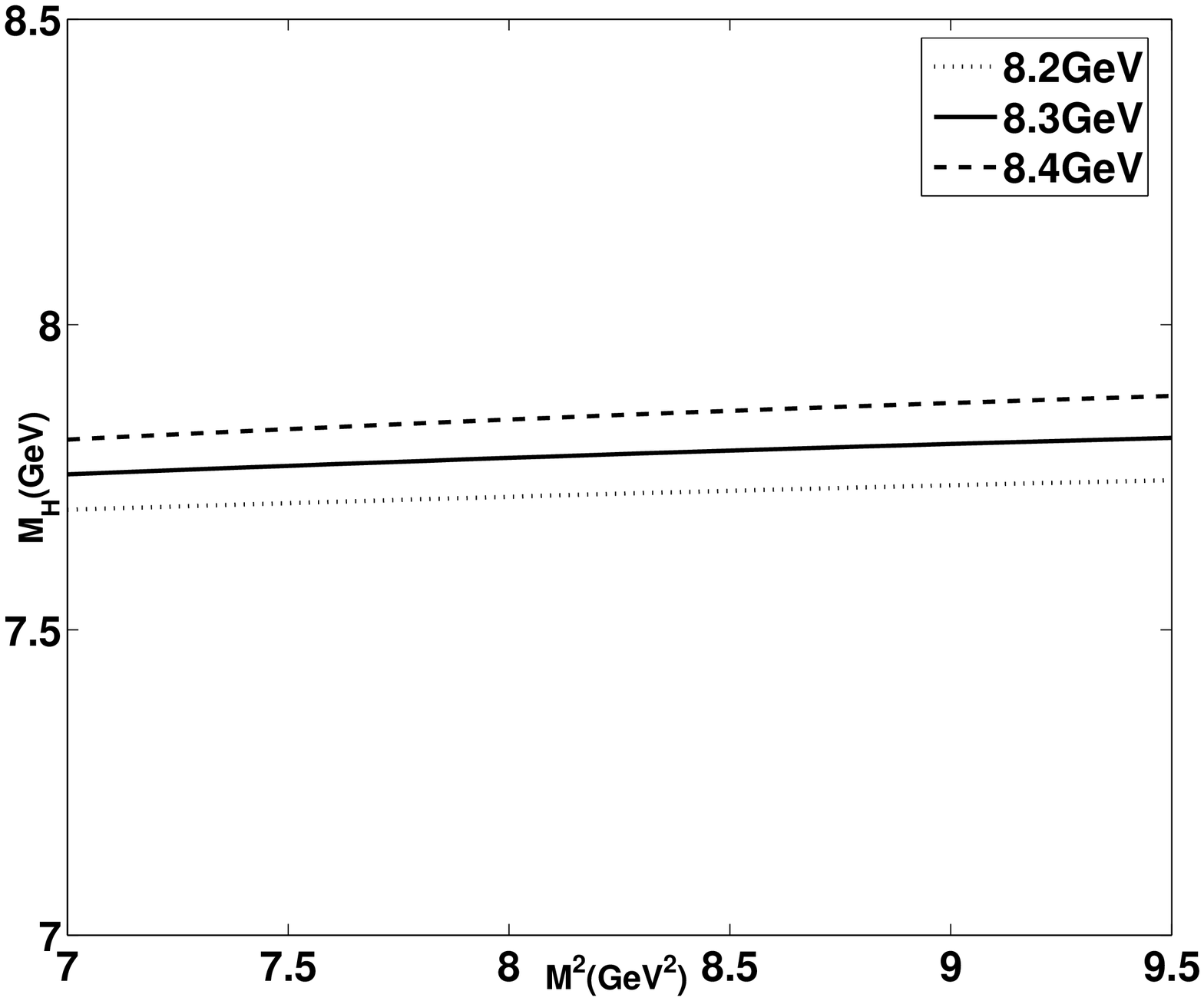}}\caption{The
dependence on $M^2$ for the masses of $D_{s}^{*}\bar{B}_{s1}$ and
$B_{s}^{*}\bar{D}_{s1}$ from sum rule (\ref{sum rule}). The
continuum thresholds are taken as $\sqrt{s_0}=8.3\sim8.5~\mbox{GeV}$
and $\sqrt{s_0}=8.2\sim8.4~\mbox{GeV}$, respectively.}
\label{fig:13}
\end{figure}

\begin{figure}
\centerline{\epsfysize=5.2truecm
\epsfbox{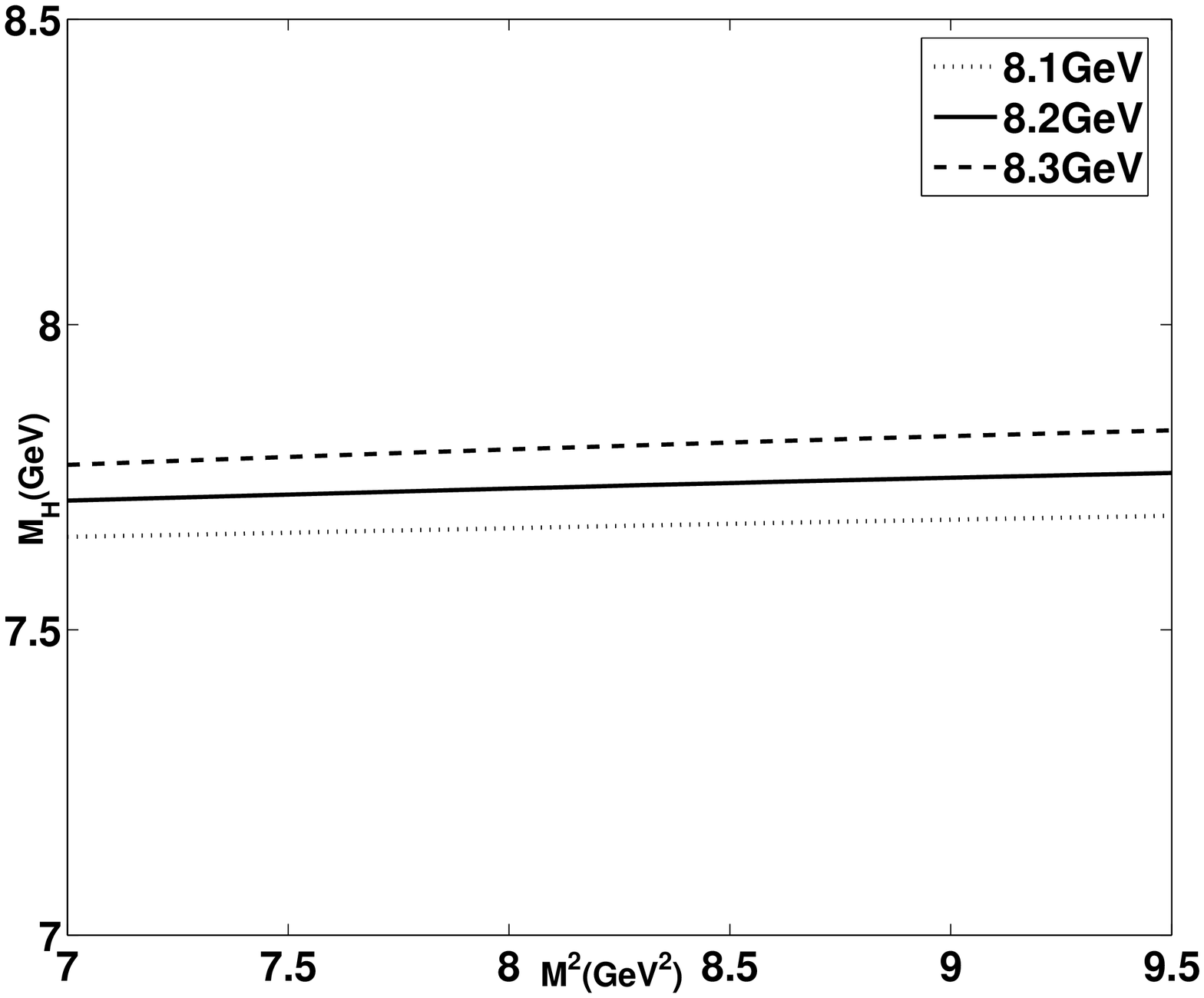}\epsfysize=5.2truecm\epsfbox{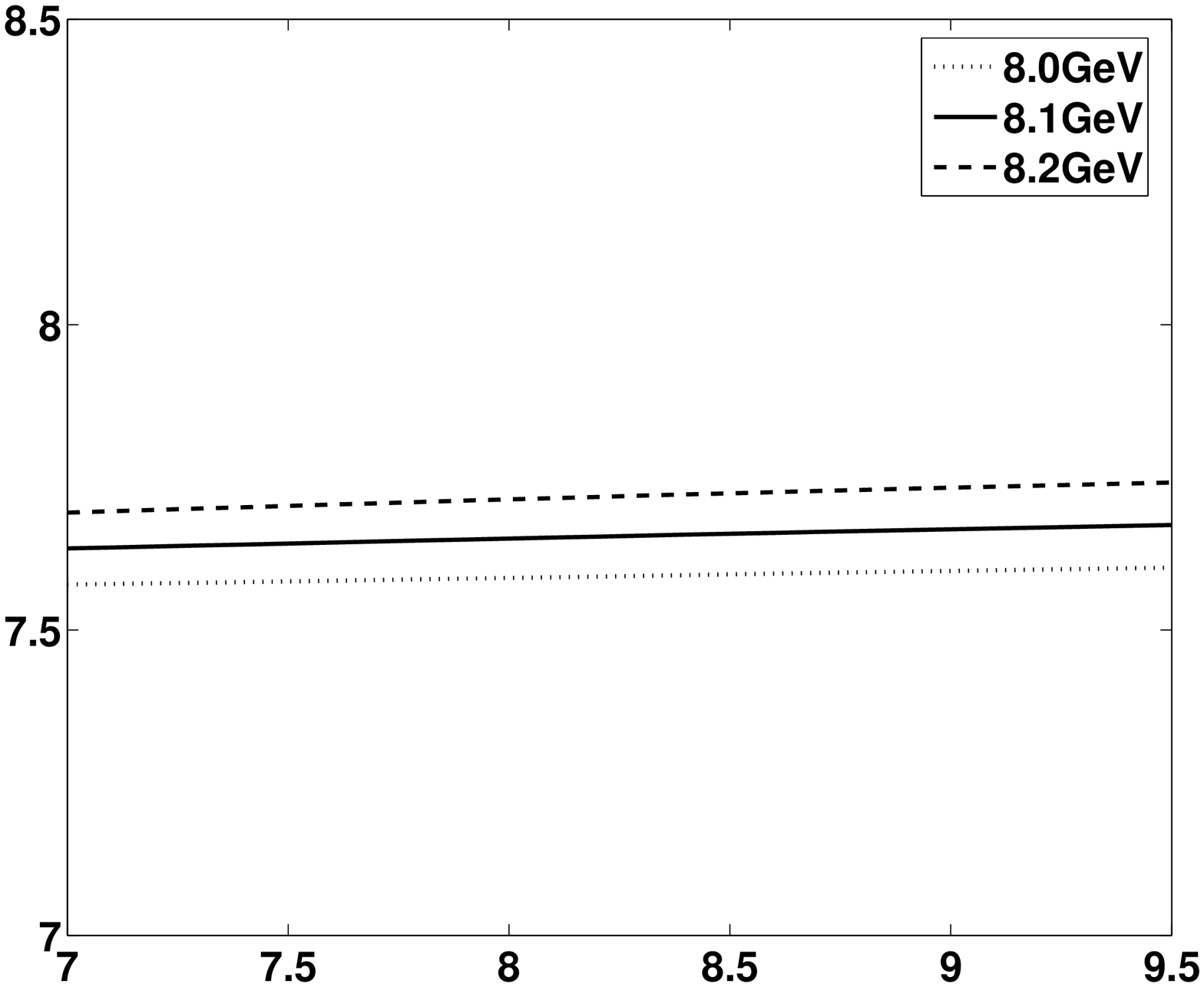}}\caption{The
dependence on $M^2$ for the masses of $D_{s}\bar{B}_{s0}^{*}$ and
$B_{s}\bar{D}_{s0}^{*}$ from sum rule (\ref{sum rule}). The
continuum thresholds are taken as $\sqrt{s_0}=8.1\sim8.3~\mbox{GeV}$
and $\sqrt{s_0}=8.0\sim8.2~\mbox{GeV}$, respectively.}
\label{fig:14}
\end{figure}

\begin{figure}
\centerline{\epsfysize=5.2truecm
\epsfbox{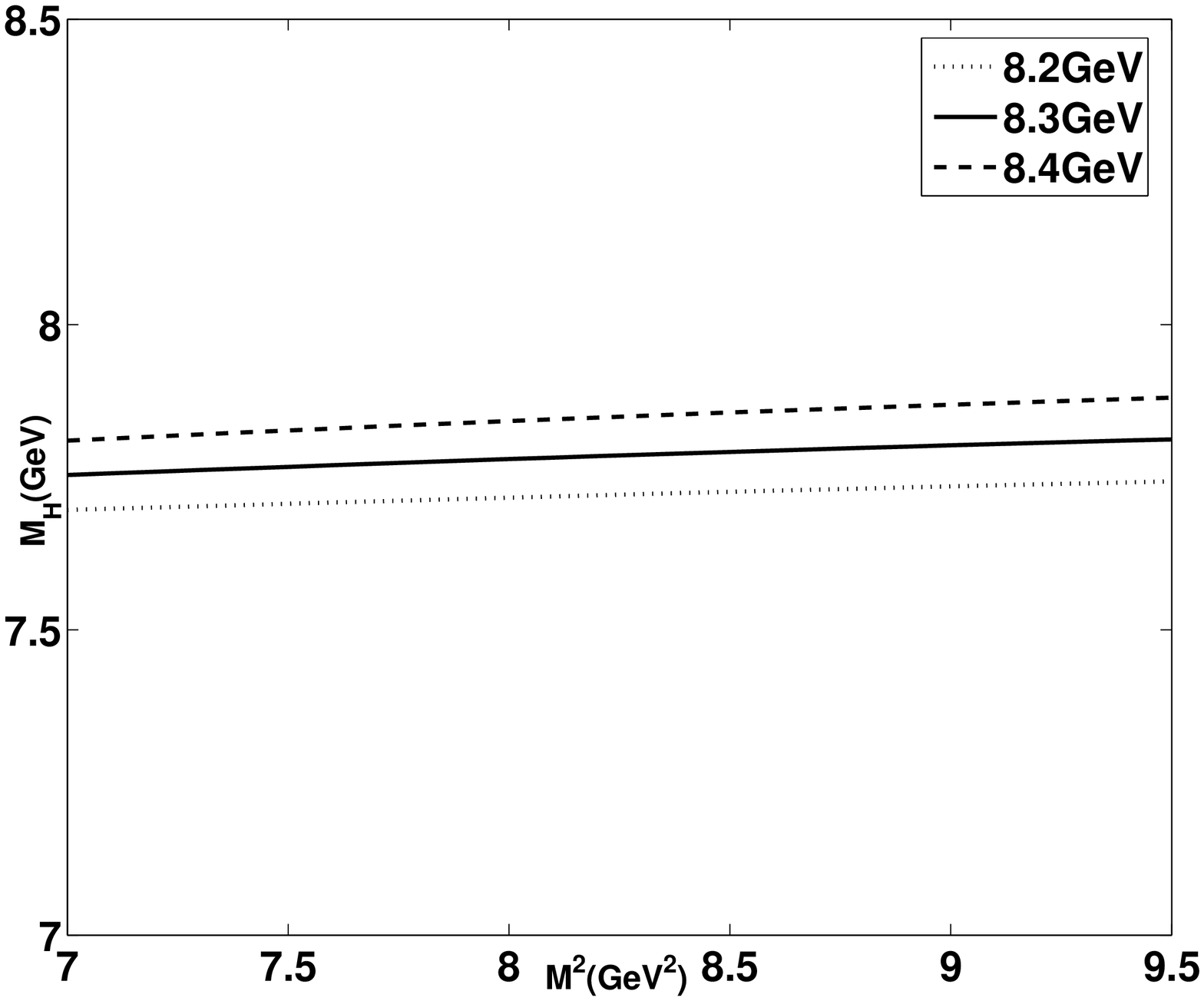}\epsfysize=5.2truecm\epsfbox{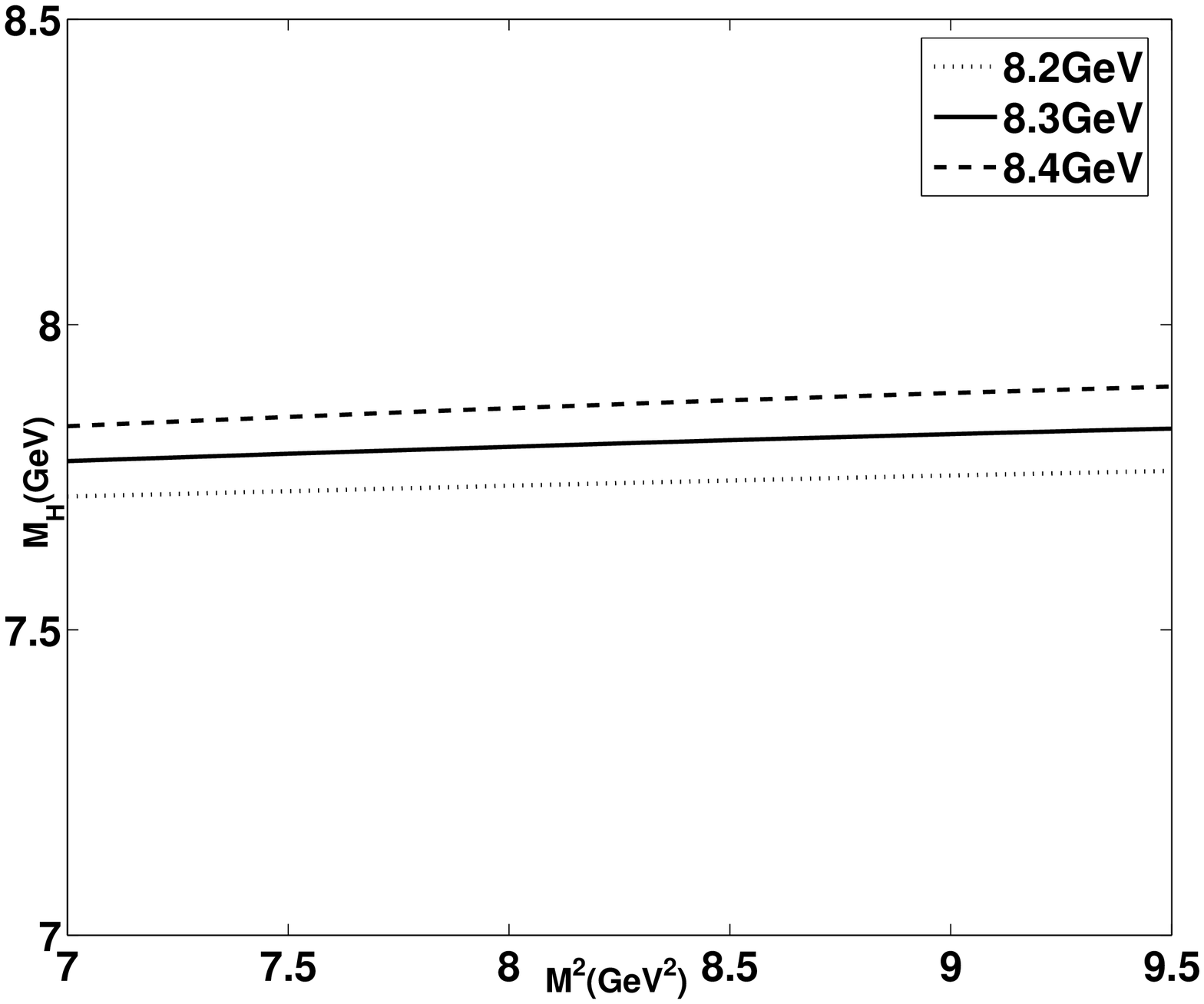}}\caption{The
dependence on $M^2$ for the masses of $D_{s1}\bar{B}_{s}$ and
$B_{s1}\bar{D}_{s}$ from sum rule (\ref{sum rule 1}). The continuum
thresholds are taken as $\sqrt{s_0}=8.2\sim8.4~\mbox{GeV}$ and
$\sqrt{s_0}=8.2\sim8.4~\mbox{GeV}$, respectively.} \label{fig:15}
\end{figure}

\begin{table}[htb!]\caption{ The mass spectra of molecular states with same heavy quarks.}
 \centerline{\begin{tabular}{c c c c c c}  \hline\hline
Hadron                            & configuration                                &     mass (GeV)                      &  Hadron                            & configuration                                &     mass (GeV)                   \\
\hline
$D_{s}\bar{D}_{s}$                &$(c\bar{s})(\bar{c}s)$                        & $3.91\pm0.10$ \cite{zhang}          &$B_{s}\bar{B}_{s}$                  &$(b\bar{s})(\bar{b}s)$                        & $10.70\pm0.10$   \cite{zhang}    \\
\hline
$D_{s}^{*}\bar{D}_{s}$            &$(c\bar{s})^{*}(\bar{c}s)$                    & $4.01\pm0.10$ \cite{zhang}          &$B_{s}^{*}\bar{B}_{s}$              &$(b\bar{s})^{*}(\bar{b}s)$                    & $10.71\pm0.11$  \cite{zhang}     \\
\hline
$D_{s}^{*}\bar{D}_{s}^{*}$        &$(c\bar{s})^{*}(\bar{c}s)^{*}$                & $4.13\pm0.10$ \cite{zhang}          &$B_{s}^{*}\bar{B}_{s}^{*}$          &$(b\bar{s})^{*}(\bar{b}s)^{*}$                & $10.80\pm0.10$  \cite{zhang}     \\
\hline
$D_{s0}^{*}\bar{D}_{s0}^{*}$      &$(c\bar{s})_{0}^{*}(\bar{c}s)_{0}^{*}$        & $4.58\pm0.10$                       &$B_{s0}^{*}\bar{B}_{s0}^{*}$        &$(b\bar{s})_{0}^{*}(\bar{b}s)_{0}^{*}$        & $11.35\pm0.09$                   \\
\hline
$D_{s1}\bar{D}_{s0}^{*}$          &$(c\bar{s})_{1}(\bar{c}s)_{0}^{*}$            & $4.64\pm0.10$                       &$B_{s1}\bar{B}_{s0}^{*}$            &$(b\bar{s})_{1}(\bar{b}s)_{0}^{*}$            & $11.38\pm0.09$                   \\
\hline
$D_{s1}\bar{D}_{s1}$              &$(c\bar{s})_{1}(\bar{c}s)_{1}$                & $4.66\pm0.12$                       &$B_{s1}\bar{B}_{s1}$                &$(b\bar{s})_{1}(\bar{b}s)_{1}$                & $11.39\pm0.13$                   \\
\hline
$D_{s}\bar{D}_{s0}^{*}$           &$(c\bar{s})(\bar{c}s)_{0}^{*}$                & $4.24\pm0.08$                       &$B_{s}\bar{B}_{s0}^{*}$             &$(b\bar{s})(\bar{b}s)_{0}^{*}$                & $11.06\pm0.10$                   \\
\hline
$D_{s1}\bar{D}_{s}$               &$(c\bar{s})_{1}(\bar{c}s)$                    & $4.37\pm0.08$                       &$B_{s1}\bar{B}_{s}$                 &$(b\bar{s})_{1}(\bar{b}s)$                    & $11.10\pm0.10$                   \\
 \hline
$D_{s}^{*}\bar{D}_{s0}^{*}$       &$(c\bar{s})^{*}(\bar{c}s)_{0}^{*}$            & $4.36\pm0.08$                       &$B_{s}^{*}\bar{B}_{s0}^{*}$         &$(b\bar{s})^{*}(\bar{b}s)_{0}^{*}$            & $11.09\pm0.10$                   \\
\hline
$D_{s}^{*}\bar{D}_{s1}$           &$(c\bar{s})^{*}(\bar{c}s)_{1}$                & $4.43\pm0.09$                       &$B_{s}^{*}\bar{B}_{s1}$             &$(b\bar{s})^{*}(\bar{b}s)_{1}$                & $11.10\pm0.10$                   \\
\hline \hline
\end{tabular}} \label{table:1}
\end{table}

\begin{table}[htb!]\caption{ The mass spectra of molecular states with differently heavy quarks.}
 \centerline{\begin{tabular}{c c c c c c}  \hline\hline
Hadron                            & configuration                                &     mass (GeV)                      &  Hadron                            & configuration                                &     mass (GeV)                   \\
$B_{s}\bar{D}_{s}$                &$(b\bar{s})(\bar{c}s)$                        & $7.31\pm0.09$                       &$B_{s}^{*}\bar{D}_{s0}^{*}$         &$(b\bar{s})^{*}(\bar{c}s)_{0}^{*}$            & $7.71\pm0.07$                    \\
\hline
$B_{s}^{*}\bar{D}_{s}$            &$(b\bar{s})^{*}(\bar{c}s)$                    & $7.37\pm0.09$                       &$B_{s}^{*}\bar{D}_{s1}$             &$(b\bar{s})^{*}(\bar{c}s)_{1}$                & $7.78\pm0.08$                    \\
\hline
$B_{s}^{*}\bar{D}_{s}^{*}$        &$(b\bar{s})^{*}(\bar{c}s)^{*}$                & $7.46\pm0.09$                       &$D_{s}^{*}\bar{B}_{s}$              &$(c\bar{s})^{*}(\bar{b}s)$                    & $7.30\pm0.09$                    \\
\hline
$B_{s0}^{*}\bar{D}_{s0}^{*}$      &$(b\bar{s})_{0}^{*}(\bar{c}s)_{0}^{*}$        & $8.07\pm0.09$                       &$D_{s1}\bar{B}_{s0}^{*}$            &$(c\bar{s})_{1}(\bar{b}s)_{0}^{*}$            & $8.07\pm0.09$                    \\
\hline
$B_{s1}\bar{D}_{s0}^{*}$          &$(b\bar{s})_{1}(\bar{c}s)_{0}^{*}$            & $8.14\pm0.09$                       &$D_{s}\bar{B}_{s0}^{*}$             &$(c\bar{s})(\bar{b}s)_{0}^{*}$                & $7.73\pm0.07$                    \\
\hline
$B_{s1}\bar{D}_{s1}$              &$(b\bar{s})_{1}(\bar{c}s)_{1}$                & $8.17\pm0.11$                       &$D_{s1}\bar{B}_{s}$                 &$(c\bar{s})_{1}(\bar{b}s)$                    & $7.78\pm0.08$                    \\
\hline
$B_{s}\bar{D}_{s0}^{*}$           &$(b\bar{s})(\bar{c}s)_{0}^{*}$                & $7.65\pm0.07$                       &$D_{s}^{*}\bar{B}_{s0}^{*}$         &$(c\bar{s})^{*}(\bar{b}s)_{0}^{*}$            & $7.79\pm0.08$                    \\
\hline
$B_{s1}\bar{D}_{s}$               &$(b\bar{s})_{1}(\bar{c}s)$                    & $7.80\pm0.08$                       &$D_{s}^{*}\bar{B}_{s1}$             &$(c\bar{s})^{*}(\bar{b}s)_{1}$                & $7.86\pm0.08$                    \\
\hline\hline
\end{tabular}} \label{table:2}
\end{table}

\end{document}